\documentclass[twocolumn]{aastex631}
\usepackage{verbatim}
\usepackage[utf8]{inputenc}
\usepackage{cellspace}
\usepackage{amssymb}
\usepackage{subfigure}
\usepackage{amsmath}

\newcommand\OIII{{\sc [O\thinspace iii]}}

\newcommand{\fesc}{${f_{\rm{esc}}^{\rm LyC}}$}

\newcommand{\mycomment}[1]{}
\newcommand{\alphared}{$\alpha_{\rm red}$}
\newcommand{\alphablue}{$\alpha_{\rm blue}$}

\begin{document}

\title{Power-law Emission-line Wings and Radiation-Driven Superwinds in Local Lyman Continuum Emitters}

\author[0000-0002-5235-7971]{Lena Komarova}
\affiliation{Departament d’Astronomia i Astrofìsica, Universitat de València, C. Dr. Moliner 50, E-46100 Burjassot, València, Spain}

\author[0000-0002-5808-1320]{M. S. Oey}
\affiliation{Astronomy Department, University of Michigan, 1085 South University Avenue, Ann Arbor, MI 48109, USA}

\author[0000-0001-8442-1846]{Rui Marques-Chaves}
\affiliation{Department of Astronomy, University of Geneva, 51 Chemin Pegasi, 1290 Versoix, Switzerland}

\author[0000-0001-5758-1000]{Ricardo Amor\'in}
\affil{Instituto de Astrof\'{i}sica de Andaluc\'{i}a (CSIC), Apartado 3004, 18080 Granada, Spain}

\author[0000-0002-6586-4446]{Alaina Henry}
\affiliation{Space Telescope Science Institute, 3700 San Martin Drive, Baltimore, MD, 21218, USA}

\author[0000-0001-7144-7182]{Daniel Schaerer}
\affiliation{Department of Astronomy, University of Geneva, 51 Chemin Pegasi, 1290 Versoix, Switzerland}

\author[0000-0001-8419-3062]{Alberto Saldana-Lopez}
\affiliation{Department of Astronomy, Oskar Klein Centre, Stockholm University, AlbaNova University Centre, SE-106 91 Stockholm, Sweden}

\author[0000-0003-1767-6421]{Alexandra Le Reste}
\affiliation{Minnesota Institute for Astrophysics, School of Physics and Astronomy, University of Minnesota, 316 Church str SE, Minneapolis, MN 55455,USA}

\author[0000-0002-9136-8876]{Claudia Scarlata}
\affiliation{Minnesota Institute for Astrophysics, School of Physics and Astronomy, University of Minnesota, 316 Church str SE, Minneapolis, MN 55455,USA}

\author[0000-0001-8587-218X]{Matthew J. Hayes}
\affiliation{Department of Astronomy, Oskar Klein Centre, Stockholm University, AlbaNova University Centre, SE-106 91 Stockholm, Sweden}

\author[0000-0003-2722-8841]{Omkar Bait}
\affiliation{SKA Observatory, Jodrell Bank, Lower Withington, Macclesfield, SK11 9FT, UK}

\author[0000-0002-2724-8298]{Sanchayeeta Borthakur}
\affiliation{School of Earth \& Space Exploration, Arizona State University, Tempe, AZ 85287, USA}

\author[0000-0003-4166-2855]{Cody Carr}
\affiliation{Center for Cosmology and Computational Astrophysics, Institute for Advanced Study in Physics
Zhejiang University, Hangzhou 310058, China}

\author[0000-0002-0302-2577]{John Chisholm}
\affiliation{Department of Astronomy, University of Texas at Austin, Austin, TX 78712, United States}

\author[0000-0001-7113-2738]{Harry C. Ferguson}
\affiliation{Space Telescope Science Institute, 3700 San Martin Drive, Baltimore, MD, 21218, USA}

\author{Vital Gutierrez Fernandez}
\affiliation{Astronomy Department, University of Michigan, 1085 South University Avenue, Ann Arbor, MI 48109, USA}

\author[0000-0002-2129-0292]{Brian Fleming}
\affiliation{Laboratory for Atmospheric and Space Physics, Boulder, Colorado, United States}

\author[0000-0002-0159-2613]{Sophia R. Flury}
\affiliation{Institute for Astronomy, University of Edinburgh, Royal Observatory, Edinburgh, EH9 3HJ, UK}

\author[0000-0002-7831-8751]{Mauro Giavalisco}
\affiliation{Department of Astronomy, University of Massachusetts Amherst, Amherst, MA 01002, United States}

\author{Andrea Grazian}
\affiliation{INAF-Osservatorio Astronomico di Padova, Vicolo dell’Osservatorio 5, I-35122, Padova, Italy}

\author[0000-0001-6670-6370]{Timothy Heckman}
\affiliation{Department of Physics and Astronomy, Johns Hopkins University, Baltimore, MD 21218, USA}

\author[0000-0002-6790-5125]{Anne E. Jaskot}
\affiliation{Astronomy Department, Williams College, Williamstown,
MA 01267, USA}

\author[0000-0001-7673-2257]{Zhiyuan Ji}
\affiliation{Department of Astronomy, University of Massachusetts Amherst, Amherst, MA 01002, United States}

\author[0000-0002-3005-1349]{Göran Östlin}
\affiliation{Department of Astronomy, Oskar Klein Centre, Stockholm University, AlbaNova University Centre, SE-106 91 Stockholm, Sweden}

\author[0000-0001-8940-6768]{Laura Pentericci}
\affiliation{INAF, Osservatorio Astronomico di Roma, via Frascati 33, I-00078 Monteporzio Catone, Italy}

\author{Swara Ravindranath}
\affiliation{Space Telescope Science Institute, 3700 San Martin Drive, Baltimore, MD, 21218, USA}

\author{Trinh Thuan}
\affiliation{Astronomy Department, University of Virginia, Charlottesville, VA 22904, USA}

\author[0000-0001-7299-8373]{Jose M. V\'ilchez}
\affiliation{Instituto de Astrof\'{i}sica de Andaluc\'{i}a (CSIC), Apartado 3004, 18080 Granada, Spain}

\author[0000-0003-0960-3580]{Gabor Worseck}
\affiliation{Institut f\"ur Physik und Astronomie, Universit\"at Potsdam, Karl-Liebknecht-Str. 24/25, D-14476 Potsdam, Germany}

\author[0000-0002-9217-7051]{Xinfeng Xu}
\affiliation{Department of Physics and Astronomy, Northwestern University, 2145 Sheridan Road, Evanston, IL, 60208, USA}


\shorttitle{Feedback in Local LCEs}
\shortauthors{Komarova et al.}

\correspondingauthor{Lena Komarova}
\email{olena.komarova@uv.es}

\begin{abstract}
We investigate broad emission-line wings, reaching $\leq 800\rm~km~s^{-1}$, observed in 26 galaxies with Lyman continuum (LyC) observations, primarily from the Low-redshift Lyman Continuum Survey (LzLCS). 
Using Magellan/MIKE, VLT/X-shooter, and WHT/ISIS high-resolution spectroscopy, we show that this fast gas appears to probe the dominant feedback mechanisms linked to LyC escape. 
We find that in 14 galaxies, the wings are best fit with power laws of slope $\alpha \sim -3.5 \text{ to } -1.6$, with four others best fit by Gaussians of width $\sigma_{\rm BW} \sim 300~\rm km~s^{-1}$; the remaining eight show ambiguous wing morphologies. Gaussian wings are found only at low $O_{32}$ =  $[\rm O~III]\lambda5007/[O~II]\lambda3726,3729$ and high metallicity, while power-law wings span the full range of these parameters. The general evidence suggests a dual-mode paradigm for LyC escape:  radiation-driven superwinds traced by power-law wings and supernova-driven feedback traced by Gaussian wings. For the former, the $<3$ Myr-old, pre-supernova stellar population correlates with more luminous, faster winds.  The data also show that radiation-driven wind parameters like wind luminosity and
power-law slope $\alpha$ depend on the UV luminosity more than the optically thick covering fraction, consistent with ``picket-fence" radiative transfer.  Observed $\alpha$ values flatten with both escaping LyC luminosity and higher extinction, while still preserving the anticorrelation between these two quantities.  Additionally, the differential between red and blue slopes 
implies that extinction and dense gas are centrally concentrated relative to the wind emission. Overall, our results show that power-law emission-line wings probe LyC-driven winds and LyC escape in metal-poor starbursts.
\end{abstract}


\section{Introduction} \label{sec:intro}
The epoch of reionization (EoR) at $z > 6$ marked a key transition in our universe's history from a largely neutral, opaque state to an ionized and transparent one. While the relative contributions of starbursts and active galactic nuclei (AGN) to reionization remain uncertain \citep{Fontanot2012, Duncan2015,  Cristiani2016, Matsuoka2018, Grazian2024}, star-forming galaxies remain a principal agent
\citep{Robertson2013, Robertson2015, Finkelstein2019, Naidu2020, Matthee2022, Mascia2023, Yeh2023, Matthee2023, Atek2024, Simmonds2024a}. However, a critical remaining unknown is the mechanisms that promote the escape of ionizing radiation, or Lyman continuum (LyC), from hydrogen-rich galaxies. On the one hand, photoionization by young massive stars may reduce the neutral hydrogen optical depth, or even result in a density-bounded medium that allows isotropic LyC escape \citep[e.g.,][]{Zackrisson2013, McClymont2024}. 
Traditionally, the primary feedback mechanism associated with LyC leakage is clearing by large-scale winds resulting from the collective feedback of
SNe and stellar winds. Such mechanical
feedback produces a hole-ridden, anisotropic ISM geometry, promoting LyC escape through the optically thin chimneys  \citep{Heckman2011, Zastrow2013, Kimm2014, Ma2015, Reddy2016, Trebitsch2017, Gazagnes2018, Steidel2018, Ma2020, Saldana-Lopez2022, Bait2023}. 

However, at low metallicity, characteristic of the early universe \citep{Curti2023, Nakajima2023, Sanders2023, Chemerynska2024, He2024}, the onset of SNe is expected to be delayed \citep{Heger2003, Sukhbold2016}.  This is because at low metallicity, most of the stars above $20-30 \rm~M_{\odot}$ \citep[e.g.,][]{OConnor2011, Sukhbold2016} fail to explode, and instead directly collapse into black holes \citep[e.g.,][]{Zhang2008, OConnor2011}.\cite{Jecmen2023} show that the majority of the mechanical energy input from SN feedback at $Z \lesssim 0.4~\rm Z_{\odot}$ or $12 + \log(\rm O/H) \lesssim 8.3$ is not released until 10 Myr into a cluster's lifetime. The retention of dense gas, in turn, promotes catastrophic cooling of any hot gas, which 
may suppress adiabatic superwinds in metal-poor starbursts with high gas densities
\citep{Silich2004, Lochhaas2017, Jaskot2017, Jaskot2019, Gray2019, Danehkar2021}. 
Therefore, the mechanism that instead may dominate in the most extreme, young, metal-poor starbursts is radiation-driven 
feedback.  Radiation can drive
superwinds, where optically thick gas and dust clumps are accelerated directly by photon momentum from star clusters or AGN \citep{Ishibashi2015, Thompson2015, Thompson2016, Krumholz2017, Tomaselli2021, Flury2023}. A prime example of such systems is the highly ionized, young, and metal-poor \citep[$12 + \log(\rm O/H) = 7.9$;][]{Izotov1997} system Mrk 71-A in NGC~2366, which shows direct evidence of catastrophic cooling \citep{Oey2017, Oey2023}. In \cite{Komarova2021}, we demonstrated that its $\sim 3000 \rm ~km~s^{-1}$ emission-line wings can only be explained by LyC and/or Ly$\alpha$ radiation pressure on tiny, dense neutral hydrogen clumps with an extremely low filling factor on the order of $10^{-3} - 10^{-2}$. As the clumps are continually driven to large distances ($\gtrsim 500$~pc) from the cluster and high speeds, LyC photons must reach distances on this order, and thus significant quantities likely escape through the inter-clump regions. 

Thus, a paradigm of two modes of LyC escape, dictated by the dominant feedback regime, begins to emerge. In the Low-redshift Lyman Continuum Survey \citep[LzLCS;][]{Flury2022a}, the largest sample of local LyC emitters (LCEs) to date, \cite{Flury2022b} see a hint of these two modes in strongly star-forming galaxies. One population 
of LCEs shows high H$\beta$ equivalent width (EW), young ages, high $O_{32} = [\rm O~III]\lambda5007/[O~II]\lambda3726,3729$, and low metallicities. The other is characterized by lower EW(H$\beta$), older ages, lower $O_{32}$, and higher metallicities. The former population may be dominated by radiation feedback, and the latter by mechanical feedback. \cite{Carr2025} and \cite{Bait2023} present further evidence for this scenario, finding that weaker leakers are preferentially older, SN-dominated systems, while stronger LCEs are younger and radiation-dominated. Furthermore, \cite{Flury2024} show that the confluence of both feedback modes, promoted by bursty star formation, may be required for the highest LyC escape fractions. 

The above insights into feedback in LzLCS LCEs were derived from 
a variety of multiwavelength observations and stellar population synthesis modeling, including optical spectroscopy.
Another important diagnostic of feedback mechanisms that can be leveraged from these data are broad emission-line wings. These broad wings are observed in 
\OIII\ $\lambda5007$ and Balmer emission lines, and reach $300-1000~\rm km~s^{-1}$ \citep{Amorin2012, Bosch2019, Hogarth2020, Amorin2024}. In addition to LzLCS objects, such wings are likewise seen in several extreme Green Peas (GPs) \citep{Jaskot2013, Izotov2016a, Izotov2016b, Izotov2018a, Izotov2018b}, which are some of the strongest known LCEs. Importantly, \cite{Amorin2024} showed that the width of the broad wing correlates with the LyC escape fraction \fesc{}, demonstrating a direct link between feedback and LyC escape.  While SN mechanical feedback or turbulent outflows have been suggested to explain these wings \citep[e.g.,][]{Hogarth2020}, the exact physical mechanisms
that cause them remain unknown. It appears that AGN feedback can be ruled out based on a systematic lack of very high-ionization markers, such as [Ne~V], Fe~VII], or X-ray emission \citep{Hogarth2020, Marques-Chaves2022, Amorin2024}. 
However, we linked similar broad wings to radiation-driven winds from a young super star cluster (SSC) in
Mrk~71, \citep{Komarova2021}, the closest GP analog. 
Thus, it is of great interest to test whether radiation driving from young SSCs is the underlying mechanism for the broad wings in the LzLCS galaxies and similar LCEs.

The morphology of the broad wings may be a clue to the underlying feedback mechanism. Indeed, \cite{Flury2023} find that the shape of line wings can be directly linked to driving mechanisms, such as ram pressure. On the one hand, the multitude of expanding shells in 
SN-driven superbubbles can produce complex, asymmetric emission-line profiles comprised of multiple Gaussian components \citep{Castaneda1990, Roy1991, Chu1994, Rosado1996}.
When integrated over the entire nebula, the resulting wing profile is a broad Gaussian \citep{Chu1994, Torres-Flores2013, Firpo2010}, as determined by the Central Limit Theorem, where the average of many independent contributions approaches a normal distribution.
On the other hand, radiation-driven superwinds can produce power-law or exponential broad wings, depending on the gas density profile \citep{Krumholz2017}. This is seen in Mrk 71-A, whose power-law  wings are linked to radiation driving, as inferred from its energy and momentum budgets \citep{Komarova2021}.  Thus, characterizing the functional form of broad wings in individual galaxies may help distinguish radiation-driven from SN-driven superwinds. 

In this paper, following up on the \cite{Amorin2024} result, we investigate the origin of the broad wings of 20 LzLCS galaxies and 6 extreme GPs \citep{Izotov2016a, Izotov2016b, Izotov2018a, Izotov2018b}. To further illuminate how stellar feedback promotes LyC escape, we use the broad-wing properties of the \OIII$\lambda5007$ emission line, in conjunction with galaxy properties, to test whether power-law emission-line wings are indeed linked to radiation-driven feedback, while Gaussian wings are linked to SN-driven feedback.
We then examine the relation of these feedback modes to LyC escape in these local LCEs.
In Section \ref{sec: sample and obs}, we describe our sample and observations, and our analysis in Section~\ref{sec: analysis}. We present our results in Section~\ref{sec: results}, discuss their implications in Section~\ref{sec: discussion}, and summarize our conclusions in Section~\ref{sec: conclusion}.

\section{Sample and Observations}
\label{sec: sample and obs}
Our sample consists of 20 star-forming galaxies at $z = 0.23 - 0.36$ from the LzLCS \citep{Flury2022a}, for which follow-up high-resolution optical spectra have been obtained, as well as 6 GPs at $z = 0.29-0.37$ \citep{Izotov2016a, Izotov2016b, Izotov2018a, Izotov2018b}. The galaxies in the LzLCS were selected from the SDSS and GALEX surveys, targeting the star-forming region of the 
Baldwin - Phillips - Terlevich (BPT) diagram \citep{Baldwin1981},
and they span a range of $O_{32}$, UV slope $\beta$, and star formation surface densities $\Sigma_{\rm SFR}$. We adopt the published data, including stellar and gas properties of these objects from the above studies and previous LzLCS works by \citet{Flury2022a} and \citet{Saldana-Lopez2022}. 
We present these properties in Table~\ref{table:sample}, where for each galaxy, 
Column~1 lists the galaxy's ID from SDSS
and Column~2 lists the
redshift $z$.  Column~3 gives the absolute magnitude at 1500~\AA~$M_{1500}$ from stellar population models, uncorrected for dust extinction.  Column~4 gives the oxygen abundance $12+\log(\rm O/H)$, derived using direct measurement of the nebular electron temperature, and 
$O_{32}$ is listed in Column~5.  The equivalent widths of H$\beta$ and Ly$\alpha$ are given in Columns~6 and 7, respectively. Column~8 provides the UV light fraction of $<3$~Myr-old stars $f_*(t < \rm 3)$ from \citep{Flury2024}, based on stellar population synthesis models by \citet{Saldana-Lopez2022}.
Columns~9 and 10 respectively show the nebular extinction $E(B-V)$ derived from the Balmer decrement and UV half-light radius $r_{\rm 50}$ measured from the COS acquisition images.
The star formation rate derived from H$\alpha$, $SFR_{H\alpha}$, is given in Column 11.  The observed LyC luminosity $L_{\rm LyC, obs}$ and LyC escape fraction \fesc\ are given in Columns 12 and 13, respectively.
For \fesc, we adopt the values presented by \citet{Flury2022a}, based on UV stellar population fits from \cite{Saldana-Lopez2022}. Our sample contains 20 confirmed LCEs, with UV LyC escape fractions \fesc\ $= 0.2\% - 62\%$, as well as 6 non-leakers with $2\sigma$ upper limits \fesc~$\leq 0.06-0.98\%$. As shown in Table \ref{table:sample}, our combined sample spans a wide range of properties: $O_{32} = 1.5 - 27$, and oxygen abundances $\log(\rm O/H) + 12 = 7.6 - 8.4$. We investigate in detail the broad wings of the \OIII\ $\lambda5007$ emission lines in this sample.

For 12 of these galaxies, we use optical spectra obtained with the X-shooter spectrograph on the Very Large Telescope, and those obtained with the Magellan Inamori Kyocera Echelle 
spectrograph \citep[MIKE;][]{Bernstein2003} on the Magellan Clay Telescope for 9 galaxies. For the remaining 5 objects, we use spectra from the Intermediate Dispersion Spectrograph and Imaging System (ISIS) on the 4.2 m William Herschel Telescope. Table \ref{table:broadwing} 
specifies the instrument used to observe each galaxy in the sample.

The MIKE observations, conducted on 2021 January $17-18$, and 2022 March $25-26$, cover the wavelength range $\sim 3300-9500$~\AA. With the 0.7\arcsec~slit and $2\times2$ pixel binning, the resolving power was $R \sim 32000$ on the blue side and $R \sim 31000$ on the red, as measured from arc lines.
The total exposure times for each object
are in the range
$3600-9000~\rm s$. We performed standard reduction steps, including flat-fielding, wavelength calibration, and extraction, using the 
\texttt{CarPy} pipeline \citep{Kelson2000, Kelson2003}. To correct for the residual blaze function \citep{Skoda2008}, we fit polynomials to each order after masking all significant features. Basic flux calibration was performed order by order, using spectrophotometric standard 
stars. We applied secondary flux corrections to the orders covering \OIII\ $\lambda5007$ 
using the SDSS spectra. 
We caution that the SDSS aperture is $3\arcsec$ while the MIKE aperture is $0.7\arcsec$, affecting the accuracy of this flux calibration adjustment; however, we do not expect this to be a significant issue for our sample, as these galaxies are very compact. Since a standard star was not observed during the 2021 run, we flux-calibrated the 2021 observations using only SDSS \OIII\ $\lambda5007$ fluxes, to calibrate the MIKE order(s) covering this line. To reduce the effect of cosmic rays and bad pixels, we additionally sigma-clipped each spectrum using a threshold of $8\sigma$ in $5$~\AA~windows. Lastly, we clipped the edges 
with low signal-to-noise (SNR)
of each order and median-combined the overlapping regions. In the reduced spectra, we detect the continuum with SNR $= 2.5-7.0$ around \OIII\ $\lambda5007$ in all but one object, J012910+145935.

The X-shooter echelle spectra in the VIS arm ($\lambda_{\rm obs} = 5595-10240$ \AA), originally presented by \cite{Marques-Chaves2022},  were obtained in 2021 (ESO Program ID 106.215K; PI: Schaerer). The 1.0\arcsec~slit was used, resulting in a resolving power $R \sim 9000$.  
For 5 of the 12 galaxies, we use the fully calibrated X-shooter dataset  detailed in \cite{Guseva2020} (ESO Program ID 102.B-0942; PI: Schaerer). 
Lastly, the ISIS observing configuration and reduction are described in detail by \cite{Hogarth2020}. The spectra (Program P27, PI: Amorín) were obtained with the 0.9\arcsec~slit.
The ISIS blue and red arms cover $\sim 4000$~\AA~to $\sim 6000$~\AA~and $\sim 7200$~\AA~to $\sim 8000$~\AA, respectively. The resulting resolving power was $R \sim 6000 - 10000$.

\section{Analysis}
\label{sec: analysis}

\subsection{Broad-Wing Morphology}

As originally demonstrated by \cite{Amorin2012}, the emission lines of GPs show complex, multi-component profiles, consisting of $1-3$ narrow cores plus broad wings. The multiple narrow components likely trace distinct star-forming knots that are separated on kpc scales, as is evident from several 2D spectra. 
To understand the physical mechanisms traced by the broad wings, we 
study
the broad-wing functional form, or morphology, in \OIII\ $\lambda$5007. 

To date, broad emission-line wings in GP-like objects
have been assumed to consist of one or multiple Gaussian components \citep{Amorin2012, Bosch2019, Hogarth2020, Amorin2024}. But from a visual inspection of the broad wings, some appear to have a power law form \citep{Komarova2021}. We therefore test whether these high-velocity wings can be described by power laws, as predicted for radiation-driven winds, or whether they are better fit by Gaussians, as may be expected for SN-driven, mechanical feedback (Section~\ref{sec:intro}).
These wings are separate components from the narrower, Gaussian cores originating from the dense starburst H~II regions.

\movetabledown=1.8in
\begin{rotatetable}
\begin{deluxetable*}{ccccccccccccc}
\tablecolumns{13}  
\tablecaption{Sample Properties
\label{table:sample}}
\tablehead{
\colhead{Galaxy} & \colhead{$z$} & \colhead{$M_{\rm 1500}$  \tablenotemark{\rm \scriptsize a}} & \colhead{$\log(\rm O/H) $  } & \colhead{$O_{32}$ \tablenotemark{\rm \scriptsize b} } & \colhead{EW(H$\beta$)} & \colhead{EW(Ly$\alpha$)} & \colhead{$f_*(t < \rm 3)$ \tablenotemark{\rm \scriptsize c}} & \colhead{$E(B-V)$ \tablenotemark{\rm \scriptsize d}} & \colhead{$r_{\rm 50}$ \tablenotemark{\rm \scriptsize e}} & \colhead{$\log(\rm SFR_{H\alpha})$} & \colhead{$L_{\rm LyC, obs}$}  & \colhead{\fesc~\tablenotemark{\rm \scriptsize f}} \\
\colhead{} & \colhead{} & \colhead{} & \colhead{+12} & \colhead{} & \colhead{\AA} & \colhead{\AA} & \colhead{} & \colhead{} & \colhead{kpc} & \colhead{$\rm M_{\odot}~yr^{-1}$}& \colhead{$~10^{41}~\rm erg/s$} & \colhead{} }
\startdata
J003601+003307 & 0.35 & $-18.1$  & $7.8$ & $13.1$ & $160$ & $93$ & $0.35$ & $0.11$ & $0.45$ & $1.2$ & $0.93$ & $<0.010$     \\
J004743+015440 & 0.35 & $-20.2$  & $8.0$ & $4.5$  & $62$ & $42$ & $0.96$ & $0.19$ & $0.62$ & $1.3$ & $13.1$ & $0.01\pm0.01$ \\
J011309+000223 & 0.31 & $-19.8$  & $8.3$ & $2.3$  & $41$ & $31$ & $0.62$ & $0.22$ & $0.63$ & $0.6$ & $0.87$ & $0.01\pm0.00$ \\
J012217+052044 & 0.37 & $-19.5$  & $7.8$ & $7.6$  & $87$ & $71$ & $0.69$ & $0.12$ & $0.71$ & $0.9$ & $1.02$ & $0.02\pm0.01$ \\
J012910+145935 & 0.28 & $-19.6$  & $8.4$ & $2.2$  & $73$ & $40$ & $0.39$ & $0.16$ & $0.64$ & $1.1$ & $0.27$ & $<0.006$   \\
J081409+211459 & 0.23 & $-20.5$  & $8.1$ & $1.6$  & $29$ & -- & $0.26$ & $0.26$ & $1.44$ & $1.2$ & $0.97$ & $<0.007$     \\
J090146+211928 & 0.30 & $-18.5$  & $8.2$ & $12.6$ & $255$ & $170$ & $0.27$ & $0.22$ & $0.18$ & $1.3$ & $0.21$ & $0.03\pm0.01$   \\
J091113+183108 & 0.26 & $-20.0$  & $8.1$ & $2.4$  & $73$ & $53$ & $0.75$ & $0.25$ & $0.44$ & $1.4$ & $3.07$ & $0.02\pm0.01$  \\
J091703+315221 & 0.30 & $-20.0$  & $8.5$ & $2.6$  & $51$ & $30$ & $0.49$ & $0.12$ & $0.41$ & $1.3$ & $4.31$ & $0.09\pm0.03$  \\
J092532+140313 & 0.30 & $-19.8$  & $7.9$ & $6.5$  & $177$ & $83$ & $0.68$ & $0.16$ & $0.40$ & $1.7$ & $2.19$ & $0.09\pm0.03$  \\
J095838+202508 & 0.30 & $-18.5$  & $7.8$ & $8.2$  & $131$ & $68$ & $0.39$ & $0.13$ & $0.49$ & $1.2$ & $0.24$ & $0.02\pm0.02$  \\
J101138+194721 & 0.33 & $-18.8$  & $8.0$ & $27.0$ & $237$ & $115$ & $0.86$ & $0.23$ & $0.15$ & $1.4$ & $1.48$ & $0.11\pm0.02$ \\
J105331+523753 & 0.25 & $-20.9$  & $8.25$& $3.4$  & $73$ & $7$ & $0.33$ & $0.20$ & $0.62$ & $1.4$ & $1.05$ & $0.01\pm0.00$  \\
J113304+651341 & 0.24 & $-19.6$  & $7.98$& $5.0$  & $64$ & $37$ & $0.50$ & $0.15$ & $0.70$ & $0.9$ & $0.90$ & $0.02\pm0.01$  \\
J115205+340050 & 0.34 & $-20.7$  & $8.00$& $5.4$  & $198$ & $79$ & $0.20$ & $0.13$ & $0.38$ & $1.4$ & $4.16$ & $0.18\pm0.06$  \\
J115449+244333 & 0.37 & $-19.2$  & $7.65$& $11.5$ & $220$ & $133$ & $0.41$ & $0.15$ & $0.64$ & $1.3$ & $5.35$ & $0.62\pm0.24$ \\
J115855+312559 & 0.24 & $-20.4$  & $8.39$& $2.4$  & $70$ & $49$ & $0.44$ & $0.19$ & $0.55$ & $1.3$ & $7.23$ & $0.06\pm0.01$ \\
J123519+063556 & 0.33 & $-18.7$  & $8.36$& $5.5$  & $86$ & $15$ & $0.17$ & $0.10$ & $0.45$ & $1.2$ & $0.08$ & $<0.005$   \\
J124423+021540 & 0.24 & $-19.8$  & $8.16$& $4.8$  & $143$ & $41$ & $0.51$ & $0.23$ & $1.00$ & $1.5$ & $0.44$ & $0.004\pm0.003$  \\
J124835+123403 & 0.26 & $-19.8$  & $8.16$& $4.6$  & $139$ & $97$ & $0.26$ & $0.10$ & $0.33$ & $1.2$ & $0.69$ & $0.04\pm0.01$  \\
J131037+214817 & 0.28 & $-19.5$  & $8.35$& $2.0$  & $52$ & $38$ & $0.79$ & $0.16$ & $0.42$ & $1.1$ & $0.98$ & $0.02\pm0.01$   \\
J131419+104739 & 0.30 & $-20.4$  & $8.33$& $1.5$  & $36$ & $7$ & $0.62$ & $0.28$ & $0.97$ & $1.3$ & $0.16$ & $<0.001$   \\
J134559+112848 & 0.24 & $-20.2$  & $8.32$& $1.5$  & $27$ & -- & $0.65$ & $0.32$ & $1.34$ & $1.2$ & $0.49$ & $<0.001$  \\
J144010+461937 & 0.30 & $-20.7$  & $8.21$& $2.4$  & $55$ & $29$ & $0.97$ & $0.24$ & $0.64$ & $1.5$ & $0.65$ & $0.002\pm0.001$   \\
J144231-020952 & 0.29 & $-20.1$  & $7.94$& $6.7$  & $312$ & $129$ & $0.32$ & $0.14$ & $0.42$ & $1.6$ & $2.34$ & $0.12\pm0.05$  \\
J164607+313054 & 0.29 & $-19.5$  & $8.10$& $5.4$  & $126$ & $44$ & $0.38$ & $0.12$ & $0.45$ & $1.3$ & $0.45$ & $0.02\pm0.01$ \\
\enddata 
~\\
\textbf{Notes.} 
\tablenotetext{}{
All values are from \cite{Flury2022a} unless otherwise noted.} 
\tablenotetext{\rm a}{Absolute magnitude at $1500$ \AA~from stellar population models, uncorrected for dust extinction.}
\tablenotetext{\rm b}{The typical uncertainty is $0.05$.}
\tablenotetext{\rm c}{Light fraction of $<3$~Myr populations from \citep{Flury2024} based on multi-component stellar populations fits by \cite{Saldana-Lopez2022}.}
\tablenotetext{\rm d}{Nebular extinction derived from the Balmer decrement.}
\tablenotetext{\rm e}{Half-light NUV radius from HST/COS. The typical uncertainty is $0.15$ kpc.}
\tablenotetext{\rm f}{LyC escape fraction, determined from UV SED fits.}
\end{deluxetable*}
\end{rotatetable}
\clearpage

\begin{figure*}[htbp]
    \centering
    \includegraphics[width=\textwidth]{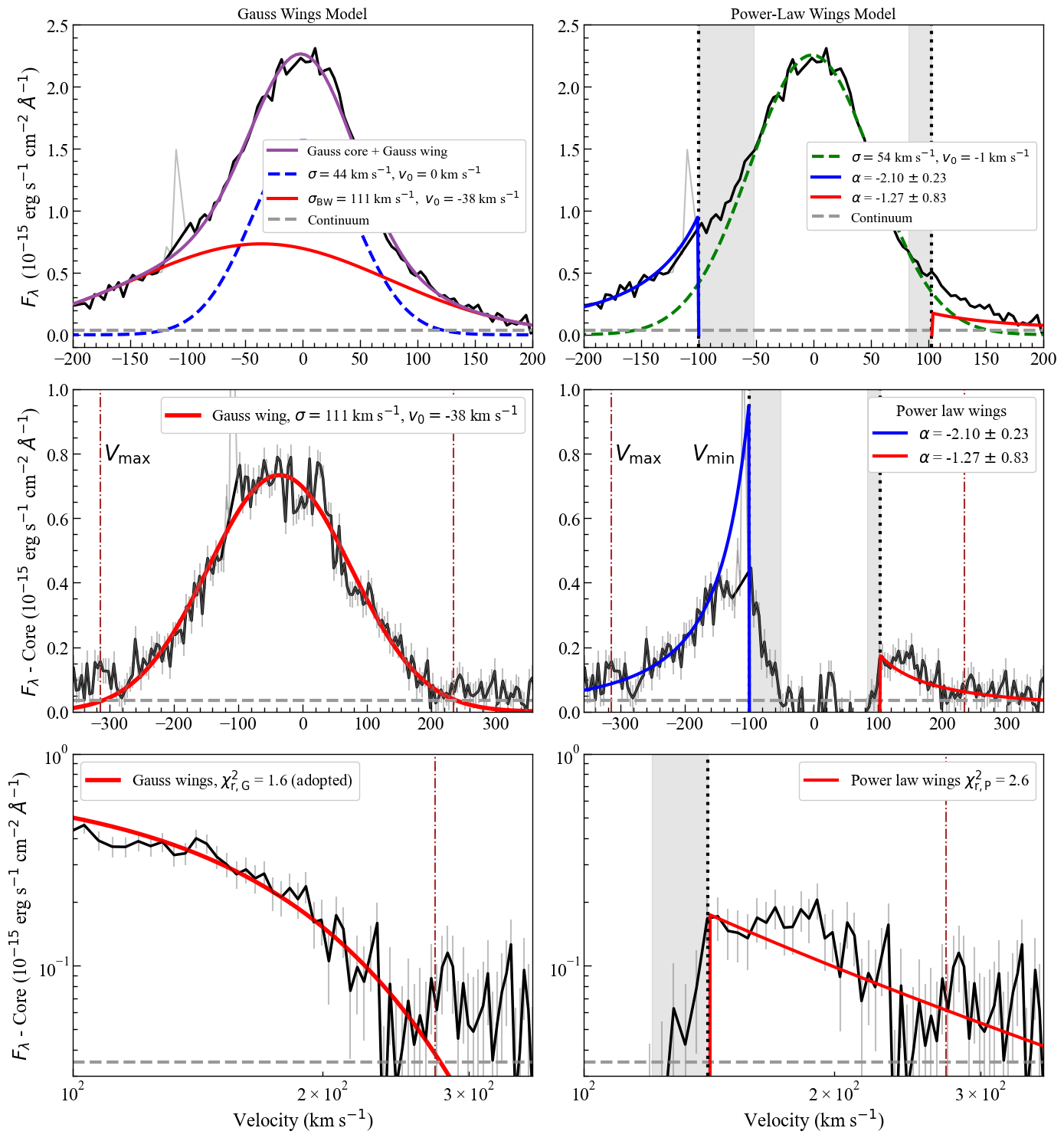}
    \caption{\textcolor{black}{Example of a Gaussian wing classifcation: J012910+145935 in \OIII\ $\lambda$5007. In the top row, we compare the Gaussian core + Gaussian wing vs. Gaussian core + power-law wing models. In the second row, we subtract each core model from the flux and overlay the best-fit wing model. In the third row, we compare Gaussian and power-law models for each set of wings in log-log space. The maximum wing velocity $V_{\rm max}$, the minimum velocity $V_{\rm min}$ in the power-law model, and the continuum fit are shown with lines as indicated in the legends. Gray regions in the power-law models represent the transition regions between the Gaussian core and power-law wings, which are not accounted for in our models. The noise spike at $V\sim-110 \rm~km~s^{-1}$ is masked prior to fitting.}
    }
\label{fig:gausswing}
\end{figure*}

\begin{figure*}[htbp]
    \centering
    \includegraphics[width=\textwidth]{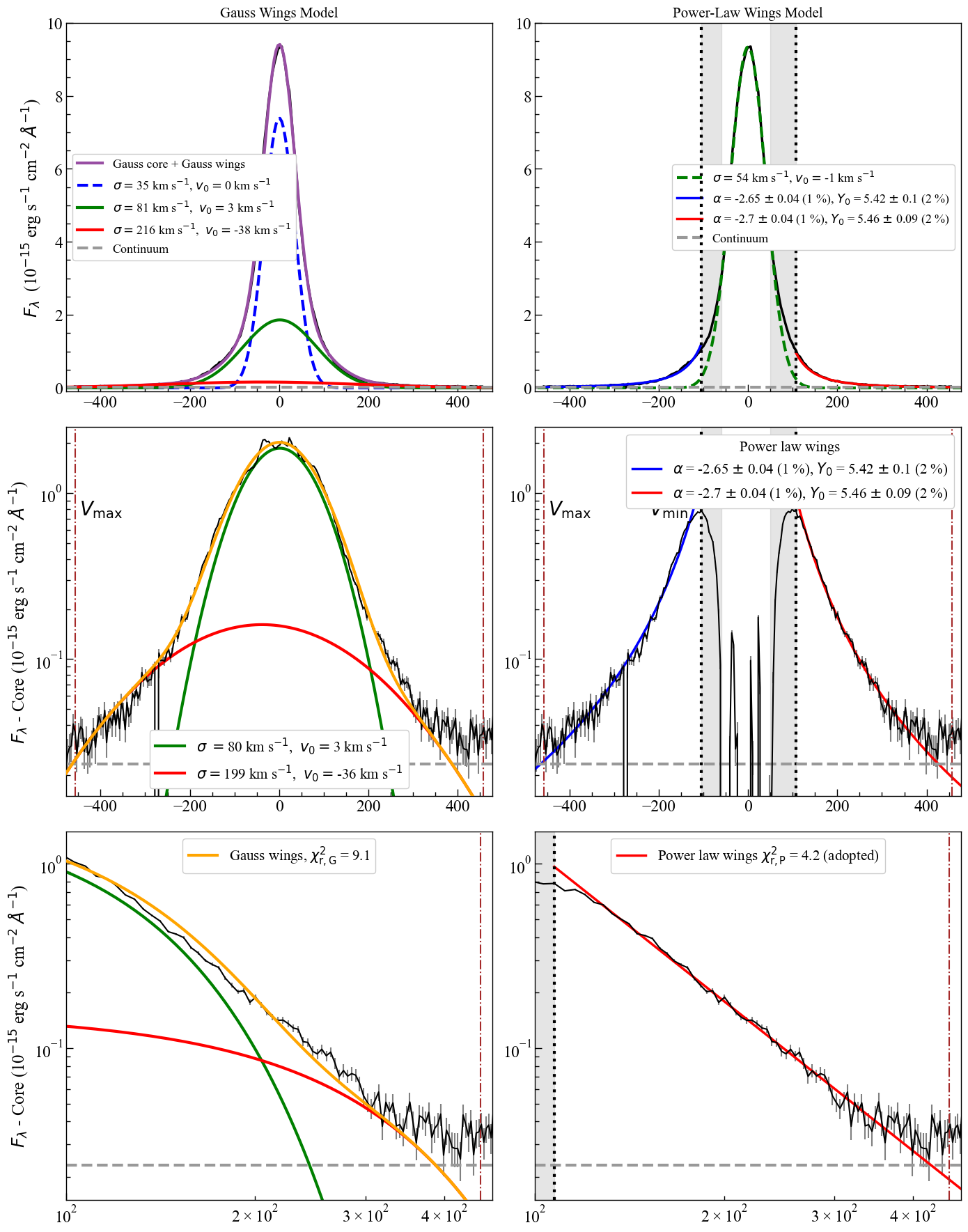}
    \caption{\textcolor{black}{Example of a power-law wing classifcation: J124835+123403 in \OIII\ $\lambda$5007. The panels are analogous to those in Figure~\ref{fig:gausswing}. 
    The feature around 500 km s$^{-1}$ is emission from Fe II $\lambda$5018, which is masked from the fits.}
    }
\label{fig:PLwing}
\end{figure*}

\begin{figure*}[htbp]
    \centering
    \includegraphics[width=\textwidth]{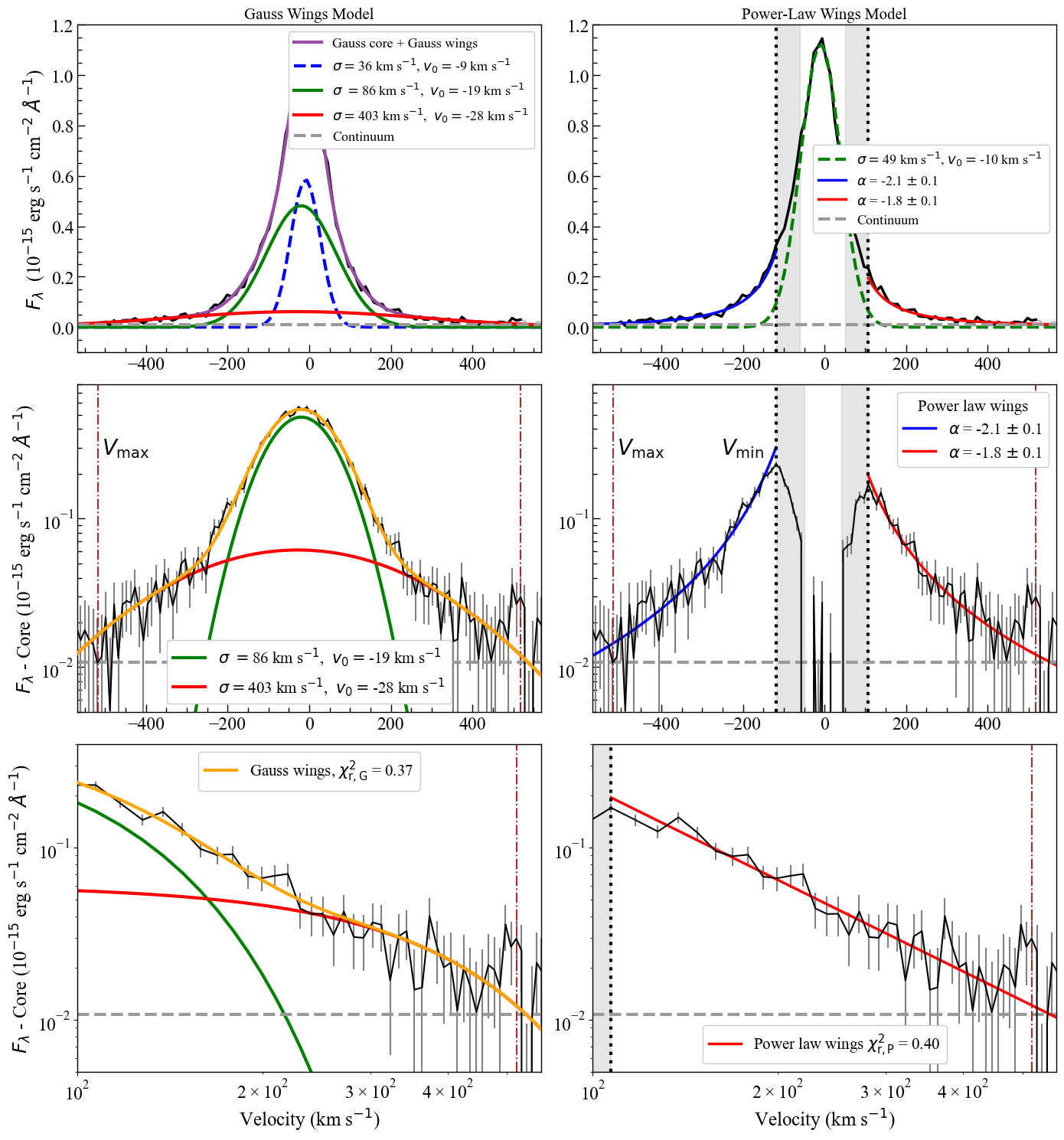}
    \caption{\textcolor{black}{Example of an ambigious wing classifcation: J144010+461937 in \OIII\ $\lambda$5007. The panels are analogous to those in Figure~\ref{fig:gausswing}. }
    }
\label{fig:Awing}
\end{figure*}

\begin{deluxetable*}{cccccccccccc}
\tablecolumns{11}
\tablecaption{\OIII\ Broad Wing Properties 
\label{table:broadwing}}
\tablehead{\colhead{Galaxy} & \colhead{Instr.\tablenotemark{\rm \scriptsize a}} & \colhead{$V_{\rm max}$} & \colhead{$V_{\rm 0, wing}$} & \colhead{$L_{\geq300}$} & \colhead{$\alpha_{\rm blue}$} & \colhead{$\alpha_{\rm red}$} & \colhead{$\sigma_{\rm BW}$} & \colhead{${\chi}^{2}_{\rm r, G}$\tablenotemark{\rm \scriptsize b}} & \colhead{${\chi}^{2}_{\rm r, P}$\tablenotemark{\rm \scriptsize c}} & \colhead{$\Delta$AIC \tablenotemark{\rm \scriptsize d}} & \colhead{Morph\tablenotemark{\rm \scriptsize e}} \\
\colhead{} & \colhead{} & \colhead{$\rm km~s^{-1}$} & \colhead{$\rm km~s^{-1}$} &\colhead{$10^{40}\rm~erg~s^{-1}$} & \colhead{} & \colhead{} & \colhead{$\rm km~s^{-1}$} & \colhead{} & \colhead{}& \colhead{}}
\startdata
J003601+003307 & Xsh  & $524\pm140$ & $-17.5\pm98.9$    & $2.99$  & $-3.1\pm0.1$ & $-3.2\pm0.3$ & --           & 2.40  & 1.50  & 78 & PL \\
J004743+015440 & Xsh  & $537\pm50$  & $18.5\pm35.4$   & $10.1$  & $-2.6\pm0.1$ & $-2.2\pm0.1$ & --           & 6.80  & 1.30  & 345 &PL \\
J011309+000223 & Xsh  & $406\pm50$  & $24.5\pm35.4$    & $2.51$  & $-2.2\pm0.1$ & $-1.6\pm0.1$ & --           & 1.70  & 1.01  & 39 &PL \\
J012217+052044 & Xsh  & $543\pm80$  & $19.0\pm56.6$    & $5.94$  & $-2.3\pm0.2$ & $-2.3\pm0.1$ & --           & 2.40  & 1.40  & 92 &PL \\
J012910+145935 & MIKE & $217\pm85$  & $-38.5\pm60.1$  & -- & --           & --           & $111\pm4$   & 1.60  & 2.60  &-57 &G  \\
J081409+211459 & Xsh  & $381\pm40$  & $8.0\pm28.3$   & $5.23$  & --           & --           & --           & 3.20  & --  &-- &A  \\
J090146+211928 & Xsh  & $608\pm40$  & $28.5\pm28.3$   & $3.21$  & $-3.5\pm0.1$ & $-2.6\pm0.1$ & --           & 5.70  & 3.50  & 124 & PL \\
J091113+183108 & Xsh  & $562\pm40$  & $33.0\pm28.3$ & $19.3$  & --           & --           & $463\pm87$  & 0.49  & --  & --& G  \\
J091703+315221 & ISIS & $594\pm50$  & $-48.0\pm35.4$ & $22.4$  & --           & --           & --           & 0.66  & 0.60  & 2.2 &A  \\
J092532+140313 & Xsh  & $594\pm70$  & $-28.5\pm49.5$  & $8.87$  & $-2.8\pm0.2$ & $-1.7\pm0.1$ & --           & 8.60  & 2.70  & 389 & PL \\
J095838+202508 & Xsh  & $376\pm40$  & $86.5\pm28.3$   & $1.14$  & $-2.7\pm0.2$ & $-2.7\pm0.2$ & --           & 2.20  & 0.73  & 60 &PL \\
J101138+194721 & Xsh  & $771\pm50$  & $48.0\pm35.4$    & $8.92$  & $-2.8\pm0.1$ & $-2.3\pm0.1$ & --           & 29.20 & 13.80 & 2050 &PL \\
J105331+523753 & ISIS & $423\pm50$  & $-29.0\pm35.4$ & $1.89$  & --           & --           & $257\pm40$  & 0.36  & --  & --& G  \\
J113304+651341 & ISIS & $413\pm50$  & $-20.5\pm35.4$    & $1.84$  & $-2.3\pm0.3$ & $-2.3\pm0.2$ & --           & 0.65  & 0.40  &14& PL \\
J115205+340050 & ISIS & $664\pm80$  & $24.0\pm56.6$    & $36.0$  & --           & --           & --           & 0.33  & 0.32  & 3.4& A  \\
J115449+244333 & Xsh  & $521\pm150$ & $101\pm106$   & $2.72$  & $-2.5\pm0.2$ & $-2.0\pm0.2$ & --           & 2.30  & 1.40  &86& PL \\
J115855+312559 & MIKE & $490\pm50$  & $-47.0\pm35.4$  & $7.49$  & --           & --           & --           & 2.37  & 2.41  & -5.1& A  \\
J123519+063556 & MIKE & $520\pm50$  & $-37.0\pm35.4$   & $8.75$  & --           & --           & --           & 3.30  & 2.80  & 4.1 & A  \\
J124423+021540 & MIKE & $470\pm100$ & $62.0\pm70.7$   & $13.3$  & --           & --           & --           & 1.65  & 1.57  & 4.0 & A  \\
J124835+123403 & MIKE & $414\pm40$  & $13.5\pm28.3$    & $3.69$  & $-2.8\pm0.1$ & $-2.6\pm0.1$ & --           & 9.10  & 4.17  &582 & PL \\
J131037+214817 & MIKE & $635\pm40$  & $-29.5\pm28.3$   & $8.11$  & $-2.2\pm0.1$ & $-2.0\pm0.1$ & --           & 4.30  & 3.50  & 130& PL \\
J131419+104739 & MIKE & $401\pm50$  & $-12.0\pm35.4$ & $6.64$  & --           & --           & $352\pm39$  & 2.20  & 3.30  &-153& G  \\
J134559+112848 & MIKE & $324\pm45$  & $30.5\pm31.8$   & $1.2$  & --           & --           & --           & 2.70  & -- & --& A  \\
J144010+461937 & ISIS & $544\pm50$  & $-4.0\pm35.4$   & $5.30$  & --           & --           & --           & 0.37  & 0.40  & 1.2 & A  \\
J144231-020952 & Xsh  & $725\pm100$ & $25.0\pm70.7$   & $12.6$  & $-2.8\pm0.1$ & $-2.2\pm0.1$ & --           & 15.30 & 3.40 & 14& PL \\
J164607+313054 & MIKE & $494\pm45$  & $5.0\pm31.8$    & $3.86$  & $-2.5\pm0.2$ & $-2.5\pm0.1$ & --           & 3.20  & 1.60  &104& PL \\
\enddata
~\\
\textbf{Notes.}
\tablenotetext{\rm a}{Spectrograph used in observations: Magellan/MIKE, VLT/X-shooter, WHT/ISIS. }
\tablenotetext{\rm b}{Reduced $\chi^2$ for the best-fitting Gaussian wing model, computed in the wings.}
\tablenotetext{\rm c}{Reduced $\chi^2$ for the best-fitting power-law wing model, computed in the wings. Missing entries represent unreliable models with dynamic range up to a factor of $< \times 2$. }
\tablenotetext{\rm d}{\textcolor{black}{Difference in the Akaike information criterion between the Gaussian and power-law wing models $\Delta$AIC = $\rm AIC_{Gauss} - AIC_{PL}$ (see text).}}
\tablenotetext{\rm e}{Wing morphology: power law (PL), Gaussian (G), or ambiguous (A). }
\end{deluxetable*}

We first obtain the best Gaussian model for each set of wings, following the fitting procedure of \cite{Amorin2024}. Before fitting the wing profiles, we mask out the neighboring contaminating line Fe~II $\lambda$5018 near \OIII\ $\lambda$5007. Using the Non-Linear Least-Squares Minimization and Curve-Fitting package \textsc{LMFIT} \citep{Newville2014}, we add additional Gaussian components to each line fit until the Akaike information criterion \citep[AIC,][]{Akaike1974} does not change significantly \citep[$\Delta$AIC $< 10$,][]{Bosch2019}. For this multi-Gaussian fit, we leave free all parameters, i.e., amplitude, centroid, and width $\sigma_{\rm BW}$. \textcolor{black}{We find that for most objects, two narrower Gaussian core components are required, with \textcolor{black}{$\sigma \leq 70~\rm km~s^{-1}$, in addition to the broad wings;} while for others, one narrow core is required, consistent with the findings of \cite{Amorin2024}}. \textcolor{black}{Together with the emission components, we simultaneously fit a constant to the local continuum in a $\pm50$~\AA~window around \OIII\ $\lambda$5007. This simple treatment is sufficient for our emission-line analysis, following \cite{Amorin2012, Hogarth2020, Amorin2024}. }
All spectral fitting is performed in the rest frame of each galaxy.

We then separately find the least-squares power-law fit to the flux $F$
for each object, of the form $F=A\times V^{\alpha}$, where $A$ is the amplitude, and $\alpha$ is the power-law slope.
Unlike in the simultaneous multi-Gaussian fit, we first isolate the narrow emission of the line core
from the wings by fitting a Gaussian for each distinct line peak.  We then subtract these core component(s) and define the broad wing as the resulting residuals above $|V| \geq |V_{\rm min}|$ as shown in Figure \ref{fig:PLwing},
where $|V_{\rm min}|$ is indicated by the vertical dotted lines.
The minimum broad-wing velocity $V_{\rm min}$ corresponds to the velocity at the peak of the core-subtracted residuals. On average,  $|V_{\rm min}| \gtrsim 150 \rm~km~s^{-1}$, though we note that $V_{\rm min}$ has large uncertainties, as shown by the gray core-wing transition zone in Figure \ref{fig:PLwing}. In objects with multiple narrow components, contamination by other star-forming regions may prevent robust disentangling of the wind component at low velocities. We then fit the wings with linear functions in log-log space. We allow the power-law amplitude and slope $\alpha$ to vary between the blue and red side of the wings, denoting these respective slopes as $\alpha_{\rm blue}$ and $\alpha_{\rm red}$. 

It is important to note that our power law models are piece-wise, and do not account for the transition region between the core(s) and the wing, shown in gray in Figure \ref{fig:PLwing}, bottom-right panel. We have explored several modifications to the pure power law function, such as truncation or exponential cut-off, but have not found a single universal analytic form that can be fitted simultaneously with the Gaussian cores. 
Most likely
the transition between slow and fast gas 
cannot be captured by a single analytic form.
This kinematic transition must be linked to the wind launch conditions and potentially other factors, such as turbulence, interaction with dense gas components, and geometric effects that differ between galaxies.
The wind emission is not expected
to extend to the lowest velocities because the 
denser, low-velocity material is close to the SSC and
cannot escape the gravitational potential. Instead, this gas 
remains bound and contributes to the narrow core near systemic velocity. As shown by \cite{Krumholz2017}, 
this effect results in line profiles consisting of discrete, high-velocity 
red- and blue-shifted components of the modelled wind emission.

On the other hand, the Gaussian wing models do extend continuously to the lowest velocities. In the SN scenario, this is physically motivated by 
the existence of multiple dense shell and kinematic structures, with much emission transverse to the line of sight and therefore at zero velocity \citep{Chu1994}.
The two different fitting methods reflect the two different physical models for the origin of the line wings:  gaussian line wings must originate from many kinematic components whose sum therefore generates a gaussian function.  Gaussians originate from the central limit theorem, thus the central part of the function is a fundamental component that must be included and accounted for.  In contrast, a power law linked to a radially accelerating wind cannot extend to zero velocity and must have a lower limit.

Our objective is to characterize the line wings with simple, physically motivated analytic models that minimize the number of parameters. We therefore select the wing model for a given galaxy with the lower reduced chi-squared $\chi^{2}_{\rm r}$, or $\chi^{2}$ per degree of freedom, computed over the full extent spanning the red and blue wings. The power-law models, fitted as linear functions in log-log space, always have four free parameters: a slope and normalization for each of the two wings. The Gaussian wing models contain three free parameters, i.e., amplitude, centroid, and width, for each fitted broad Gaussian \textcolor{black}{component. To quantify the statistical significance of the difference between these the power-law and gaussian models, we compute the difference in the Akaike information criterion $\Delta$AIC = $\rm AIC_{Gauss} - AIC_{PL}$, where PL stands for power law. The model with a lower AIC is preferred, and the extent to which a model is favored is determined by the magnitude of $\Delta \rm AIC$, with $6 < |\Delta \rm AIC| < 10$ corresponding to a significant difference, and $|\Delta \rm AIC| \geq 10$ to a strong difference \citep[e.g.,][]{dBurnham2004, Wei2016}.}
One caveat is that we reject power-law models with a velocity dynamic range spanning less than a factor of two, regardless of $\chi^{2}_{\rm r}$ and $\Delta \rm AIC$. We consider this range to be too short to reliably distinguish between these analytic forms. 

\textcolor{black}{Figures \ref{fig:gausswing} -- \ref{fig:Awing} show the fitted components for example Gaussian, power-law and ambiguous wing morphology classifications.
Cases where the Gaussian and power-law models yield similar $\chi^{2}_{\rm r}$, and $\Delta \rm AIC \leq 6$, or neither form satisfactorily reproduces the wing shape, we classify as ``ambiguous''.
For such cases, usually $1-2$ broad Gaussians can be equivalent in fit quality to power-law wings.
However, we emphasize that for most objects classified as power-law, the line wings cannot be reasonably fit with Gaussian models, even with
an arbitrary number of Gaussian components. We show all of the best fits for each galaxy in our sample in Appendix~\ref{sec:app_gauss}-\ref{sec:App_ambig}. }

\textcolor{black}{
Table~\ref{table:broadwing} gives the resulting morphology classifications for our sample objects, together with the   $\chi^{2}_{\rm r}$ for Gaussian and power-law models, and $\Delta \rm AIC$.}
A complete analysis of line profiles using all observed optical lines will be presented in Amor\'in et al. (in preparation).

\subsection{Broad-Wing Parameters}
In addition to the morphology, we determine the broad-wing centroid, maximum velocity, and luminosity in each galaxy. 
We determine the maximum detected broad-wing velocity $V_{\rm max}$ as the velocity at which the best wing model intersects the local measured continuum on each side of the wing (Figures~\ref{fig:gausswing}-\ref{fig:Awing}).
The values are measured relative to the systemic velocity of the line core as fitted by the gaussian component.  The brightest core is adopted for the systemic velocity if there is more than one core. We then compute the broad-wing centroid velocity $V_{\rm 0, wing}$ as the mid-point of the two $V_{\rm max}$ measurements in the red and blue wings. The instrumental and thermal broadening are $\sim 4-10 \rm~km~s^{-1}$ and $\sim 2 \rm~km~s^{-1}$, respectively for \OIII\, which is small compared to our estimated $V_{\rm max}$ and its uncertainties of $\geq 40\rm~km~s^{-1}$. Therefore, we do not correct for these small effects. Lastly, we compute the high-velocity luminosity  $L_{\geq 300}$ in the wings, by integrating the flux in the range
$300\rm~km~s^{-1}\it \leq |V| \leq V_{\rm max}$, i.e., the red and blue emission above $300\rm~km~s^{-1}$. This velocity threshold ensures a robust, model-independent separation of the wings from the core(s), with the latter extending to $\sim 200 \rm~km~s^{-1}$ in most objects. $L_{\geq 300}$ is undefined for one object in our sample, J012910+145935, which shows $V_{\rm max} \sim 220\rm~km~s^{-1}$.  These broad-wing parameters are also listed in Table~\ref{table:broadwing}. 

\subsection{Statistical Tests}

For relating broad-wing parameters to galaxy properties, we use the Kendall's $\tau$ rank correlation test as implemented by \citet{Flury2022b} based on \citet{Akritas1996}. This code accounts for censored data, as required for treatment of upper limits, 
and estimates confidence intervals for the correlation coefficient $\tau$, based on uncertainties in the correlated parameters, using Monte Carlo sampling. The $\tau$ coefficient is a non-parametric measure of the strength and direction of association between two ranked variables, based on the relative ordering of data pairs. It ranges from $-1$ (perfect anti-correlation) to $+1$ (perfect correlation), with 0 indicating no association. For our full sample of 26 galaxies, we consider a correlation to be
significant if the resulting $ |\tau| \geq 0.28, p \leq 0.05$, and tentative for $ |\tau| \geq 0.24, p \leq 0.1$. When considering the sub-sample of 14 galaxies with power-law wings (Section \ref{sec: results}), these respective criteria are $ |\tau| \geq 0.41, p \leq 0.05$, $ |\tau| \geq 0.36, p \leq 0.1$.

For comparing properties of sub-samples of galaxies, we perform the Kolmogorov-Smirnov (K-S) test, implemented in {\tt SciPy}. From this test, we quote $D$, quantifying the maximum difference between two distribution functions, ranging between 0 (no difference) and 1 (maximum difference).

\section{Results}
\label{sec: results}

\begin{figure}
    \centering
    \subfigure{\includegraphics[width=\columnwidth]{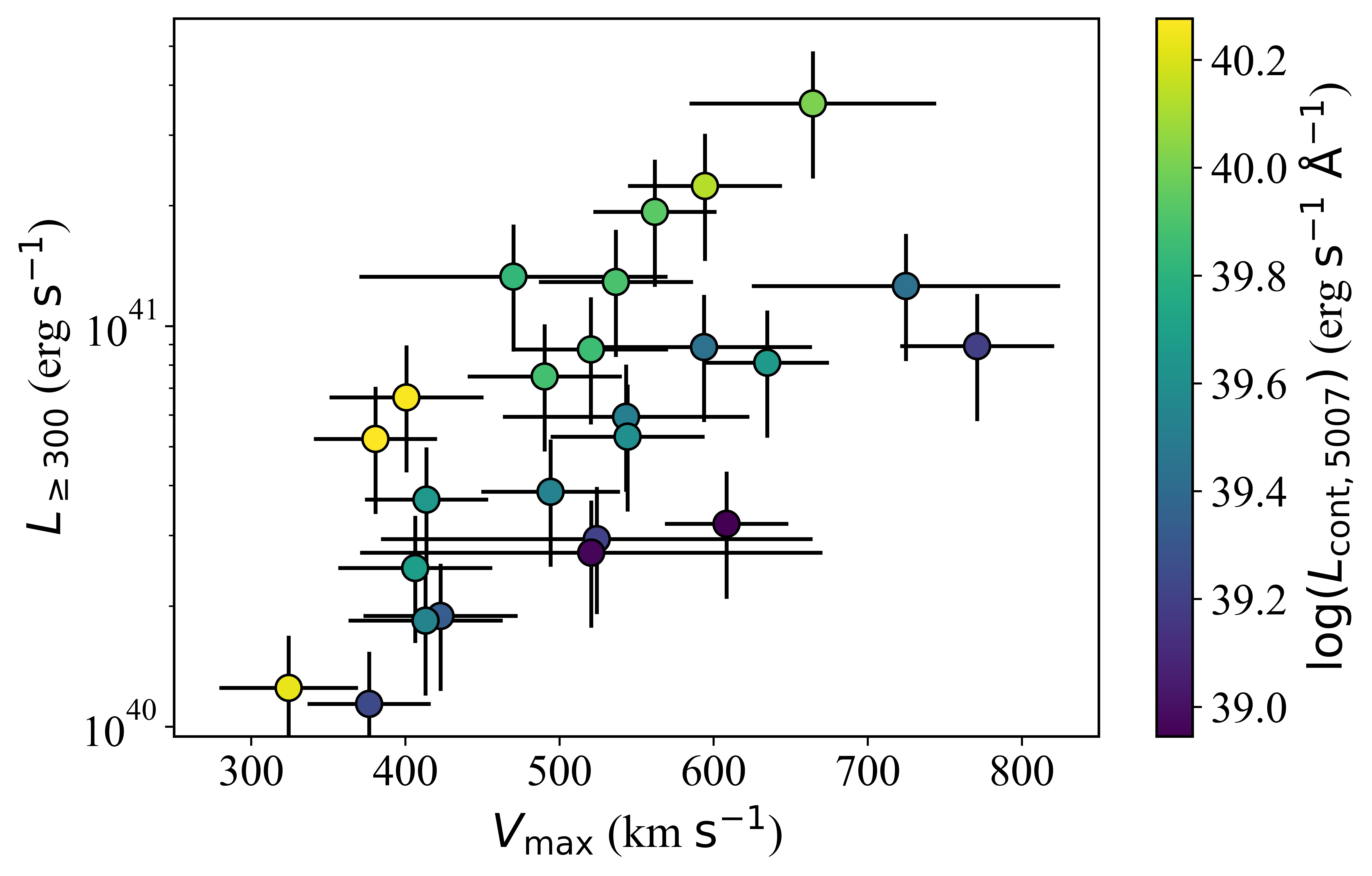}}\\
    \caption{Measured broad wing parameters from \OIII\ $\lambda$5007: maximum broad-wing velocity $V_{\rm max}$ vs. luminosity $L_{\geq300}$, color-coded by $5007$~\AA~median continuum luminosity. }
    \label{fig:vmax_L300_O3cont__orslope}
\end{figure}

\begin{figure}
\includegraphics[width=\columnwidth]{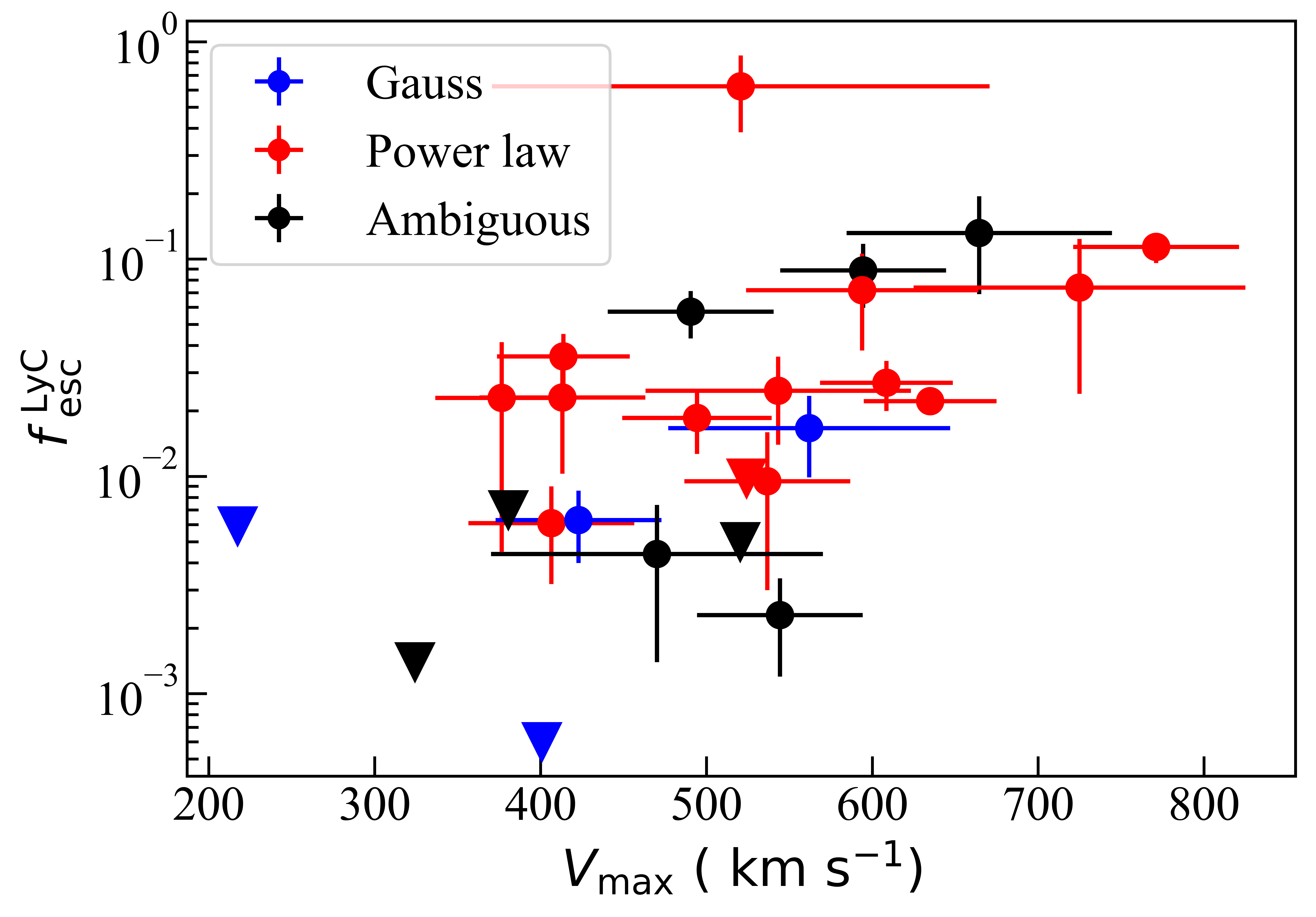} 
\caption{LyC escape fraction \fesc~vs. maximum broad-wing velocity $V_{\rm max}$ in [O~III]$\lambda5007$, color-coded by wing morphology. The triangles represent upper limits on \fesc.}
\label{fig:fesc_vmax}
\end{figure}

\begin{figure*}[t]
\centering
\includegraphics[width=0.9\textwidth]{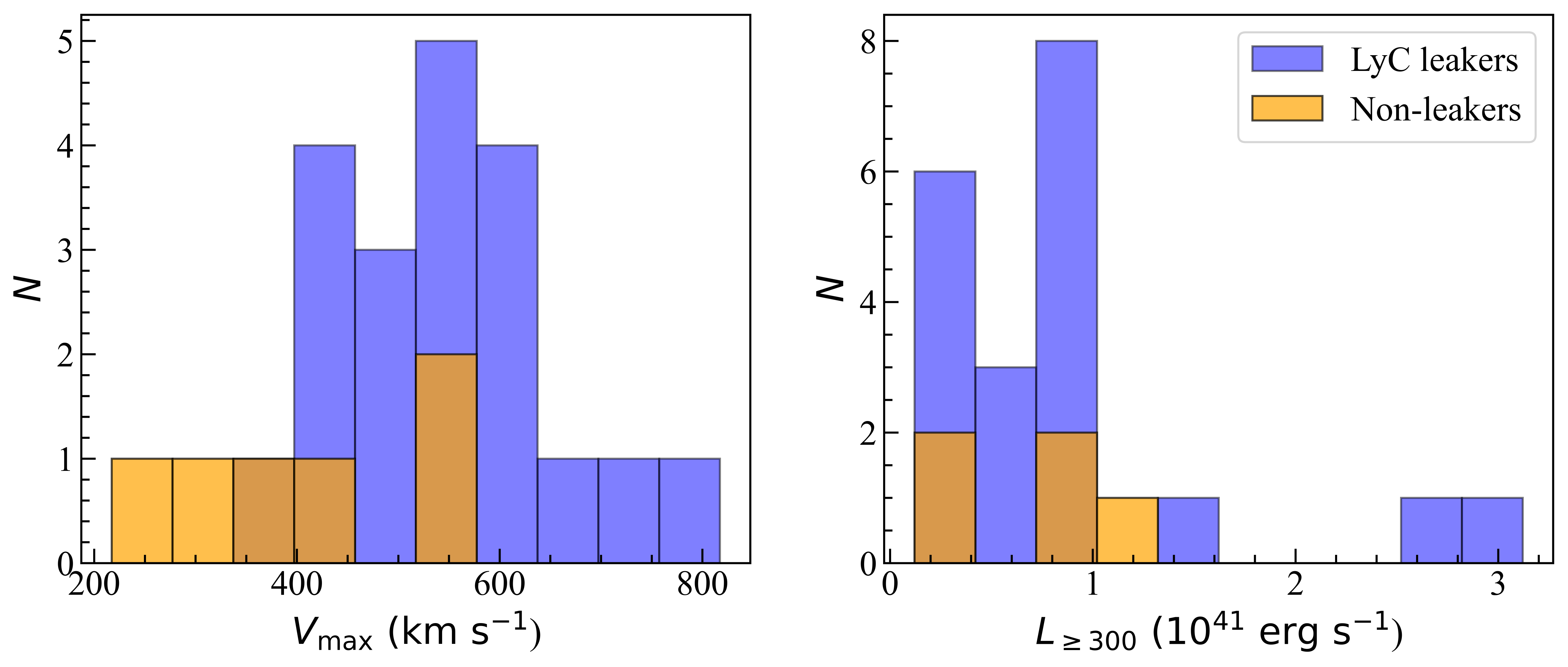} 
\caption{Comparison of $V_{\rm max}$ and $L_{\geq 300}$ between LyC leakers (blue) and non-leakers (orange).}
\label{fig:LCE_nonLCE}
\end{figure*}

We first analyze the full sample of broad
emission-line wings, regardless of wing morphology.
We consider the broad wing parameters  $V_{\rm max}$, the maximum broad-wing velocity, and $L_{\geq300}$, the high-velocity luminosity. 
In what follows, we quote the measurements for each parameter as: median (median absolute deviation). 
The maximum broad-wing velocities $V_{\rm max}$ have a range of $217-771~\rm km~s^{-1}$ with the sample median of $V_{\rm max} = 520~(93)~\rm km~s^{-1}$. We obtain broad wing luminosities at $|V| > 300 \rm~km~s^{-1}$ in the range $1.2 \times 10^{40} - 2.8 \times 10^{41}~\rm erg~s^{-1}$, corresponding to $0.6\%-12.4\%$ of the total line luminosity. 

To examine the relationship between our measured emission-line wing parameters, we compare $V_{\rm max}$, $L_{\geq300}$, and rest-frame $5024-5040$~\AA~continuum luminosity $L_{\rm cont, 5007}$ in Figure \ref{fig:vmax_L300_O3cont__orslope}. 
The continuum, linked to the galaxy stellar luminosity, is
a variable independent of the wind-driving mechanism,
but affects our measurements of $V_{\rm max}$.
This SNR effect is seen in Figure~\ref{fig:vmax_L300_O3cont__orslope}, where objects with higher $L_{\rm cont, 5007}$ tend to have lower $V_{\rm max}$ detections at a given $L_{\geq 300}$.

With the full sample of galaxies considered in this work, we start by reproducing the correlation between the broad-wing velocity and the LyC escape fraction \fesc~established by \cite{Amorin2024}, but now with the additional six objects. As shown in Figure \ref{fig:fesc_vmax}, we indeed recover this correlation, with our Kendall rank correlation test yielding $\tau = 0.44^{+0.08}_{-0.09},\ p = 1.6\times10^{-3}$. Although we assume varying wing functional forms and use the maximum detected broad-wing velocity $V_{\rm max}$ rather than the Gaussian width $\sigma_{\rm BW}$, our results are consistent with the correlation coefficient obtained by \cite{Amorin2024}, $\tau = 0.32^{+0.10}_{-0.12},\ p = 5.2\times10^{-2}$, indicating a correlation of moderate strength. 

Our sample contains 20 confirmed LCEs and 6 galaxies without significant ($> 2\sigma$) LyC detections. We compare the maximum broad-wing velocity $V_{\rm max}$ and high-velocity luminosity $L_{\geq 300}$ for leakers and non-leakers in Figure \ref{fig:LCE_nonLCE}. We find that the sub-sample of LCEs shows higher $V_{\rm max}$ than the non-LCEs, with the respective median values of $540~(69)\rm~km~s^{-1}$ and $390~(98)\rm~km~s^{-1}$. The K-S test yields $D = 0.62,\ p = 3.5\times10^{-2}$, rejecting, at the $95\%$ confidence level, the null hypothesis that the two sub-samples originate from the same distribution. This is expected from the $V_{\rm max}$-\fesc~correlation discussed above. However, there is no significant difference in the broad-wing luminosity $L_{\geq 300}$ between LCEs and non-LCEs, with the respective median values of $7.6~(3.0)\times10^{40}\rm~erg~s^{-1}$ and $7.7~(3.2)\times10^{40}\rm~erg~s^{-1}$ and the K-S test resulting in $D = 0.20,\ p = 9.9\times10^{-1}$. 

We now examine the broad wing morphology. We classify the emission-line wings for 14 of the objects as power-law in form, 4 objects as Gaussian, and 8 as ambiguous. The power-law wings are characterized by slopes $\alpha = -3.5$ to  --1.6, $V_{\rm max} = 376-771 \rm~km~s^{-1}$, and Gaussian ones by width $\sigma_{\rm BW} = 118-352 \rm~km~s^{-1}$, $V_{\rm max} = 217-562 \rm~km~s^{-1}$. Among the 8 sets of wings with ambiguous wing morphologies, 6 are reasonably well fitted with either model. In the remaining 2 cases, J134559+112848 and J081409+211459, neither power-laws nor Gaussians fit the full extent of the wings well.

Our goal now is to understand the physical origin of the broad emission. The functional form of the broad wings may trace the underlying feedback mechanism responsible for the observed high-velocity gas in each galaxy. In particular, as discussed in Section \ref{sec:intro}, we suggest that Gaussian broad wings may originate from conventional, SN-driven feedback, and power-law  wings from radiation-driven superwinds. To test this hypothesis, we evaluate whether galaxy properties associated with each of the two broad-wing morphologies are consistent with this scenario.

\subsection{Power-law vs. Gaussian Emission-line Wings}
\label{sec:PLvG}

\begin{figure}
\includegraphics[width=\columnwidth]{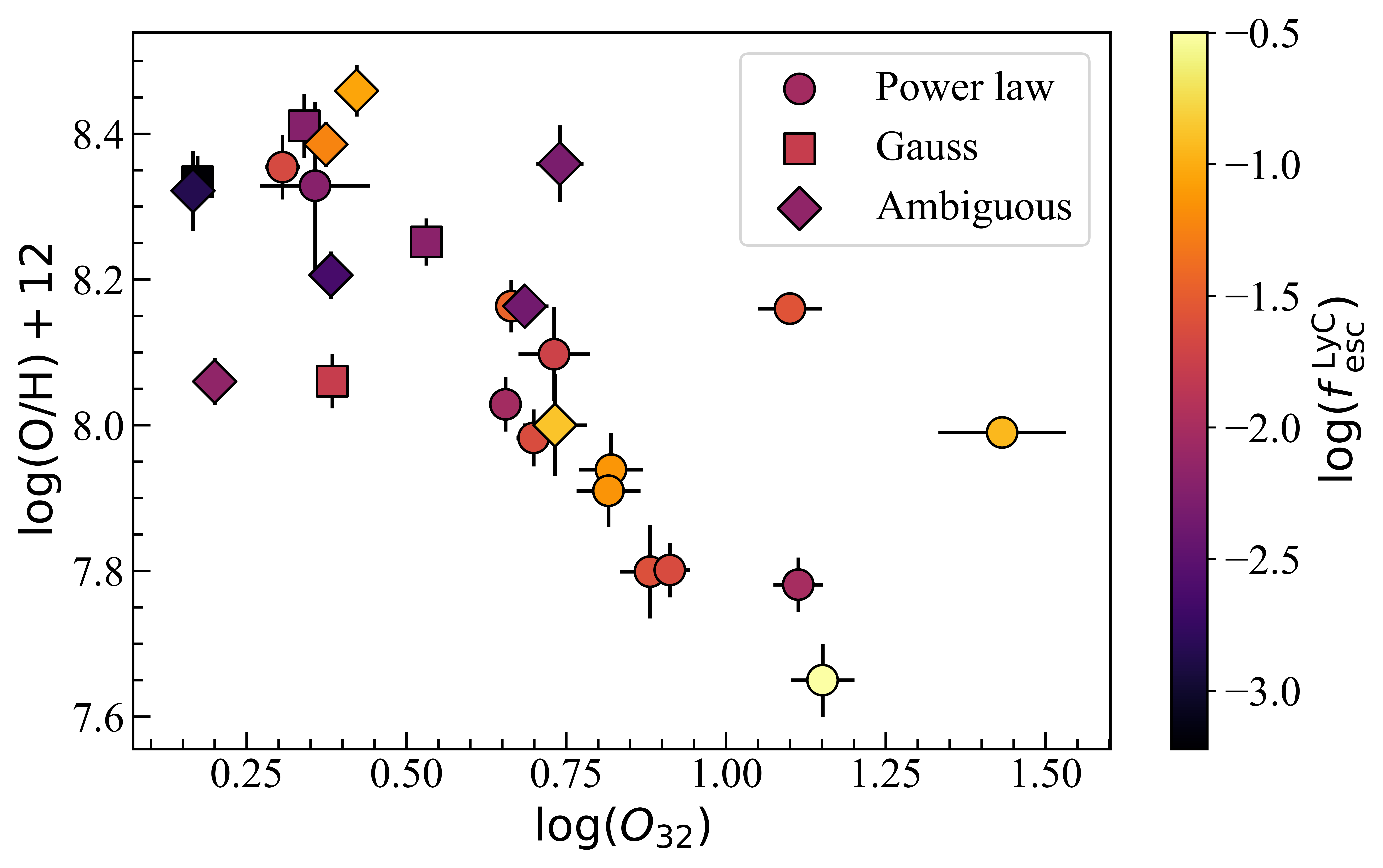} 
\caption{Oxygen abundance vs. $\log(O_{32})$ of the galaxies in our sample, color-coded by \fesc, with different symbols representing the wing morphologies. }
\label{fig:O32_metal}
\end{figure}

\begin{figure}
    \centering    \includegraphics[width=\columnwidth]{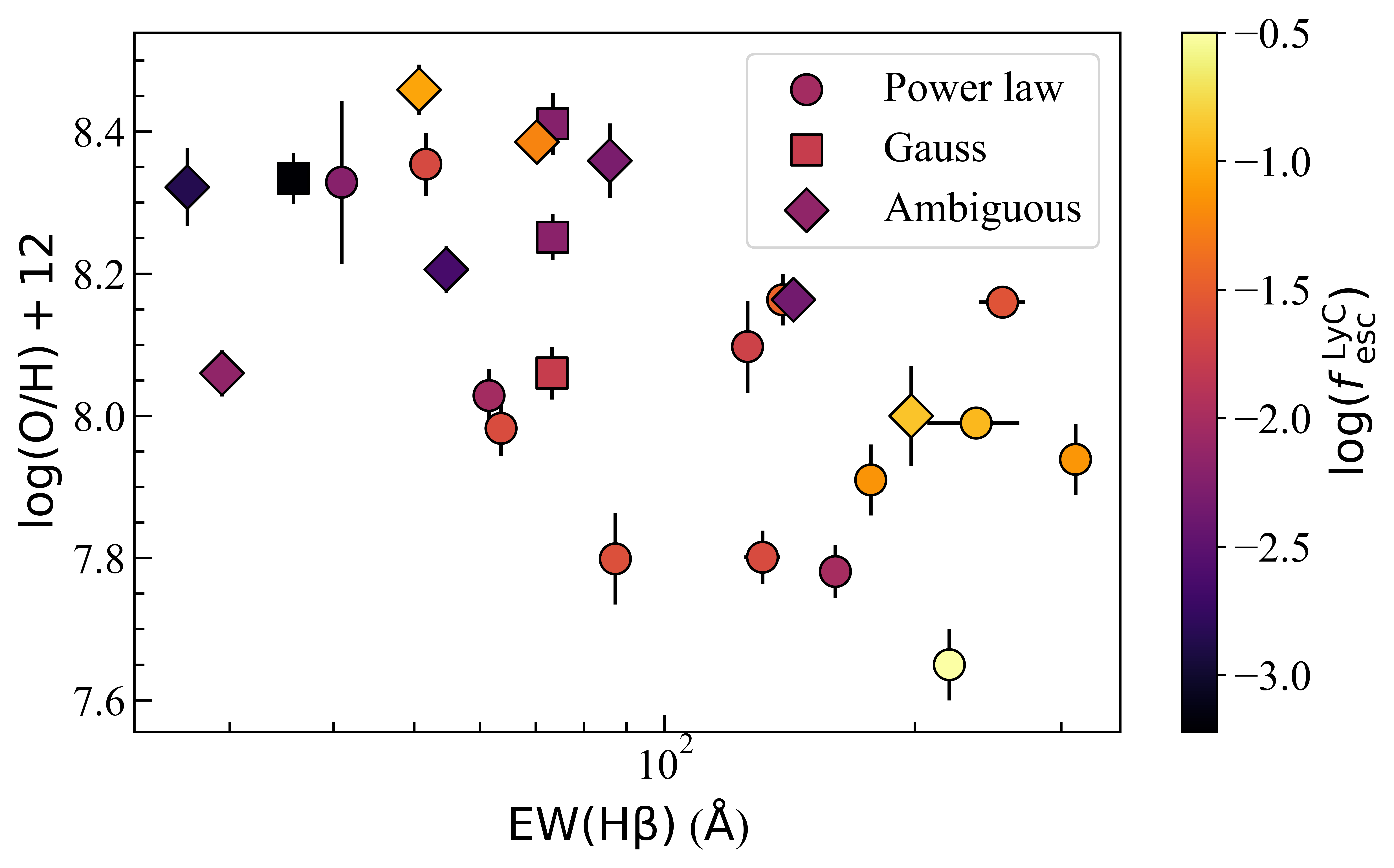}
    \caption{Oxygen abundance vs. EW(H$\beta$), color-coded by \fesc, with different symbols representing the wing morphologies.}
   \label{fig:age_metal}
\end{figure}

In Figures \ref{fig:O32_metal} and \ref{fig:age_metal}, we show how the broad wing morphology relates to $O_{32}$, EW(H$\beta$), and metallicity. We use the oxygen abundance $12 + \log(\rm O/H)$ to trace metallicity, and the H$\beta$ equivalent width EW(H$\beta$) to characterize stellar age. The EW(H$\beta$) traces the contribution of young massive stars producing ionizing emission, relative to the more evolved population dominating the optical continuum. Radiation-driven feedback is expected to be prevalent in younger ($\lesssim 3$~Myr) environments with larger EW(H$\beta$), as well as at higher ionization parameters traced by $O_{32}$, since the ionization parameter probes the ratio of radiation to gas pressure. Radiation feedback is also expected to dominate for longer periods and thus larger population ages for low metallicities, because the onset of SNe is delayed \citep{Jecmen2023}. On the other hand, SN feedback is expected to dominate at older stellar ages and higher metallicities, and to generate lower ionization parameters, since gas clearing in the inner region acts to reduce the observed ionization parameter.

We find that Gaussian emission-line wings are found exclusively at lower $O_{32} < 3.5$, higher metallicity $12+\log(\rm O/H) > 8.0$, and lower EW(H$\beta$) $\leq 75$~\AA. In contrast, galaxies with power-law wings span the full range of values of these parameters. Importantly, the most extreme starbursts, with $O_{32} > 5.5$, exclusively show power-law wings. These findings are consistent with power-law broad wings being associated with radiation-driven superwinds, and Gaussian wings with SN-driven superwinds, 
as suggested in Section~\ref{sec:intro}. 
We do see power-law wings in objects with higher metallicity and low $O_{32}$, suggesting that radiation-dominated feedback is not limited to low-metallicity conditions, and may be driven primarily by very young, pre-SN ages and intense starbursts.  In fact, higher metallicities promote catastrophic cooling \citep[e.g.,][]{Danehkar2021}, which is often linked to radiation-dominated feedback.

The galaxies with ambiguous wing morphologies
span an intermediate range of properties,
showing values similar to those 
of objects
with Gaussian wings, but reaching higher $O_{32}$ and EW(H$\beta$), with $O_{32} \sim 5$ and EW(H$\beta$) $=200$~\AA. Notably, the two galaxies for which neither model satisfactorily reproduces the wing shape, J081409+211459 and J134559+112848, have the lowest $O_{32}$ 
and EW(H$\beta$) in the sample, with values of $\sim 1.5$ and $\sim30$~\AA, respectively.
As the two weakest line emitters, the low SNR limits our analysis of these objects.

We stress that our sample consists of
unresolved galaxies, each likely containing multiple large star-forming regions with varying physical conditions and stellar populations, which may include both radiation- and SN-dominated regions. 
This may explain the overlap in properties of galaxies with Gaussian versus power-law wings. For instance, while all objects with $O_{32}>5$ show power-law wings and are likely radiation-dominated, we 
may be seeing
a mix of feedback modes at lower $O_{32}$ values, 
that are driven by different SSCs in the same aperture.

Such diversity of stellar populations and feedback mechanisms
is evident in the spatially resolved observations of the closest known LCE, Haro 11, where three star-forming knots with widely differing stellar populations, gas properties, and corresponding 
\textcolor{black}{
feedback and LyC properties are observed \citep{Komarova2024, Kunth2003, Adamo2010, Keenan2017, Ostlin2021, LeReste2024}.  
If spatially unresolved, }
Haro~11 would be observed in the ``composite" region of the parameters space in $O_{32}$, $\log(\rm O/H)+12$, and EW(H$\beta$), despite hosting two extreme, radiation-dominated knots.

\subsection{Radiation Driving of Power-Law Wings}
\label{sec:Rad-driving}

\begin{figure}
\includegraphics[width=\columnwidth]{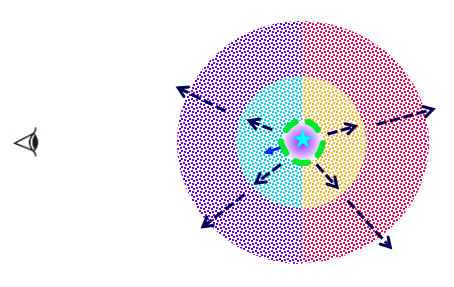} 
\caption{
LyC photons escape through gaps in the dense, fragmented shell (green) surrounding the parent cluster and drive a superwind composed of tiny neutral gas clumps with a low filling factor.  These radiation-driven wind clumps continue to absorb photons and accelerate, reaching higher speeds at  larger radial distances (arrows), corresponding to respectively larger Doppler shifts (red and blue color coding).  
The dense clouds have a slow, momentum-conserving outward motion driven by the weak stellar winds. 
}
\label{fig:cartoon}
\end{figure}

\citet{Komarova2021} used spatially resolved observations of the nearby, radiation-dominated starburst Mrk 71-A to establish the existence of a very thin superwind driven by LyC and/or Ly$\alpha$ photons.  
\textcolor{black}{It has a radially increasing velocity profile extending to hundreds of pc from the parent SSC, and appears to consist of neutral, dense ``bullets" with a very low filling factor $\sim 10^{-3} - 10^{-2}$.  This very young cluster still retains dense molecular clouds within 10 pc, and thus the wind is able to escape through low-density gaps between them. This implies that some LyC and/or Ly$\alpha$ photons also escape, to continue accelerating the wind \citep{Komarova2021}. 
We propose that the radiation-driven winds that we associate with the power-law line wings in our sample can be described by a similar model (Figure \ref{fig:cartoon}).}

Stellar radiation feedback dominates in starbursts with young ages and high ionization parameters, where
values of $\log(U)\gtrsim -2$
signal that radiation pressure can drive outflows \citep[e.g.,][]{Yeh2012}. 
Mrk 71-A, characterized by an age of $\sim 1$~Myr, EW(H$\beta$) = $505$~\AA, and $O_{32} = 23$ \citep{Micheva2017}, is a strong example of such conditions. 
\textcolor{black}{The SSC is compact 
and hosts very massive stars (VMS, $> 100~\rm M_{\odot}$) \citep{Smith2023}, making it especially relevant for
radiation-driven feedback during its first few million years.
We therefore expect the properties of our objects dominated by radiation feedback to have similar properties.}

\begin{figure}
\includegraphics[width=\columnwidth]{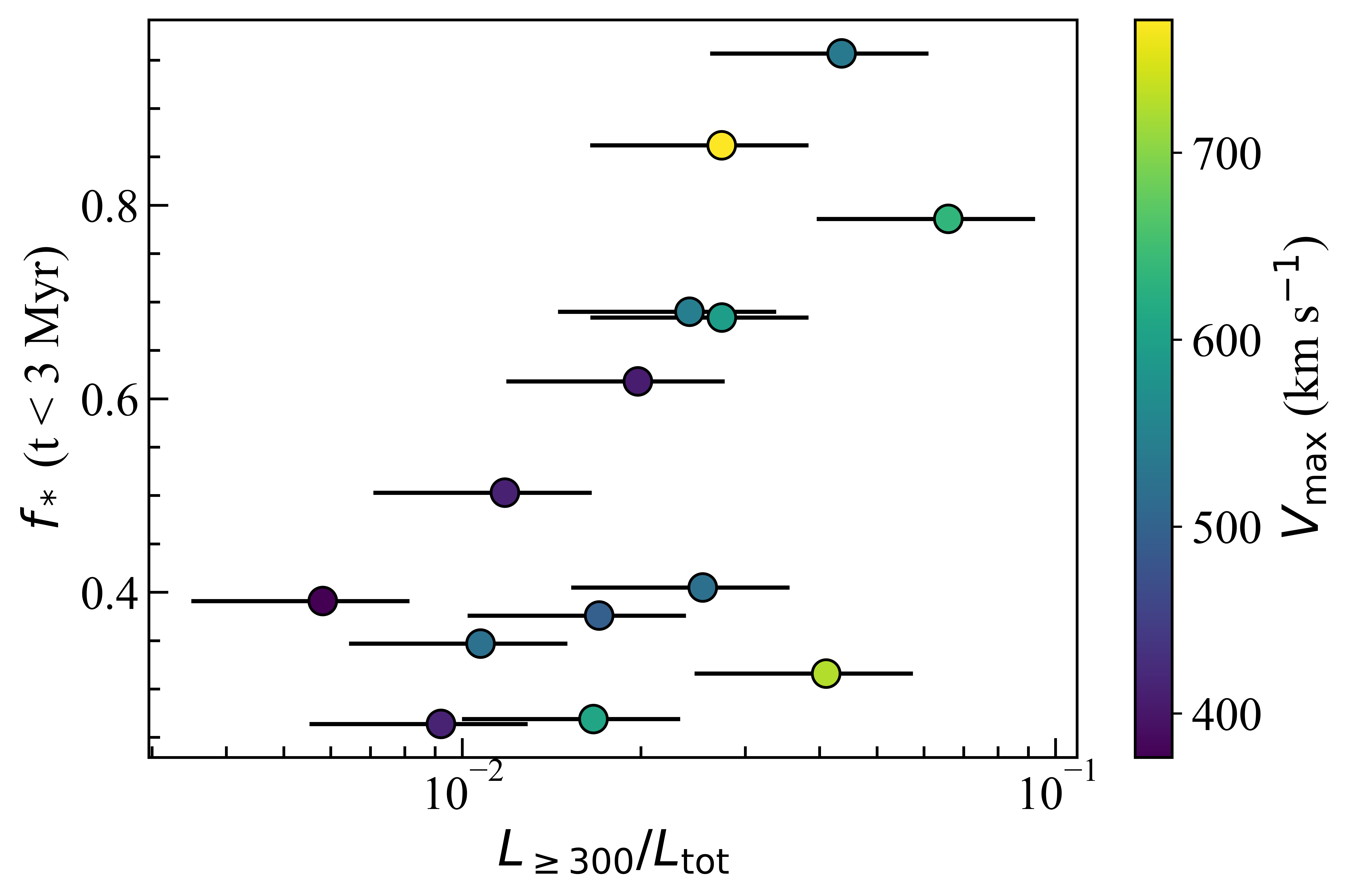} 
\caption{Normalized broad-wing luminosity vs. light fraction of $< 3$~Myr-old stellar populations, color-coded by $V_{\rm max}$.}
\label{fig:LF3_L300}
\end{figure} 

We list the correlations between galaxy properties and power-law wing parameters in Table \ref{table:O3corr}.
As shown in Figure~\ref{fig:LF3_L300}, we see a significant correlation of relatively brighter wings for 
objects increasingly dominated by the young, $<3$~Myr-old, population
($\tau = 0.49,~p = 1.4\times10^{-2}$). Given that this population is too young to be SN-dominated, this link serves as evidence that the high-velocity gas is due to radiation driving. At subsolar metallicities, where the onset of SNe is delayed and stellar winds are weak, mechanical feedback cannot provide momentum comparable to that from a LyC-driven superwind. From Starburst99 \citep{Leitherer2014} simulations of a $10^6 \rm~M_{\odot}$ cluster at $Z = 0.004$, close to the mean metallicity of our sample, we see that at 2~Myr, the mechanical momentum input $dp/dt$ is $\sim 5 \times$ lower than that from LyC luminosity,based on the mass-loss prescription by \cite{Vink2001}. 
However, more recent studies show that these values are more likely to be overestimated by one to two orders of magnitude \citep{Ramachandran2019, Rickard2022, Bjorklund2023}.
Our observations are therefore consistent with LyC radiation dominating the feedback from
the young populations observed in this low-metallicity sample, and the scenario that the broad, power-law line wings are a signature of momentum-driven winds.

\movetabledown=2.8in
\begin{rotatetable*} 
\begin{deluxetable*}{lcccccccccccc}
\tablecaption{Correlations of Broad-Wing Parameters with Galaxy Properties in Objects with Power-Law \OIII\ Wings\label{table:O3corr}}
\tablehead{
    \colhead{} & \multicolumn{2}{c}{$V_{\rm max}$} & \multicolumn{2}{c}{$L_{> 300}$} & 
    \multicolumn{2}{c}{$L_{> 300}/L_{\rm tot}$} & \multicolumn{2}{c}{$\alpha_{\rm blue}$} & \multicolumn{2}{c}{$\alpha_{\rm red}$} &
    \multicolumn{2}{c}{$\Delta\alpha$} \\
    \colhead{} & \colhead{$\tau$} & \colhead{$p$} & \colhead{$\tau$} & \colhead{$p$} & 
    \colhead{$\tau$} & \colhead{$p$} & \colhead{$\tau$} & \colhead{$p$} & 
    \colhead{$\tau$} & \colhead{$p$} & 
    \colhead{$\tau$} & \colhead{$p$}
}
\startdata
\fesc & $\mathit{0.34_{-0.15}^{+0.15}}$ & $\mathit{9.0\times10^{-2}}$ & $0.19_{-0.15}^{+0.13}$ & $3.5\times10^{-1}$ & $0.12_{-0.15}^{+0.15}$ & $5.5\times10^{-1}$ & $-0.24_{-0.15}^{+0.13}$ & $2.3\times10^{-1}$ & $0.10_{-0.15}^{+0.11}$ & $6.2\times10^{-1}$ & $0.27_{-0.15}^{+0.15}$ & $1.7\times10^{-1}$ \\
${f_{\rm{esc}}^{\rm Ly\alpha}}$ & $0.05_{-0.13}^{+0.15}$ & $7.8\times10^{-1}$ & $0.16_{-0.11}^{+0.09}$ & $4.1\times10^{-1}$ & $0.23_{-0.13}^{+0.13}$ & $2.5\times10^{-1}$ & $0.18_{-0.13}^{+0.11}$ & $3.8\times10^{-1}$ &  $\mathit{0.38_{-0.11}^{+0.09}}$  & $\mathit{5.5\times10^{-2}}$ & $0.30_{-0.13}^{+0.13}$ & $1.4\times10^{-1}$ \\
$L_{\rm LyC, obs}$ & $0.30_{-0.15}^{+0.15}$ & $1.4\times10^{-1}$ & $\mathbf{0.43_{-0.15}^{+0.13}}$ & $\mathbf{3.3\times10^{-2}}$ & $\mathbf{0.54_{-0.15}^{+0.13}}$ & $\mathbf{7.3\times10^{-3}}$ & $0.07_{-0.13}^{+0.13}$ & $7.4\times10^{-1}$ & $\mathbf{0.58_{-0.11}^{+0.11}}$& $\mathbf{3.7\times10^{-3}}$ &$\mathbf{0.45_{-0.15}^{+0.13}}$ & $\mathbf{2.5\times10^{-2}}$ \\
$L_{\rm LyC, abso}$ & $-0.03_{-0.13}^{+0.11}$ & $8.7\times10^{-1}$ & $0.30_{-0.11}^{+0.11}$ & $1.4\times10^{-1}$ & $\mathit{0.34_{-0.11}^{+0.13}}$ & $\mathit{8.9\times10^{-2}}$ & $\mathbf{0.50_{-0.06}^{+0.09}}$ & $\mathbf{1.2\times10^{-2}}$ & $\mathbf{0.43_{-0.11}^{+0.11}}$ & $\mathbf{3.3\times10^{-2}}$ & $0.08_{-0.13}^{+0.13}$ & $7.0\times10^{-1}$ \\
EW(Ly$\alpha$) & $0.32_{-0.11}^{+0.11}$ & $1.1\times10^{-1}$ & $0.12_{-0.09}^{+0.11}$ & $5.5\times10^{-1}$ & $-0.08_{-0.13}^{+0.13}$ & $7.0\times10^{-1}$ & $\mathbf{-0.57_{-0.02}^{+0.02}}$ & $\mathbf{4.4\times10^{-3}}$ & $-0.19_{-0.11}^{+0.11}$ & $3.5\times10^{-1}$ & $0.21_{-0.13}^{+0.13}$ & $3.0\times10^{-1}$ \\
EW(H$\beta$) & $\mathit{0.36_{-0.06}^{+0.09}}$ & $\mathit{7.1\times10^{-2}}$ & $0.12_{-0.07}^{+0.09}$ & $5.5\times10^{-1}$ & $-0.08_{-0.09}^{+0.11}$ & $7.0\times10^{-1}$ & $\mathbf{-0.66_{-0.02}^{+0.02}}$ & $\mathbf{1.0\times10^{-3}}$ &  $-0.27_{-0.09}^{+0.09}$ & $1.7\times10^{-1}$ & $0.16_{-0.13}^{+0.13}$ & $4.1\times10^{-1}$ \\
SFR & $\mathbf{0.43_{-0.04}^{+0.06}}$ & $\mathbf{3.3\times10^{-2}}$ & $\mathbf{0.47_{-0.04}^{+0.04}}$ & $\mathbf{1.8\times10^{-2}}$ & $0.16_{-0.07}^{+0.07}$ & $4.1\times10^{-1}$ & $\mathbf{-0.48_{-0.02}^{+0.02}}$ & $\mathbf{1.6\times10^{-2}}$ & $0.10_{-0.04}^{+0.07}$ & $6.2\times10^{-1}$ & $\mathbf{0.41_{-0.10}^{+0.10}}$ & $\mathbf{4.3\times10^{-2}}$ \\
$\Sigma_{\rm SFR, H\beta}$ & $0.32_{-0.11}^{+0.11}$ & $1.1\times10^{-1}$ & $\mathit{0.36_{-0.11}^{+0.11}}$ & $\mathit{7.1\times10^{-2}}$ & $0.14_{-0.15}^{+0.15}$ & $4.8\times10^{-1}$ & $\mathit{-0.37_{-0.04}^{+0.06}}$ & $\mathit{6.3\times10^{-2}}$ & $-0.10_{-0.15}^{+0.15}$ & $6.2\times10^{-1}$  & $0.03_{-0.18}^{+0.15}$ & $8.7\times10^{-1}$ \\
$M_{1500}$ & $-0.01_{-0.11}^{+0.13}$ & $9.5\times10^{-1}$ & $\mathbf{-0.41_{-0.09}^{+0.11}}$ & $\mathbf{4.3\times10^{-2}}$ & $\mathit{-0.34_{-0.11}^{+0.13}}$ & $\mathit{8.9\times10^{-2}}$ & $-0.20_{-0.11}^{+0.15}$ & $3.2\times10^{-1}$ & $\mathbf{-0.43_{-0.11}^{+0.11}}$ & $\mathbf{3.3\times10^{-2}}$ & $-0.30_{-0.13}^{+0.15}$ & $1.4\times10^{-1}$ \\
$\log(\rm O/H)$ & $0.05_{-0.13}^{+0.13}$ & $7.8\times10^{-1}$ & $0.12_{-0.11}^{+0.11}$ & $5.5\times10^{-1}$ & $0.27_{-0.15}^{+0.13}$ & $1.7\times10^{-1}$ & $0.20_{-0.13}^{+0.11}$ & $3.2\times10^{-1}$ & $0.12_{-0.09}^{+0.09}$ & $5.5\times10^{-1}$ & $0.16_{-0.15}^{+0.11}$ & $4.1\times10^{-1}$ \\
$O_{32}$ & $0.19_{-0.13}^{+0.13}$ & $3.5\times10^{-1}$ & $-0.10_{-0.11}^{+0.09}$ & $6.2\times10^{-1}$ & $-0.30_{-0.13}^{+0.15}$ & $1.4\times10^{-1}$ & $\mathbf{-0.44_{-0.04}^{+0.02}}$ & $\mathbf{2.8\times10^{-2}}$ & $-0.32_{-0.15}^{+0.13}$ & $1.1\times10^{-1}$ & $-0.14_{-0.18}^{+0.15}$ & $4.8\times10^{-1}$ \\
$f_*(t < \rm 3)$ & $0.16_{-0.09}^{+0.09}$ & $4.1\times10^{-1}$ & $0.27_{-0.11}^{+0.11}$ & $1.7\times10^{-1}$ & $\mathbf{0.49_{-0.13}^{+0.11}}$ & $\mathbf{1.4\times10^{-2}}$ & $\mathit{0.37_{-0.10}^{+0.10}}$ & $\mathit{6.3\times10^{-2}}$ & $\mathit{0.36_{-0.11}^{+0.11}}$ & $\mathit{7.1\times10^{-2}}$ & $0.05_{-0.15}^{+0.11}$ & $7.8\times10^{-1}$ \\
$r_{50, \rm UV}$ & $\mathit{-0.38_{-0.13}^{+0.13}}$ & $\mathit{5.5\times10^{-2}}$ & $-0.27_{-0.13}^{+0.13}$ & $1.7\times10^{-1}$ & $-0.08_{-0.18}^{+0.18}$ & $7.0\times10^{-1}$ & $\mathbf{0.48_{-0.11}^{+0.11}}$ & $\mathbf{1.6\times10^{-2}}$ & $0.08_{-0.15}^{+0.18}$ & $7.0\times10^{-1}$ & $-0.27_{-0.18}^{+0.18}$ & $1.7\times10^{-1}$ \\
E(B-V) & $0.27_{-0.15}^{+0.13}$ & $1.7\times10^{-1}$ & $0.16_{-0.15}^{+0.13}$ & $4.1\times10^{-1}$ & $\mathit{0.38_{-0.18}^{+0.15}}$ & $\mathit{5.5\times10^{-2}}$ & $0.04_{-0.15}^{+0.18}$ & $8.3\times10^{-1}$ & $\mathbf{0.43_{-0.18}^{+0.15}}$ & $\mathbf{3.3\times10^{-2}}$ &  $\mathbf{0.47_{-0.18}^{+0.18}}$ & $\mathbf{1.8\times10^{-2}}$ \\
\enddata
\tablecomments{For each combination of correlated parameters 
for the sample of 14 objects, we show the Kendall rank correlation coefficient $\tau$ and $p$ values. 
Tentative correlations with $ 0.05 < p \leq 0.1$ are italicized, and 
significant correlations with $p \leq 0.05$ are boldfaced.}
\end{deluxetable*}
\end{rotatetable*}

\begin{figure*}
\centering
\includegraphics[width=0.8\textwidth]{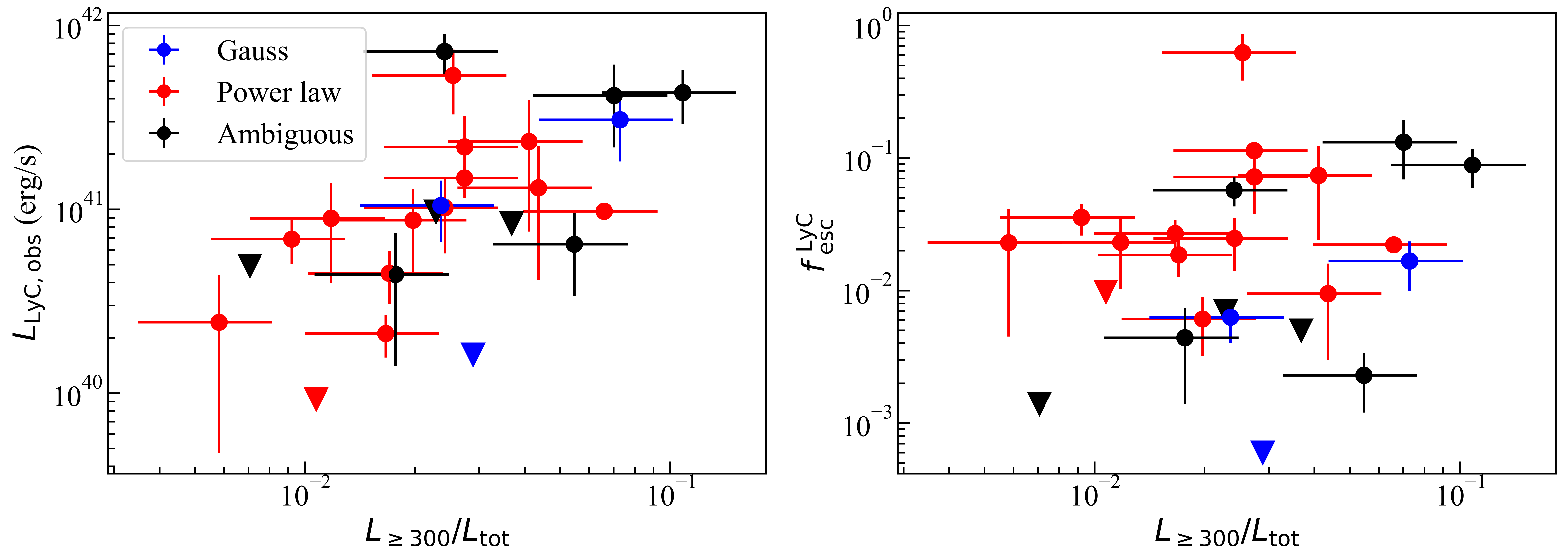} 
vc b\caption{Left: observed LyC luminosity vs.  normalized broad-wing luminosity $L_{\rm \geq 300}/L_{\rm tot}$. Right: LyC escape fraction \fesc~vs. $L_{\rm \geq 300}/L_{\rm tot}$.}
\label{fig:highvel_LLyC}
\end{figure*}

The radiative transfer of LyC for radiation-dominated feedback is expected to occur via the ``picket-fence" geometry \citep{Heckman2001, Heckman2011, Gazagnes2018, Steidel2018, Jaskot2019, Gazagnes2020}, where LyC escapes via optically thin windows among optically thick clouds (Figure~\ref{fig:cartoon}). In the LzLCS sample, stronger LyC leakers show higher HI and absorption line residual fluxes for low ionization species (LIS), indicating that LyC likely escapes via low-density paths intermixed with higher-density channels \citep{Saldana-Lopez2022}. We find that the normalized wing luminosity $L_{\geq300}/L_{\rm tot}$ correlates with the leaked LyC luminosity $L_{\rm LyC, obs}$ ($\tau = 0.54,~p = 7.3\times10^{-3}$); however, it does not significantly correlate with the LyC escape {\it fraction} \fesc~($\tau = 0.12,~p = 5.5\times10^{-1}$; Figure \ref{fig:highvel_LLyC}).
This suggests that the wind emission emerges via the same optically thin channels as the LyC radiation in these leakers. The lack of correlation with \fesc\ indicates that the emerging wind luminosity 
is more closely linked to the LyC luminosity than 
the covering fraction of dense gas,
consistent with the LyC-driven model for the wind. Indeed,
we find no significant correlations of the fractional wing luminosity with the H~I covering fractions $C_{\rm HI}$ derived by \cite{Saldana-Lopez2022} ($\tau = 0.09,~p = 4.9\times10^{-1}$). 
Thus, these results are fully consistent with LyC driving the wind through the picket-fence geometry, where both emerge between clouds of dense gas.

\begin{figure*}
\centering 
\includegraphics[width=0.7\textwidth]{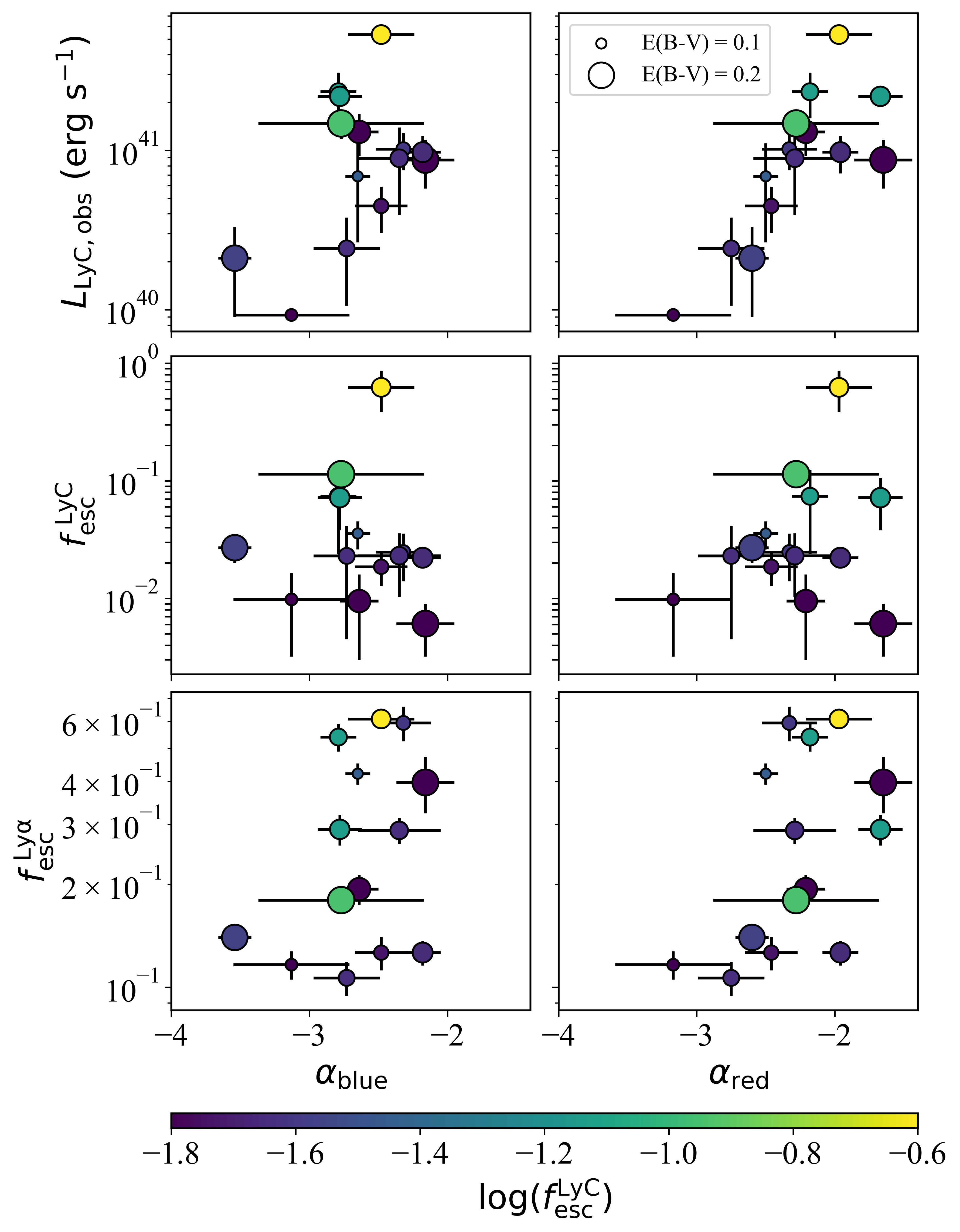} 
\caption{Power-law slopes $\alpha_{\rm blue}$ and $\alpha_{\rm red}$ of the broad emission-line wings  vs observed LyC luminosity (top), LyC escape fraction (middle), and Ly$\alpha$ escape fraction (bottom). The points are color-coded by \fesc\ and sized by $E(B-V)$. }
\label{fig:slopes_LyC}
\end{figure*}

\subsection{Broad-wing power-law slope and LyC escape
\label{sec:slopeLyC}
}

In addition to the broad-wing velocity and luminosity, a fundamental parameter characterizing the power-law wings is the power-law slope $\alpha$, which also
shows trends linked to LyC escape.
As we show below, it appears to represent the combination of the intrinsic wind velocity profile and the optical depth structure. 
As described in Section~\ref{sec: analysis}, we measure $\alpha$ independently in the blue and red wings as $\alpha_{\rm blue}$ and $\alpha_{\rm red}$, respectively. 
The blue wing originates from the near-side, approaching part of the wind, and the red one from the far side, as illustrated in Figure~\ref{fig:cartoon}. 
 
Both blue and red slopes appear to be linked to the youngest, UV-dominant population.
Both \alphablue\ and \alphared\ show a tentative correlation with the UV light fraction of $<3$~Myr-old stars $f_*(t < \rm 3)$ (blue: $\tau = 0.37,~p = 6.3\times10^{-2}$; red: $\tau = 0.36,~p = 7.1\times10^{-2}$; 
Table~\ref{table:O3corr}). Both slopes are shallower for higher fractions of $<3$~Myr stars, suggesting that the wind velocity profile is faster for populations increasingly dominated by the youngest stars.
This shared trend therefore suggests that the wing profiles 
are likely determined largely by the intrinsic wind velocity structure linked to radiation-driving.

Figure \ref{fig:slopes_LyC} shows the relations of  \alphablue\ and \alphared\ to \fesc, as well as to  
the observed LyC luminosity $L_{\rm LyC, obs}$, and $f_{\rm esc}^{\rm Ly\alpha}$. The figure is color-coded by \fesc, and we see a possible trend where, for a given \fesc, galaxies with stronger $L_{\rm LyC, obs}$ tend to have higher, flatter values of \alphablue; however, there is no such trend between \fesc\ and \alphablue\ (see also Table~\ref{table:O3corr}).  This pattern is similar to that seen in Figure~\ref{fig:highvel_LLyC}.
It therefore suggests that $\alpha$ is linked directly to the wind luminosity, implying that more luminous winds also have velocity profiles weighted more toward high velocities.
This further supports
our suggestion in the preceding section that the wind emission is fundamentally linked to the LyC luminosity, rather than the escape fraction, and consistent with picket-fence radiative transfer. 

Thus, our data suggest that stronger, more luminous leakers may tend to have shallower slopes.
Such a scenario may be expected in the radiation-driven wind case, since
more escaping photons allow prolonged acceleration of the wind to larger velocities, which  
may lead to an observed velocity profile weighted toward higher values, and hence, a shallower slope.
It is less clear whether there is a similar trend with higher $f_{\rm esc}^{\rm Ly\alpha}$ (Figure \ref{fig:slopes_LyC}).
This may imply that LyC, rather than Ly$\alpha$ opacity, dominates the wind acceleration.

\subsection{
Broad-wing power-law slope and extinction
}\label{sec:slope_dust}

\begin{figure*}
\centering
\includegraphics[width=\textwidth]{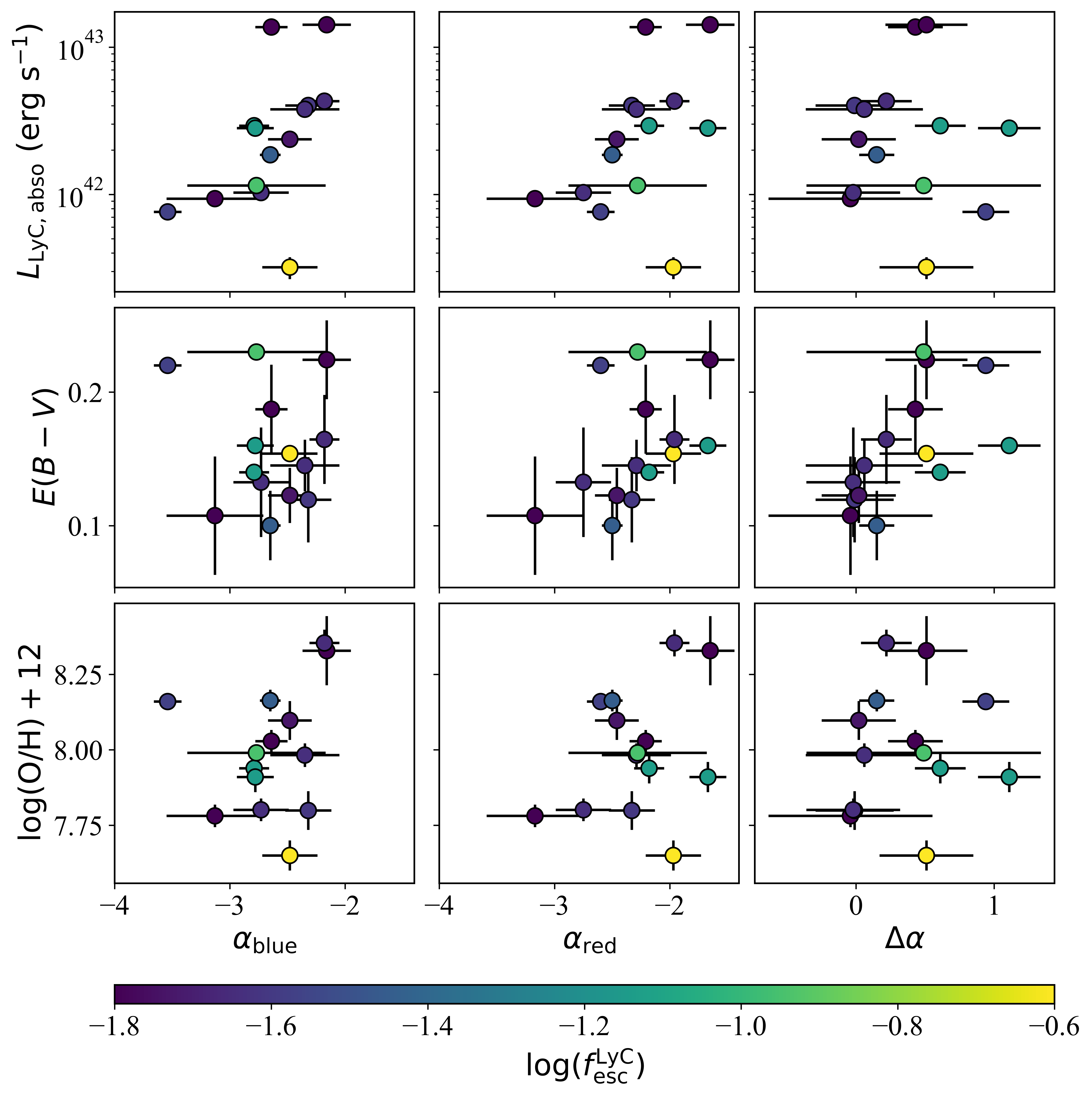} 
\caption{Power-law slopes $\alpha_{\rm blue}$ and $\alpha_{\rm red}$, and the slope difference $\Delta\alpha$ vs absorbed LyC luminosity (top), $B-V$ color excess (middle), and oxygen abundance (bottom). The points are color-coded by \fesc. }
\label{fig:slopes_EBV_metal}
\end{figure*}

\begin{figure*}
\includegraphics[width=\columnwidth]{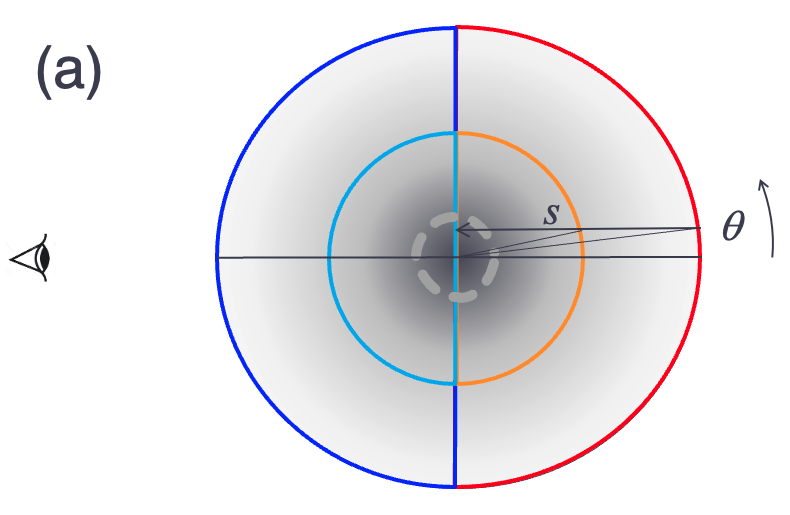}
\hspace*{0.2in}
\includegraphics[width=\columnwidth]{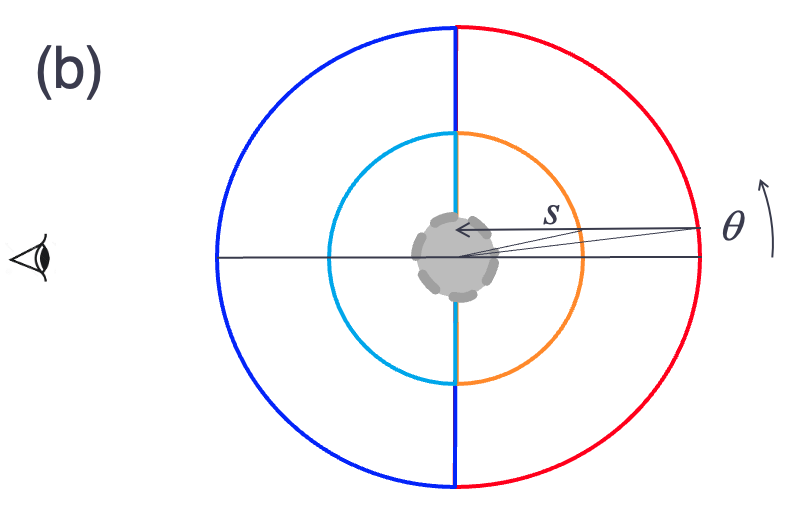}
\includegraphics[width=\columnwidth]{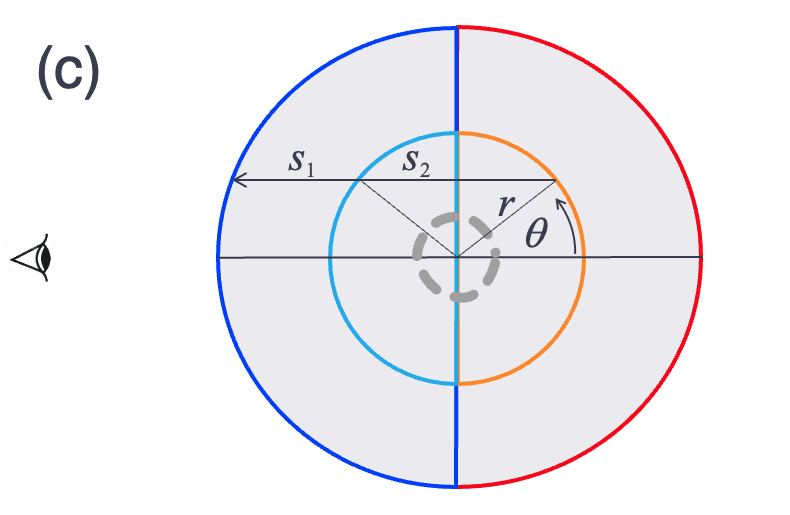}
\hspace*{0.3in}
\includegraphics[width=\columnwidth]{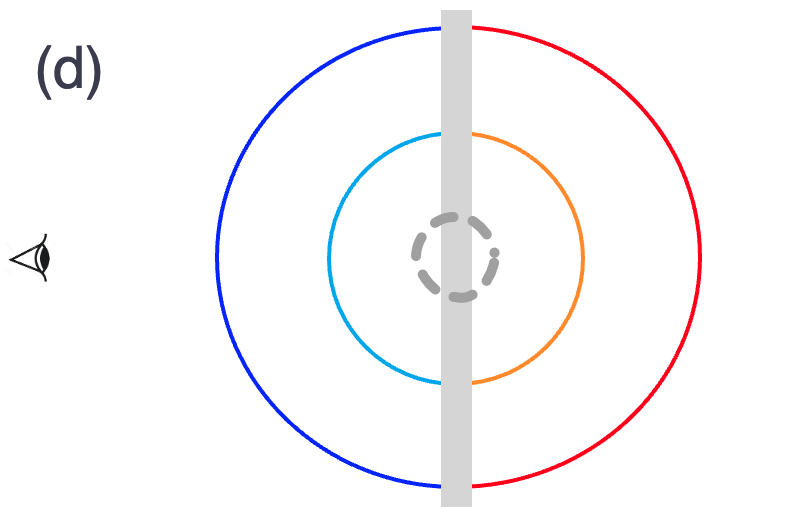}
\caption{Sketches of different dust configurations (shaded gray). 
Panel (a) shows a model for centrally concentrated extinction from \cite{Carr2021}, and Panel (b) shows one in which extinction originates exclusively from the central dense zone, which is much smaller than the radial extent of the emitting wind.  
In these two panels, the shown sightline $s$ includes photons emitted from slower gas at smaller radius (orange) and fast gas at large radius (red); these originate from excitation by UV radiation from the SSC emitted over a larger range of $\theta$ for slow gas relative to fast gas.  Panel (c) shows a model for dust distributed uniformly within the wind volume.  For the shown sightline (straight arrow), the blueshifted path length is given by $s_1$ and the redshifted one by $s1+2s_2$.  For the model in Panel (d), the extinction originates from a slab across the center of the system.
See text for details on how these models affect the emission-line profiles.
}
\label{fig:cartoon2}
\end{figure*}

\begin{figure*}
\centering
\includegraphics[width=0.6\textwidth]{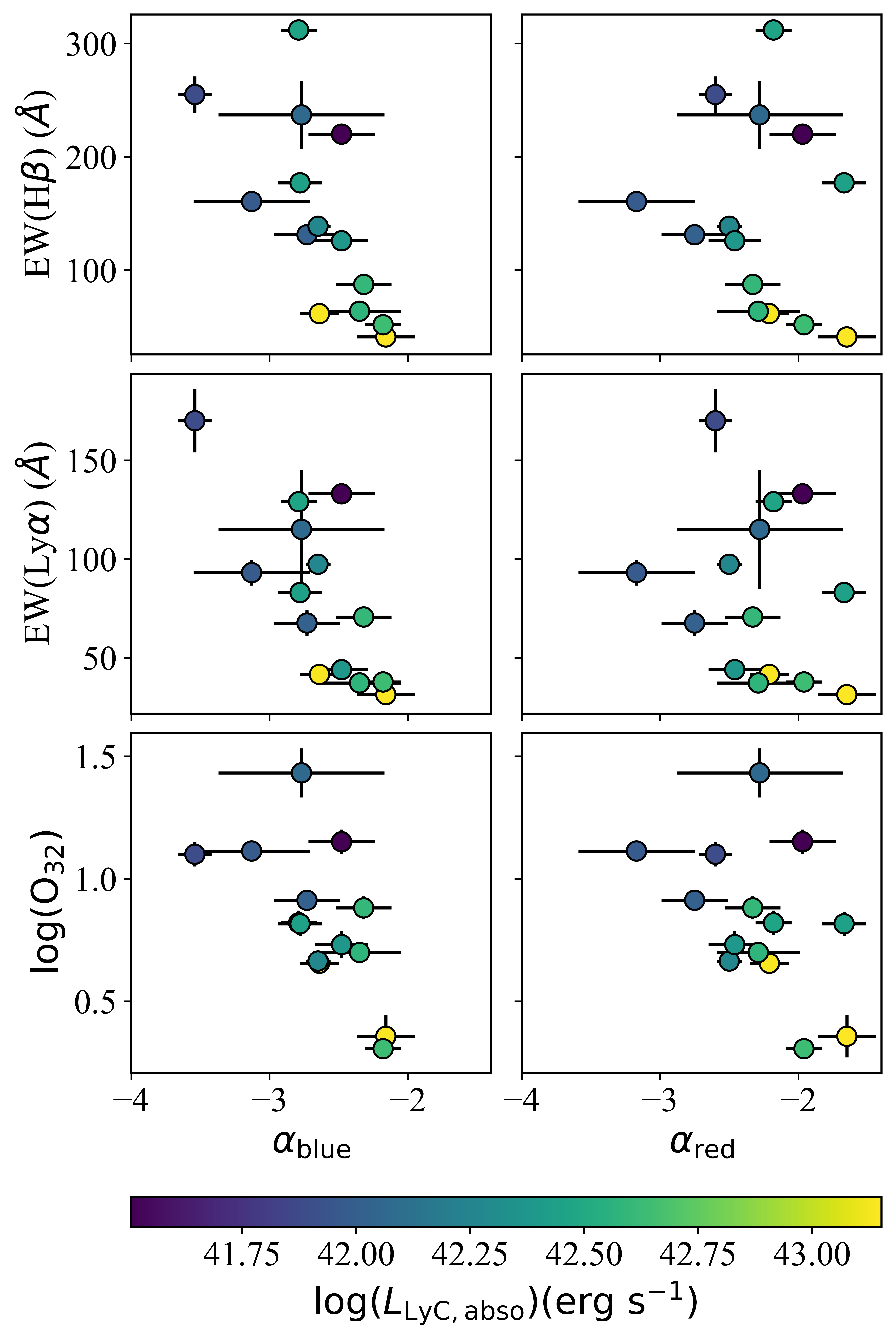} 
\caption{Blue and red wing power-law slopes \alphablue\ and \alphared\
vs  EW(H$\beta$) (top),  EW(Ly$\alpha$) (middle), and $O_{32}$ (bottom). The points are color-coded by the absorbed LyC luminosity.}
\label{fig:slope_Blue}
\end{figure*}

We find that the red wing is systematically shallower than the blue wing (Table \ref{table:broadwing}), with measured  slope differences $\Delta\alpha = \alpha_{\rm red}-\alpha_{\rm blue}$ in the range $-0.04$ to $1.1$. 
As shown by \citet{Carr2021}, this can be explained by dust extinction.
Figure~\ref{fig:slopes_EBV_metal} shows the relations of \alphared, \alphablue, and $\Delta\alpha$ with the absorbed LyC luminosity $L_{\rm LyC, abso} = L_{\rm LyC, int}-L_{\rm LyC, obs}$, where $L_{\rm LyC, int}$ and $L_{\rm LyC, obs}$ are the intrinsic and observed LyC luminosities, respectively; and also
with $E(B-V)$ and $12+\log$(O/H).  Trends emerge for \alphared\ when considering a given \fesc, keeping in mind that \fesc\ anticorrelates with extinction \citep[e.g.,][]{Saldana-Lopez2022}. 
The correlations are much weaker for \alphablue, which is consistent with expectation, given the longer path lengths for redshifted vs blueshifted emission.
We find that the slope difference $\Delta\alpha$ also increases with higher dust extinction $E(B-V)$ ($\tau = 0.47,~p = 1.8\times10^{-2}$), as shown in Figure \ref{fig:slopes_EBV_metal} and Table~\ref{table:O3corr}, consistent with this scenario.
We note that $E(B-V)$ is derived from the nebular Balmer lines rather than UV extinction, and so it may not be especially sensitive to line-of-sight effects probed here.

The relations in Figure~\ref{fig:slopes_LyC} show significantly stronger trends for \alphared\ than \alphablue, especially that for $L_{\rm LyC, obs}$ (see also Table~\ref{table:O3corr}).  
In Figure~\ref{fig:slopes_LyC}, the size of the points increases with higher $E(B-V)$, demonstrating that the correlations are enhanced in the \alphared\ plots largely by the shifted positions of galaxies with higher $E(B-V)$.  Thus, the prominent correlations for \alphared\ may be somewhat misleading since they are driven in part by extinction.

The fact that \alphared\ is shallower with extinction is an important diagnostic of the dust geometry relative to the wind.
In Figure~\ref{fig:cartoon2}, we consider four different configurations for the radiative transfer.  Their effects on the emission-line profile rely on the expectation that the radiation-driven wind velocity increases with radius, as discussed above (Section~\ref{sec:Rad-driving}).
Figure~\ref{fig:cartoon2}a shows a radially decreasing power-law density distribution in which the dust is cospatial with the emitting wind.
\citet{Carr2021} show that such a model results in a flattening of the red emission-line wing.  Figure~\ref{fig:cartoon2}b considers a model where all the extinction originates from the central, compact dusty region, whose radius is much smaller than the extent of the wind.  This model is also expected to flatten the slope of the red wing:  
in both of these models, the centrally concentrated extinction disproportionately attenuates redshifted wind emission originating at smaller, slower radial zones (orange) 
compared to that from larger, faster ones (red).  For a given sightline $s$, these smaller radial zones correspond to emission excited by UV originating from a larger range of angles $\theta$, when compared to the contribution from larger radial zones.  This results in a greater relative attenuation in luminosity from lower velocity gas, flattening the red slope.  

On the other hand, Figure~\ref{fig:cartoon2}c shows dust that is uniformly distributed within the wind volume, rather than centrally concentrated.  Here, the extinction is determined only by the path length in the line of sight.  As shown in Figure~\ref{fig:cartoon2}c, for a given sightline, the components contributing from the same radial zone are specified by $|\theta|$.  The blueshifted and redshifted path lengths are given by $s_b = s_1$ and $s_r = s_1 + 2s_2$, respectively. We see that the difference $s_r - s_b = 2r\cos\theta$ is largest when $\cos\theta = 1$, and decreases with $\cos\theta$. Therefore, the largest projected velocities experience the greatest attenuation at all radial zones, implying that the slope of the red emission-line wing steepens.
Alternatively, a uniform dust screen through the center of the system and perpendicular to the line of sight (Figure~\ref{fig:cartoon2}d) would attenuate all emission from the red wing uniformly.  This would simply reduce the normalization of the red wing luminosity and it would not affect the slope.

Therefore, the observed trend of flatter \alphared\ with higher dust extinction points to 
centrally concentrated extinction as in Figures~\ref{fig:cartoon2}a and \ref{fig:cartoon2}b.  This is consistent with the nature of Green Peas as extremely compact systems where the gas remains close to the ionizing SSC, with LyC radiation escaping through a picket-fence configuration for the dust \citep[e.g.,][]{Jaskot2019}.  
However, we cannot easily distinguish between the two models in Figures \ref{fig:cartoon2}a and \ref{fig:cartoon2}b from the red wing flattening alone.

Instead, the behavior of the blue wing slope can help discriminate between these two models.
For the blueshifted emission in Figure~\ref{fig:cartoon2}b, we see that the attenuation is generated only at very small radii, and therefore it is limited to the very lowest velocities.  Therefore, the blue slope is largely unchanged. In contrast, in Figure~\ref{fig:cartoon2}a, the dust density tracks the wind density, which would cause some attenuation and some slope flattening for all blueshifted wind material, although not as much as for the redshifted side.

Parameters with strong correlations for \alphablue\ are mostly linked to the stellar continuum of the young, UV-bright population, which is directly accessible to our line of sight; this is the side of the wind probed by \alphablue.
There is a general trend that less extreme starbursts with dustier, weaker UV-emitting populations, and which have less LyC leakage in our line of sight, have flatter blue slopes; this trend is opposite to that in Section~\ref{sec:slopeLyC} where more luminous LyC emitters show flatter slopes.
Table~\ref{table:O3corr} indicates that \alphablue\ shows trends with various galaxy 
parameters that are all consistent with the blue slope flattening with higher extinction and metallicity (Figure~\ref{fig:slope_Blue}):
\alphablue\ flattens for lower EW(H$\beta$) ($\tau = -0.66,~p = 1.0 \times10^{-3}$) and EW(Ly$\alpha$) ($\tau = -0.57,~p = 4.4 \times10^{-3}$), 
and it is also shallower for lower $O_{32}$ ($\tau = -0.44,~p = 2.8\times10^{-2}$)
and lower star formation rates ($\tau = -0.48,~p = 1.6\times10^{-2}$).
These are all properties that scale with metallicity (e.g., Figures~\ref{fig:O32_metal} and \ref{fig:age_metal}). 
That this is driven by galaxy scaling relations is supported by
the correlation of \alphablue\ with larger UV half-light radii $r_{50}$ ($\tau = 0.48,~p = 1.6\times10^{-2}$), confirming that larger galaxies have flatter slopes.
Similarly, the red slope \alphared\ anticorrelates with $M_{1500}$ ($\tau = -0.43,~p = 3.3\times10^{-2}$).
As shown in Figure~\ref{fig:slopes_EBV_metal},
the lack of significant correlations between the wing slopes and $12 + \log(\rm O/H)$ suggests that metallicity itself is not the principal driver. 
Instead, as shown above, total extinction appears to be more relevant.
Figure \ref{fig:slope_Blue} shows that objects with low $O_{32}$ and H$\beta$ equivalent width have absorbed LyC luminosities that are an order of magnitude higher than the ones with high $O_{32}$ and EW(H$\beta$).
Indeed, both the blue and red slopes have significant correlations with 
$L_{\rm LyC, abso}$ (blue: $\tau = 0.50,~p = 1.2\times10^{-2}$; red: $\tau = 0.43,~p = 3.3\times10^{-2}$)  
as seen in Figure~\ref{fig:slopes_EBV_metal}.  

The sensitivity of the \alphablue\ to extinction therefore favors the model for dust distribution in Figure~\ref{fig:cartoon2}a \citep{Carr2021} rather than Figure~\ref{fig:cartoon2}b.
{\it The dust appears to be distributed with a centrally concentrated density profile that is consistent with occupying much of the wind volume.}  It may even be that the wind material itself is responsible for self-absorption of its emission; further study is needed to determine whether this is reasonable.

Overall, we therefore see that flatter power-law slopes correspond to both stronger radiation-driven winds linked to higher LyC escape and also higher dust extinction.  However, it is important to note that this does {\it not} mean that dustier objects have stronger LyC escape, which would contradict findings in the literature \citep[e.g.,][]{Saldana-Lopez2022, Flury2022b}.  The color coding in Figure~\ref{fig:slopes_EBV_metal} shows that for any given value of $\alpha$ and $\Delta\alpha$, \fesc\ tends toward lower values of $E(B-V)$ and $12+\log$(O/H), as expected.

Extinction can also explain why many parameters that correlate with \alphablue\ in Table~\ref{table:O3corr} do not correlate with \alphared, and vice versa: underlying correlations for \alphablue\ are distorted by extinction for \alphared, while non-correlations in \alphablue\ are often driven to a correlation in \alphared\ that reflects the amount of extinction. 
As noted above, Figures~\ref{fig:slopes_LyC} and \ref{fig:slopes_EBV_metal} show that the points shifting the most between plots for \alphablue\ vs \alphared\ are those with high extinction.

\begin{figure}
\includegraphics[width=\columnwidth]{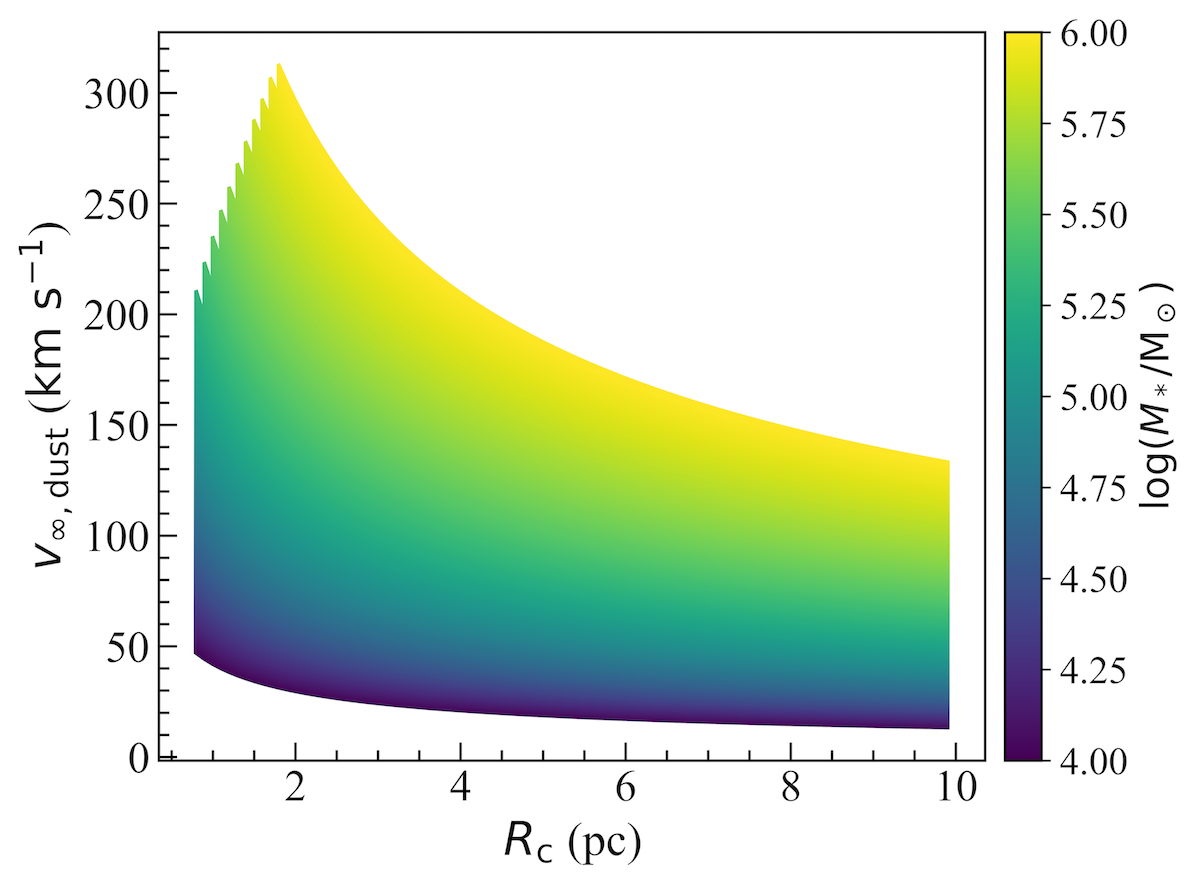} 
\caption{Terminal velocity $v_{\infty, \rm dust}$ of a radiation-driven superwind, accelerated by a cluster with radius $R_{\rm c}$, mass $M_*$, and $M_*/L_*$ ratio of Mrk~71-A, assuming dust opacity $\kappa = 10^3~\rm cm^2~g^{-1}$ (Eq. \ref{eqn:vterm}). We limit the displayed cluster parameter combinations to those that do not exceed the maximum stellar density $\Sigma< 10^5~\rm M_\odot~pc^{-2}$.}
\label{fig:vterm}
\end{figure}

\subsection{Opacity of Radiation-Driven Superwinds}
\label{sec:opacity}
Constraining the driving opacity of radiation-driven superwinds in our sample in principle requires knowledge of individual SSC and feedback 
parameters within each galaxy. In particular, the terminal velocity $v_{\infty}$ of  
the radiation-driven superwind produced by a cluster with mass $M_*$, luminosity $L_*$, and escape velocity $v_{\rm esc}$ is given by \citep{Krumholz2017, Komarova2021}:

\begin{equation}
    v_{\infty} = \sqrt{\frac{L_* \kappa}{4\pi G M_* c} - 1} \times v_{\rm esc} \quad ,
\label{eqn:vterm}
\end{equation}
where $\kappa$ is the opacity of the wind material. Thus, from known cluster properties, the observational lower limit on $v_{\infty} \sim V_{\rm max}$ can be linked to $\kappa$.

We evaluate the possibility of dust driving by comparing the predicted dust-driven velocity to the observed values. Assuming the cluster $L_*/M_* \sim 260~L_\odot/M_\odot$ of Mrk 71-A \citep{Komarova2021}, 
and the UV dust opacity $\kappa = 10^3~\rm cm^2~g^{-1}$ \citep{Draine2003}, we obtain the range of maximum dust velocities shown in Figure \ref{fig:vterm}. We consider the parameter space with cluster radii $R_{\rm c} = 1-10$~pc and cluster masses $M_* = 10^4 -10^6~\rm M_\odot$, limited by the upper limit on stellar density  $\Sigma< 10^5~\rm M_\odot~pc^{-2}$, beyond which radiation pressure will disrupt the cluster \citep{Crocker2018}. The highest velocity that dust can produce in this cluster parameter space, even assuming such extreme radiation fields as in Mrk 71-A, is 
$\sim 300 \rm~km~s^{-1}$. However, all the galaxies we classify as radiation-dominated show $V_{\rm max} \geq 376~\rm km~s^{-1}$ (Table \ref{table:broadwing}).  
While there are some circumstances in which UV dust opacity may reach values up to $10-100\times$ higher than assumed in our fiducial model \citep{Draine2003}, the link between
the power-law wing slope and luminosity to the absorbed LyC luminosity
argues that LyC driving is more likely.

\subsection{SN-driven feedback}  

As we have seen, our results show that SN feedback is more likely to dominate in the subset of our sample galaxies with Gaussian broad wings (Section~\ref{sec:PLvG}).
These objects have 
higher metallicities and lower ionization parameters,
consistent with expectations that SNe are more prevalent at higher metallicity and older ages \citep[e.g.,][]{Jecmen2023}.
Mechanical feedback is known to generate gaussian line profiles in observed systems, e.g., 30 Dor \citep{Chu1994}, which originate from the superposition of many expanding shells. 

While our results support the existence of UV-driven superwinds in the objects with power-law emission-line wings,
such line profiles may not necessarily be exclusively linked to radiation driving.
Moreover, observed core nebular line profiles from energy-driven feedback are often known to be irregular and more strongly asymmetric, as is also observed for some of our power-law sample \citep[e.g.,][]{Shopbell1998, Xu2024}
For example, multiphase, high-resolution hydrodynamic simulations show that $10^4$ K gas entrained in a hot, energy-driven wind can similarly show radially increasing velocities in a mature, $\sim30$ Myr-old, M82-like starburst  \citep[e.g.,][]{Schneider2020, Schneider2024}.  It is possible that such gas could also show broad, power-law wings similar to those in our sample.
The preceding sections demonstrate that the ages, momentum budget, and other parameters preclude this scenario in some of our power-law sample, but it remains possible that the power-law wings for some higher-metallicity galaxies could originate from mechanical, rather than radiative, feedback.
Another possible mechanism for obtaining a power-law profile for SN-driven feedback is turbulent mixing layers
on the surfaces of cool gas clumps interfacing with a hot adiabatic wind \citep{Binette2009, Eggen2021}.
But in general, velocities in excess of $\sim 200 \rm~km~s^{-1}$ are not expected from turbulence, unless driven by AGN \citep{Ulivi2024}, and the same age restrictions for SN-driven superwinds apply.

\subsection{Ly$\alpha$ Emission and Feedback}

We use measurements of the spatially resolved Ly$\alpha$ emission from the Lyman-alpha and Continuum Origins Survey \citep[LaCOS;][]{LeReste2025}. The LaCOS survey is an HST/ACS imaging program (Program ID 17069, PI: Hayes; archival data from Program ID 14131, PI: Orlitov\'a and Program ID 11107, PI: Heckman) for 42 of the 66 total LzLCS galaxies, in a filter set that includes SBC/F150LP and F165LP. In the redshift range $0.23 < z < 0.32$, these two filters cover Ly$\alpha$ line emission and local UV continuum, respectively. The aim of the survey is to map ISM properties and Ly$\alpha$ radiative transfer in LyC leakers and non-leakers.  For \textcolor{black}{15 galaxies cross-matched between LaCOS and our sample, we consider the Ly$\alpha$ parameters presented by \cite{SaldanaLopez2025_lacos}: radii $r_{\rm 20, Ly\alpha}$ and $r_{\rm 90, Ly\alpha}$ containing $20\%$ and $90\%$ of the total Ly$\alpha$ flux, respectively, were measured from continuum-subtracted Ly$\alpha$ images, with the total Ly$\alpha$ flux defined to be integrated over 8\arcsec~apertures; and Ly$\alpha$ halo fraction HF was obtained from 2D fitting of the Ly$\alpha$ emission with a core + halo model and defined as the halo contribution to the total Ly$\alpha$ luminosity. We relate these Ly$\alpha$ parameters to our broad-wing parameters for all wing morphologies.
Since there are only seven power-law objects with LaCOS measurements, we also include galaxies classified to have ambiguous wing morphology when considering power-law slopes.}

\begin{figure}
    \centering
    \includegraphics[width=\columnwidth]{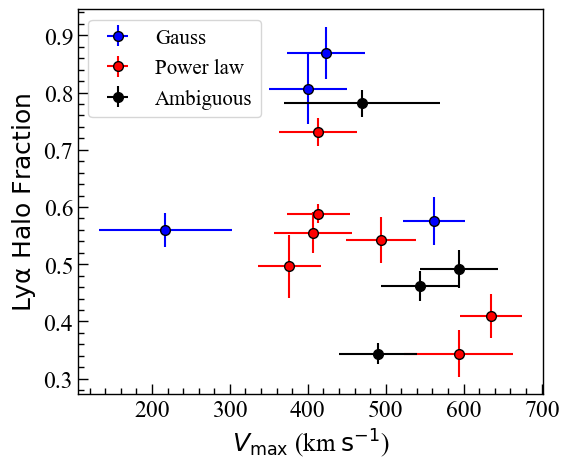}
    \caption{\textcolor{black}{$V_{\rm max}$ vs. Ly$\alpha$ halo fraction (see text). The points are color-coded by wing morphology.} }
    \label{fig:LaCOSHF}
\end{figure}

\textcolor{black}{As shown in Figure \ref{fig:LaCOSHF},
objects with a lower HF tentatively show higher $V_{\rm max}$ ($\tau = -0.33,\ p = 8.3\times10^{-2}$). \cite{SaldanaLopez2025_lacos} find HF to anti-correlate with \fesc, such that stronger leakers have less H~I to scatter Ly$\alpha$ photons. 
Since $V_{\rm max}$ correlates with \fesc\ (Figure~\ref{fig:fesc_vmax}), an anticorrelation between $V_{\rm max}$ and HF is consistent with their findings. 
Figure~\ref{fig:LaCOSHF} shows that this relation is dominated by the offset of objects with Gaussian wings to higher HF:
$0.70\pm0.14$, compared to $0.52\pm0.12$ in objects with power-law wing morphologies or $0.50\pm0.13$ for the combined sample of power-law and ambiguous morphologies. This is in line with Gaussian wings being associated with, on average, lower LyC escape fractions and lower $V_{\rm max}$ (Figure~\ref{fig:fesc_vmax}), as expected if they correspond to older systems dominated by SN feedback.  For such objects, the gas has been dispersed from the parent SSC, while at the same time, a lower fraction of the Ly$\alpha$ is due to the older central starburst; both of these effects increase the HF. Possibly there is also a modest trend among the power-law and ambiguous objects maintaining the link between higher $V_{\rm max}$ and lower HF, although we caution that the trend is not statistically significant ($p = 0.14$). This would again be consistent with a similar trend in Figure~\ref{fig:fesc_vmax} between $V_{\rm max}$ and \fesc.}

\begin{figure}
    \centering
    \includegraphics[width=\columnwidth]{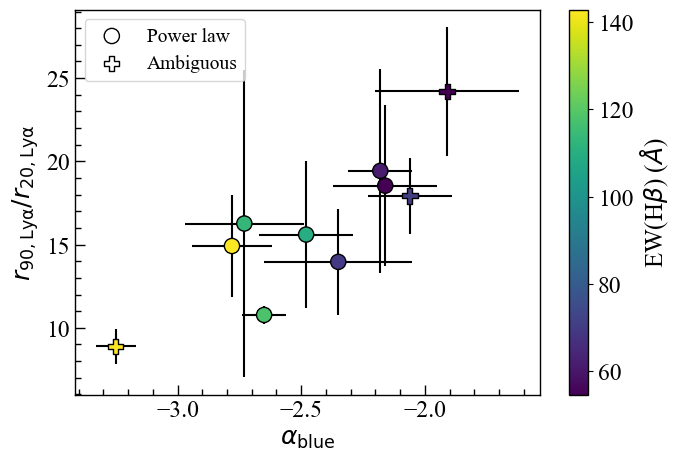}
    \caption{\textcolor{black}{Blue slope $\alpha_{\rm blue}$ vs. $r_{90, \rm Ly\alpha}/r_{20, \rm Ly\alpha}$ for objects with measured power-law wing slopes. The symbols indicate the wing morphology as shown in the legend, and the points are color-coded by EW(H$\beta$).} }
    \label{fig:LacosC}
\end{figure}

\textcolor{black}{As shown in Figure~\ref{fig:LacosC}, the blue power-law slope is shallower for a higher Ly$\alpha$ concentration parameter 
$C_{\rm Ly\alpha} = r_{\rm 90, Ly\alpha}/r_{\rm 20, Ly\alpha}$ \citep{SaldanaLopez2025_lacos}, with 
$\tau = 0.60,\ p = 1.6\times10^{-2}$. 
The correlation is only significant 
when the three objects with ambiguous wing morphologies are also included in the analysis.
For these three objects, a power-law fit to the wing is not significantly preferred over a multi-Gaussian one
(Appendix~\ref{sec:App_ambig}). 
However, these objects may be dominated by the same physical process as 
those with power-law wings, i.e., radiation driving (Section~\ref{sec:PLvG}).  The $C_{\rm Ly\alpha} - \alpha_{\rm blue}$ relation shown by this combined sample suggests that shallower blue wings may be found in 
objects with more concentrated Ly$\alpha$ morphologies. 
The color-coding in Figure~\ref{fig:LacosC} demonstrates that objects with shallower blue slopes and higher $C_{\rm Ly\alpha}$ also have lower EW(H$\beta$).  As noted in Section~\ref{sec:slope_dust}, the relation between \alphablue\ and EW(H$\beta$) appears to be driven by galaxy scaling relations, and the trend with $C_{\rm Ly\alpha}$ therefore may be caused by the effect that larger, more luminous star-forming galaxies are more extended for a central starburst of a given size, proportionately increasing $r_{\rm 90, Ly\alpha}$ relative to $r_{\rm 20, Ly\alpha}$. }

\textcolor{black}{Our findings thus allow a first glance into the possible link between Ly$\alpha$ morphology and stellar feedback modes in confirmed local LCEs. 
Both Ly$\alpha$ halo fraction and concentration parameter appear to be most easily interpreted as driven by the extended emission.  Both parameters are high for objects with older, more metal-rich populations. }

\section{Discussion}
\label{sec: discussion}
\subsection{A New Paradigm for LyC Escape}

In the classical paradigm of LyC escape, SN feedback promotes the leakage of ionizing photons by expelling neutral gas and clearing out optically thin pathways in the ISM \citep[e.g.,][]{Heckman2011, Zastrow2013, Kimm2014, Ma2015, Trebitsch2017, Steidel2018, Barrow2020, Saldana-Lopez2022}. 
However, in the early universe and other metal-poor environments, radiation-driven feedback may instead facilitate LyC escape.
The onset of SNe is expected to be delayed until $\sim 10$~Myr after a star cluster's birth \citep{Jecmen2023, Sukhbold2016}, 
whereas the majority of known LCEs show $2-6$ Myr-old stellar populations \citep{Izotov2016a, Izotov2018b, Saldana-Lopez2022}. 
Stacking analysis by \cite{Flury2024} reveals that while older ($> 8$~Myr) stellar populations are ubiquitous among LCEs, but very young ($<3$~Myr) populations are most important in the most prodigious LCEs.
Galaxies with higher Ly$\alpha$ escape fractions may have slower velocities in LIS absorption lines tracing cold, neutral gas \citep{Henry2015, Jaskot2017, Carr2021, Carr2025}, suggesting weaker or suppressed 
mechanical feedback. Indeed, \cite{Carr2025} find that the strongest LyC leakers have lower mass, momentum, and energy loading in neutral gas.
Stronger LCEs furthermore show flatter radio spectral indices indicative of a lack of non-thermal emission from SN feedback \citep{Bait2023}. In addition, the compactness and high gas densities characteristic of LyC-leaking starbursts may induce catastrophic radiative gas cooling, further diminishing the efficiency of any adiabatic superwinds \citep{Silich2004, Gray2019}. Direct evidence of catastrophic cooling such as nebular C~IV $\lambda1550$ emission is indeed seen to be
associated with the extremely young SSC in Mrk~71-A \citep{Oey2023}, 
which additionally exhibits
slow shell expansion velocities of $\leq 10 \rm~km~s^{-1}$ \citep{Oey2017, Komarova2021}; as mentioned earlier, this object is 
the nearest Green Pea analog and LCE candidate. Moreover, \cite{Pascale2023} present evidence for catastrophic cooling in the LyC-leaking knot of the Sunburst arc, where \cite{Mainali2022} report a wind component in nebular lines. \cite{Smith2023} suggest this knot is analogous to Mrk 71-A. Thus,
radiation-driven feedback, rather than
SN feedback, seems to be the dominant mode of LyC escape 
in the youngest, most extreme starbursts.  

Instead, a new framework is emerging for LyC escape, based on studies of the largest sample of 
confirmed LCEs at low redshift, the LzLCS. There appear to be two classes of LCEs: 
one sub-sample shows relatively older ages, lower ionization, and higher metallicity \citep{Flury2022b}.
The second group is characterized by extremely young ages, high $O_{32}$, and low metallicity, and on average higher \fesc~than the first group. These properties correspond to the radiation-dominated feedback described above. Indeed, stacking analysis of the LzLCS, in combination with 23 additional LCEs from the literature, shows that, while mechanical feedback is enhanced in LCEs compared to non-LCEs, the strongest leakers are dominated by $<3$~Myr populations, too young for SN dominance \citep{Flury2024}. 
There is thus multifaceted evidence to support a new paradigm in which there are two regimes 
promoting the escape of ionizing radiation from starbursts:
one driven by SN feedback, and another by radiation-driven feedback.

Our study is consistent with this dual-mode paradigm for LyC escape. We suggest an additional way of distinguishing these two feedback modes by the morphology of the broad, $\lesssim 800\rm~km~s^{-1}$ emission-line wings that directly trace high-velocity ionized gas 
in very young, metal-poor systems. We present evidence showing that radiation-driven superwinds
may be characteristic of the radiation-dominated mode, and are likely a key LyC escape mechanism in 
such objects as GPs, which are extremely young, highly ionized, and metal-poor starbursts. 
Conventional, SN-driven feedback still likely promotes LyC escape in older, more metal-rich populations.
It remains to be determined whether superwinds generated by mechanical feedback can exhibit power-law nebular line wings, and if so, under what circumstances.  But it is likely to be at significantly older ages in metal-poor galaxies \citep{Jecmen2023}.
In addition, prior SN feedback could still be important in reshaping the ISM and promoting LyC escape, 
even for the radiation-driven mode \citep{Flury2024}. 

Radiation-driven superwinds may in fact be common at higher redshifts, where more starbursts are
super-Eddington, promoting LyC and/or Ly$\alpha$ escape \citep{Ferrara2024}. Such feedback may explain ``blue monsters'' at $z > 10$, where radiation-driven outflows clear dust efficiently \citep{Ferrara2023, Ziparo2023}. These massive early galaxies are predicted to have high star-formation efficiencies and thus 
lack mechanical feedback
due to their free-fall times ($\sim 1$~Myr) being shorter than the time required to develop stellar winds and SNe \citep{Li2024, Dekel2023}.

It is important to note that LyC escape may still occur 
via feedback modes not discussed in this work,
such as in the case of the extreme LCE J1316+2614 with \fesc$\sim90\%$ \citep{Marques-Chaves2024}, which shows no signatures of outflows. Instead, the efficient LyC escape may be explained by the high star-formation efficiency of $\geq 70\%$ in this object
leading to density-bounded ionization.
 
\subsection{Green Pea-Like Galaxies in the EoR}
Our findings linking radiation-dominated feedback to LyC escape point to the possibility of a
key role for GP-like galaxies in cosmic reionization. GPs are in fact defined by their compactness and exceptionally high ionization parameters \citep{Cardamone2009, Amorin2012GTC, Fernandez2022}, which are directly linked to radiation-dominated gas kinematics \citep{Yeh2012}.

The first results from JWST are revealing that such extreme starbursts are abundant in the early universe \citep{Rhoads2023, Mascia2023, Sanders2023, Simmonds2024a, Simmonds2024c, Saxena2024}, with even higher ionizing photon production efficiencies and lower metallicities \citep{ArellanoCordova2022, Schaerer2022, Matthee2023, Llerena2024, Castellano2024, Topping2024metal}, as well as lower dust content \citep{Cullen2024}, than observed at low redshifts \citep{Schaerer2022}. In addition to enhanced ionizing photon production efficiencies, a stack of 1000 galaxies at $z = 4-10$ shows multiple signatures of strong LyC escape \citep{Hayes2024}. With higher ionization but delayed SN feedback due to the extremely low metallicities, these high-redshift galaxies likely have radiation-dominated stellar feedback.
Indeed, broad emission-line wings have been detected in some of these objects with NIRCam wide field slitless spectroscopy and JWST/NIRSpec MSA spectroscopy \citep{Xu2023, Matthee2023, Carniani2024, Saldana2025}.

\section{Conclusion}
\label{sec: conclusion}

Local LyC emitters, and in particular the most extreme starbursts such as Green Peas, exhibit high-velocity gas reaching $800\rm~km~s^{-1}$ {in HWZI, observed as broad wings in emission lines of \OIII\ $\lambda5007$ and H$\alpha$. The velocity of this gas increases with the LyC escape fraction \citep{Amorin2024}, suggesting a direct link between stellar feedback and LyC leakage. 
These broad line wings may be able to identify the presence
of radiation-driven superwinds, as seen in the nearest Green Pea analog Mrk~71-A, in which LyC photons impart their momentum on tiny, dense neutral knots, and a fraction of LyC escapes 
through their low filling factor \citep{Komarova2021}.
We use our sample of 20 galaxies from the LzLCS survey \citep{Flury2022a} and 6 GPs from the literature \citep{Izotov2016a, Izotov2016b, Izotov2018a, Izotov2018b} 
to determine the origin of the feedback traced by the broad emission-line
wings.  In particular, we test whether radiation-driven winds like that in Mrk~71-A are also responsible for the broad emission-line wings observed in these reionization-era analogs and whether they indeed promote LyC escape.

With spectroscopy from Magellan/MIKE, VLT/X-shooter, and WHT/ISIS, we determine the functional form or morphology of the broad-wing profile in \OIII\ $\lambda$5007, as well as the wing
maximum velocity, luminosity, systemic velocity, and other kinematic parameters,
in each galaxy. We assume that the broad-wing morphology traces the dominant feedback regime in each starburst.  As noted above, radiation-driven superwinds produce line wings with power-law or exponential profiles \citep{Krumholz2017, Komarova2021}, whereas Gaussian profiles in young objects may be observed from unresolved regions dominated by mechanical feedback from SNe, where a large number of kinematic components obeys the central limit theorem \citep{Chu1994}. 

We find that 14 out of the 26 sets of broad wings in our sample are better fit by power laws with slopes of $-3.5$ to $-1.6$.  Four other galaxies show Gaussian wings, and the remaining eight have ambiguous wing morphologies. We see that 
both Gaussian and power-law wing morphologies are found at lower $O_{32} < 3.5$, higher metallicities $12+\log(\rm O/H) > 8.0$, and moderate EW(H$\beta$) $ < 75$~\AA.
However, the galaxies with higher $O_{32}$, lower $12+\log(\rm O/H) $, and higher EW(H$\beta$) all show non-Gaussian, power-law wings. In these extreme systems, such indicators of radiation feedback
are linked to both the observed galaxy-wide properties and those of the broad-wing profiles such as wing luminosity and $V_{\rm max}$.
Objects with ambiguous wing morphologies show properties intermediate between those of the Gaussian and power-law classes.

We perform Kendall rank correlation tests for broad-wing parameters and galaxy properties in the power-law group. We find that the normalized wing luminosity at $\geq300~\rm~km~s^{-1}$ $L_{\rm \geq300}/L_{\rm tot}$ increases with the light fraction of stars younger than 3~Myr, $f_*(t < \rm 3~Myr)$, suggesting that these youngest populations are responsible for radiation driving of the power-law wings, since the onset of SNe is not expected for ages $<3$~Myr, even at solar metallicity. 
Moreover, the power-law slope $\alpha$ is
shallower for higher $f_*(t < \rm 3~Myr)$, pointing to faster winds in younger populations.
We also find that the normalized wing luminosity and $\alpha$ 
both correlate with the observed LyC luminosity $L_{\rm LyC, obs}$ but not the LyC escape fraction \fesc.  This suggests that {\it tracers of radiation-driven winds depend more on the driving UV luminosity than the covering fraction.}  
This is consistent with the wind and LyC photons emerging through the same optically thin channels in a picket-fence geometry \citep{Heckman2001, Heckman2011, Gazagnes2018, Steidel2018, Jaskot2019, Gazagnes2020, Saldana-Lopez2022}. It also supports the scenario that radiation-driven superwinds modulate the LyC escape in the galaxies with power-law emission-line wings, and that
higher LyC luminosities can accelerate faster winds.

We find that for galaxies with power-law slopes, the value of $\alpha$ appears to be determined not only by the intrinsic wind velocity profile as noted above, but also by extinction.
We find that \alphared\ is systematically shallower than \alphablue, with $\Delta\alpha\sim 0.5$ to 1.0, with \alphared\ becoming flatter for higher extinction.
This implies a centrally concentrated dust distribution, either one that is spatially extended and strongly decreasing with radius,
or a compact central dust zone limited to a smaller radius. Both of these configurations lead to preferential attenuation of the lower-velocity redshifted wind emission originating at smaller radii, which flattens the red slope. 
Additionally, \alphablue\ trends with galaxy scaling relations, being flatter for galaxies that are, e.g., more metal-rich, and have lower $O_{32}$, SFR, EW(H$\beta$) and EW(Ly$\alpha$).  
The correlation of \alphablue\ with absorbed LyC and extinction-linked properties is consistent with the first dust model and not the second.  

Thus, both intrinsic wind structure and extinction appear to drive the observed values of \alphablue\ and \alphared, with extinction probing the geometry of the central, dusty region with respect to the wind velocity profile.
We stress that extinction does {\it not} correlate with LyC escape in our data.  Instead, there is a multi-dimensional relation whereby for a given \fesc, $\alpha$ flattens with extinction, while still showing trends of increasing LyC escape and flatter $\alpha$ with lower extinction.  Our data remain consistent with previous findings that extinction anticorrelates with LyC escape.

We show that observed velocities in excess of $300 \rm~km~s^{-1}$ are difficult to explain with dust opacity. Therefore, these winds are most likely accelerated by LyC and Ly$\alpha$ photons,
as seen in Mrk 71-A \citep{Komarova2021}. 
The weak correlation between $f_{\rm{esc}}^{\rm Ly\alpha}$ and power-law slope suggests that LyC, rather than Ly$\alpha$, radiation may dominate this process.  This is reasonable, since Ly$\alpha$ is a resonant line that quickly rescatters, diluting its directional impact.

\textcolor{black}{Using new Ly$\alpha$ imaging of a sub-sample of our galaxies
from the LaCOS survey \citep{LeReste2025}, we link the spatially resolved Ly$\alpha$ properties to broad-wing parameters. We find tentative evidence linking Ly$\alpha$ morphology to the dominant feedback mode in our considered LCEs. Objects with higher Ly$\alpha$ halo fractions show lower $V_{\rm max}$ and Gaussian wing profiles. This is consistent with older, SN-dominated systems with lower LyC escape, where gas is dispersed to larger distances, and the central, aged starburst emits relatively less in Ly$\alpha$. In contrast, galaxies with power-law wings show lower halo fractions and higher $V_{\rm max}$, consistent with more compact gas morphologies in the radiation-dominated mode and enhanced LyC escape. We also uncover a 
possible relation 
between the Ly$\alpha$ concentration parameter and 
\alphablue, 
which may be linked to galaxy scaling relations.
Together, these results provide first evidence that Ly$\alpha$ morphology and broad-wing properties may jointly trace the age and feedback mechanisms in LCEs. } 

The galaxies with Gaussian broad wings, associated with higher metallicities and lower $O_{32}$ values, are likely dominated by mechanical feedback from SNe. 
As evidenced by their lower $O_{32}$ values and higher, though still subsolar, metallicities, these objects tend to be somewhat older, and more consistent with such a scenario. 

Overall, our findings are consistent with the bimodal paradigm for LyC escape \citep[e.g.,][]{Flury2024}, now reinforced and clarified by observed emission-line wing profiles. These appear to be a powerful tracer of feedback and LyC escape mechanisms in very young starbursts: power-law wings mostly trace radiation-dominated superwinds, and Gaussian wings trace SN feedback.  For systems with lower ionization parameters and higher metallicity, it remains somewhat unclear whether power-law wings could also result from SN-driven superwinds; in such cases, the timescale for developing a full-fledged superwind is critical.
Another caveat is that our unresolved galaxies likely contain a multiplicity of star-forming regions with varying conditions, which limits the connections we can make between observed feedback properties and specific SSC parameters. These regions likely span a range of stellar ages, ionization parameters, metallicities, and both feedback modes may be present simultaneously. 

We have shown that power-law emission-line wings
have the potential to reveal fundamental parameters for the radiation-dominated conditions and feedback in individual objects.
As JWST is uncovering extremely young, blue, metal-poor starbursts at $z > 6$, follow-up spectroscopic observations of such galaxies 
and their emission-line wings will help reveal the importance of radiation-driven feedback to LyC escape and to cosmic reionization.

\begin{acknowledgements}
    We thank Mark Krumholz and Claus Leitherer for useful discussions. We also thank the anonymous referee for useful suggestions that improved this manuscript. This work was supported by NASA grants HST-GO-16261.001 and HST-GO-17069.003 to M.S.O. 
    R.A. and J.M.V. acknowledge financial support from the Severo Ochoa grant CEX2021-001131-S funded by MCIN/AEI/10.13039/501100011033.  R.A. and J.M.V. acknowledge funding from projects PID2023-147386NB-I00  and PID2022- 136598NB-C32 “Estallidos8”, respectively. M.J.H. is fellow of the Knut \& Alice Wallenberg Foundation. 
    This work is based on observations with the NASA/ESA Hubble Space Telescope, which is operated by the Association of Universities for Research in Astronomy, Incorporated, under NASA contract NAS5-26555. Support for Program numbers HST-GO-16261 and HST-GO-17069 was provided through grants from the STScI under NASA contract NAS5-26555.
        
\end{acknowledgements}

\facilities{HST/ACS,COS, Magellan/MIKE, VLT/X-shooter, WHT/ISIS}
\software{ Astropy \citep{astropy:2022}, {\tt SciPy} \citep{Scipy}, \textsc{lmfit} \citep{Newville2016}, Matplotlib \citep{Hunter:2007}, {\tt kendall} \citep{Flury2022b}.}

\clearpage

\appendix
For each galaxy in the sample, we present our adopted model for the [O~III]$\lambda$5007 broad emission-line wings: Gaussian or power-law. 
Galaxies are classified as having ambiguous morphology if neither of these models is statistically preferred; in these cases, we show both tested models.  See Section~\ref{sec: analysis} for details regarding the fitting procedure.

\section{Gaussian Wings}
\label{sec:app_gauss}
In Figure~\ref{fig:gauss_fits}, we show our adopted emission-line wing models and parameters for the four galaxies classified with Gaussian broad wing morphologies: J012910+145935, J091113+183108, J105331+523753, and J131419+104739. These models consist of Gaussian core(s) and Gaussian wings fitted simultaneously. 

\begin{figure*}[h]
    \centering
    \gridline{\fig{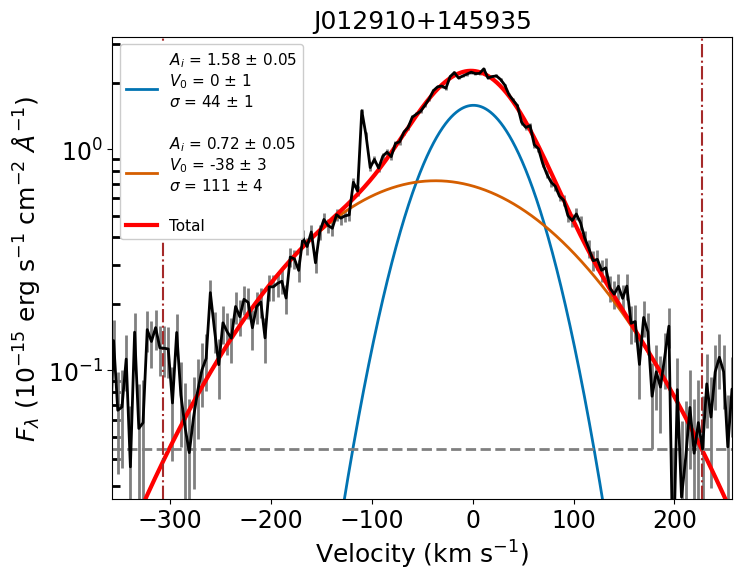}{0.48\textwidth}{}
              \fig{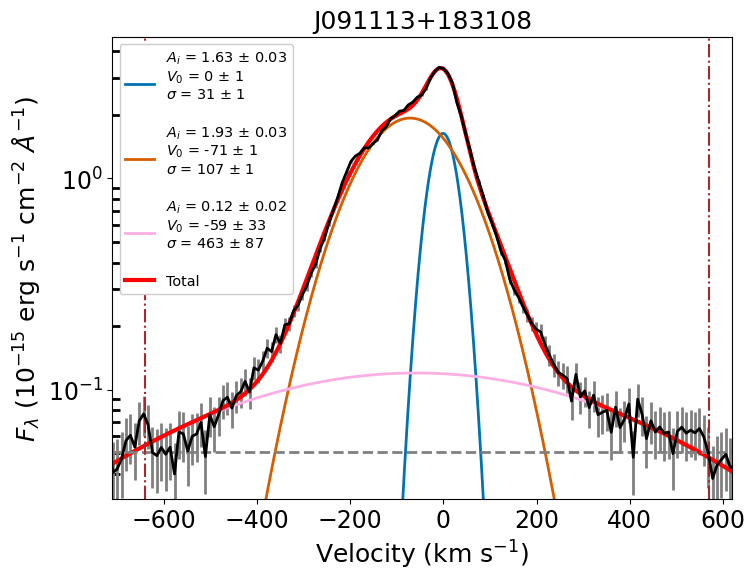}{0.48\textwidth}{}}
    \gridline{\fig{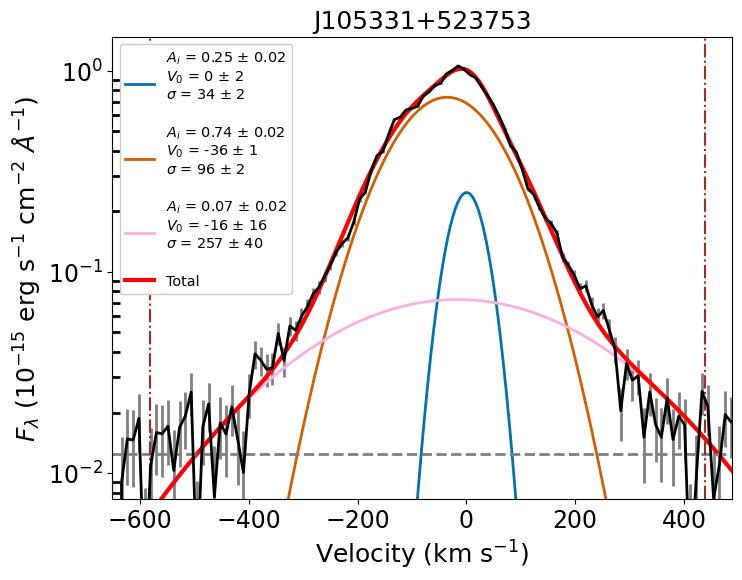}{0.48\textwidth}{}
              \fig{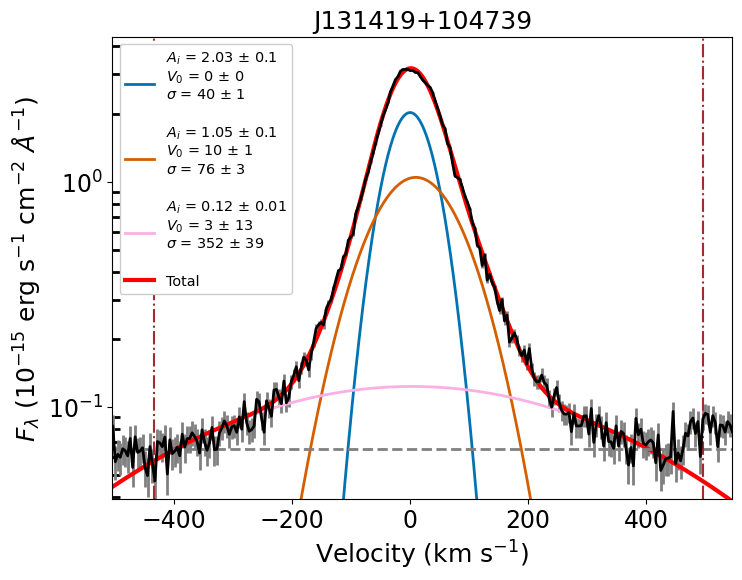}{0.48\textwidth}{}}
    \caption{Multi-Gaussian [O~III]$\lambda$5007 fits for the four galaxies classified as showing Gaussian wing morphologies. The Gaussian parameters are indicated in the legends with the following units: amplitudes $A$ in $10^{-15}$ erg s$^{-1}$ cm$^{-2}$ \AA$^{-1}$, centroids $V_0$ and widths $\sigma$ in km~$\rm s^{-1}$. The best-fit local continuum is shown as a gray dashed line, and the maximum wing velocity $V_{\rm max}$ as dash-dotted brown lines.   }
    \label{fig:gauss_fits}
\end{figure*}
\clearpage

\section{Power Law Wings}
In Figure~\ref{fig:PL_A}, we show our adopted wing models for the 14 galaxies classified with power-law morphologies: J003601+003307, J004743+015440, J011309+000223, J012217+052044, J090146+211928, J092532+140313, J095838+202508, J101138+194721, J113304+651341, J115449+244333, J124835+123403, J131037+214817, J144231-020952, and J164607+313054. As detailed in Section~\ref{sec: analysis}, the red and blue wings are fitted with separate power-laws after subtracting the narrower Gaussian core(s). 

\begin{figure*}
\centering
\gridline{\fig{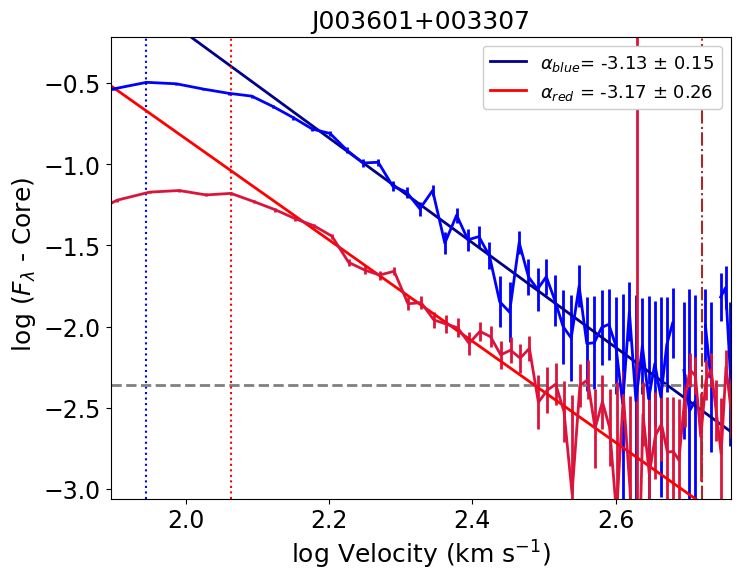}{0.48\textwidth}{}%
          \fig{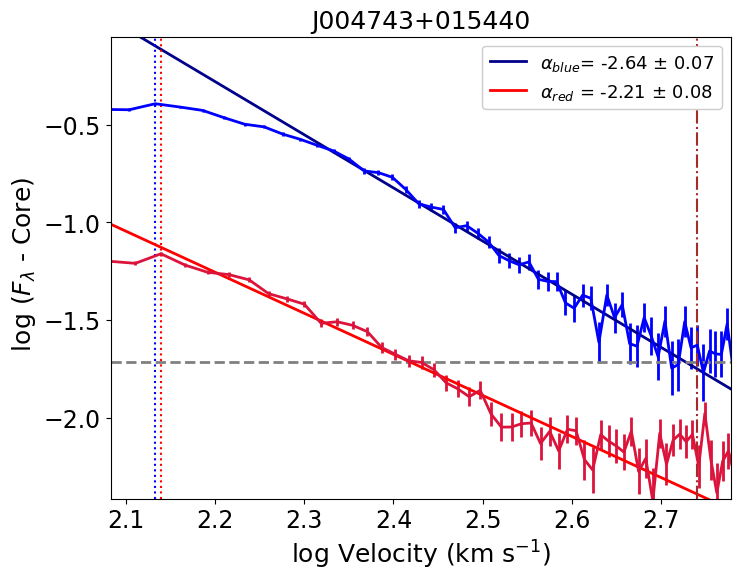}{0.48\textwidth}{}}
\gridline{\fig{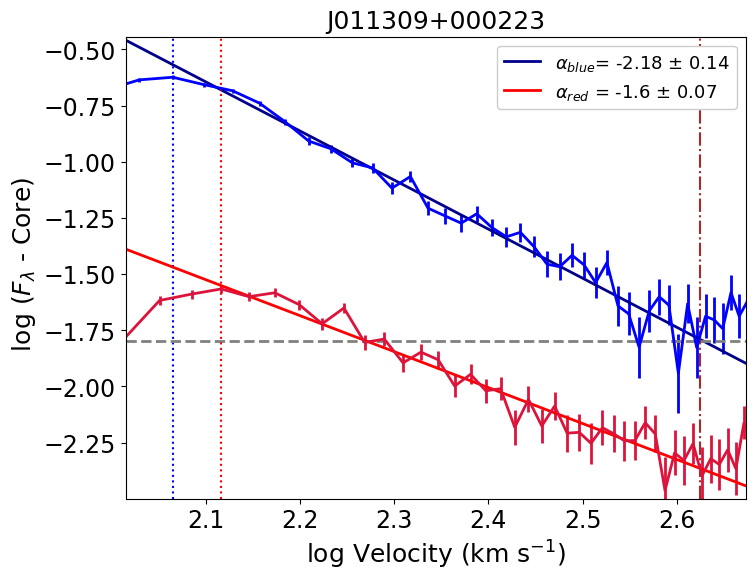}{0.48\textwidth}{}%
          \fig{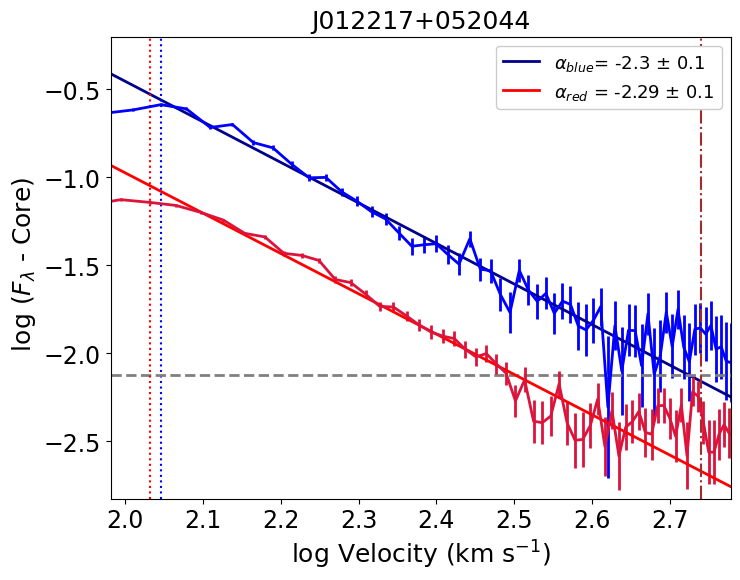}{0.48\textwidth}{}}
\gridline{\fig{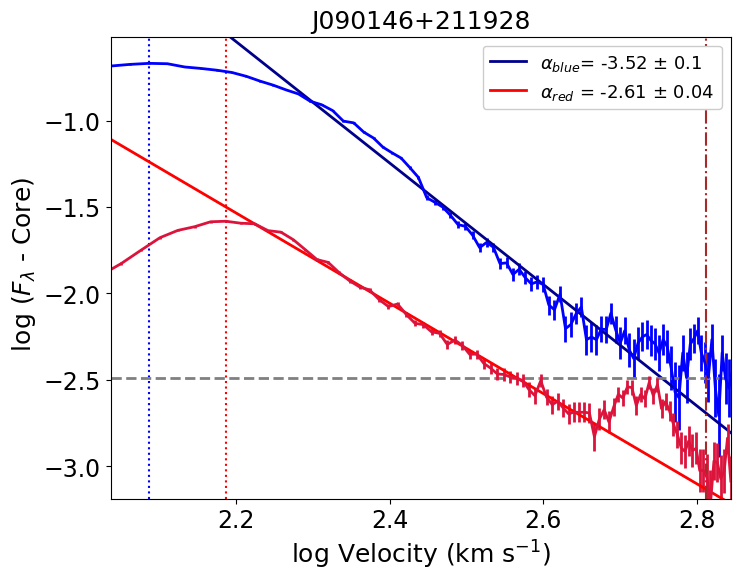}{0.48\textwidth}{}%
          \fig{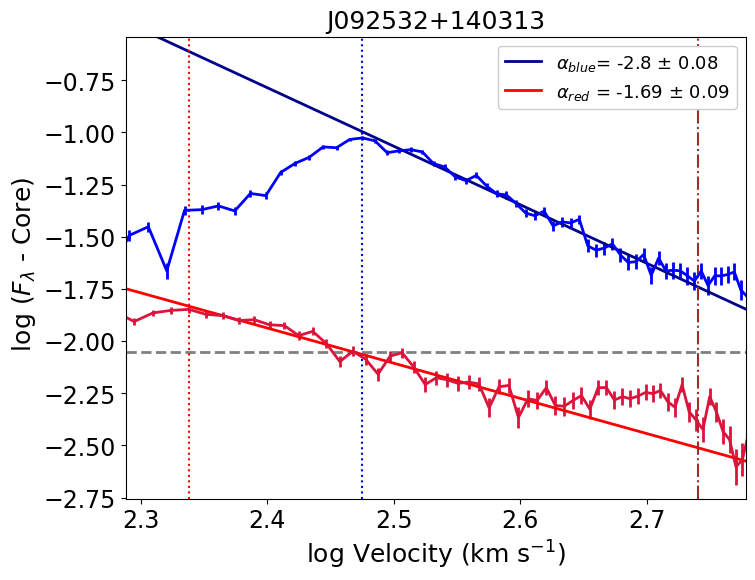}{0.48\textwidth}{}}
\caption{Power-law [O~III]$\lambda$5007 fits for galaxies classified with power-law wing morphologies. The blue and red wing fits are shown with their corresponding colors, and the respective best-fit slopes are shown in the legend. The red wing is arbitrarily offset 
on the $y$-axis for display purposes. The best-fit local continuum is shown as a gray dashed line, and the maximum wing velocity $V_{\rm max}$ as dash-dotted brown lines. }
\label{fig:PL_A}
\end{figure*}
\clearpage

\setcounter{figure}{19}
\begin{figure*}
\centering
\gridline{\fig{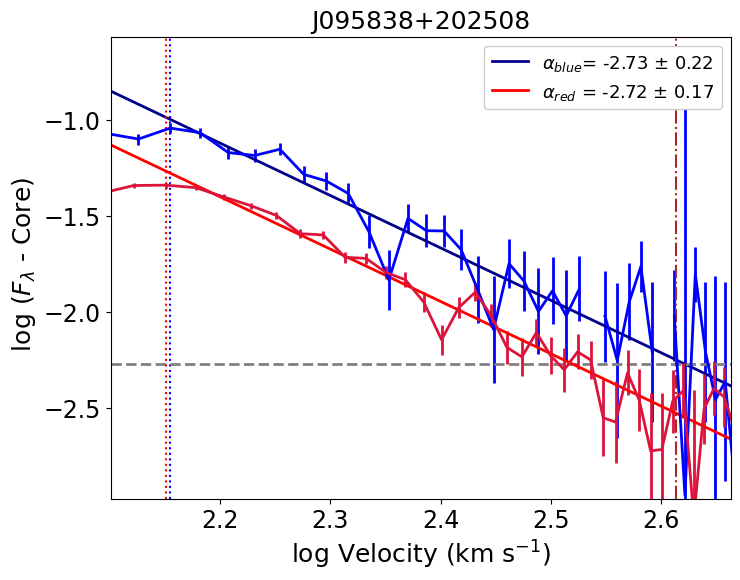}{0.48\textwidth}{}%
          \fig{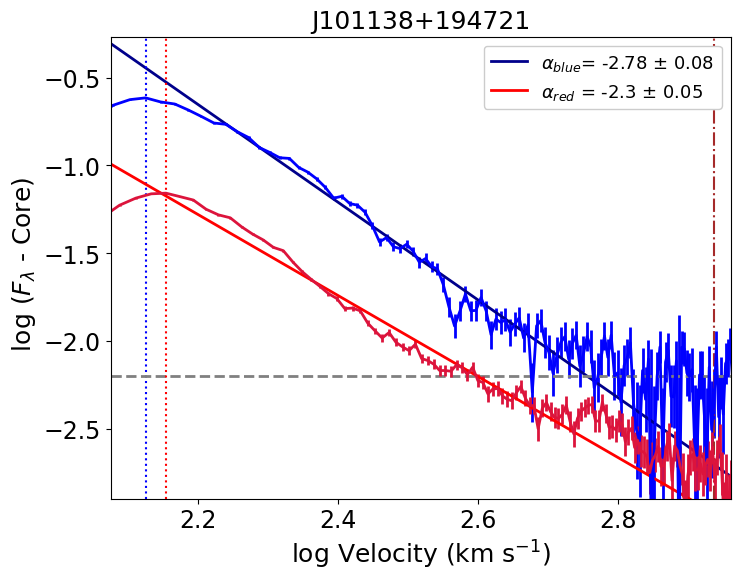}{0.48\textwidth}{}}
\gridline{\fig{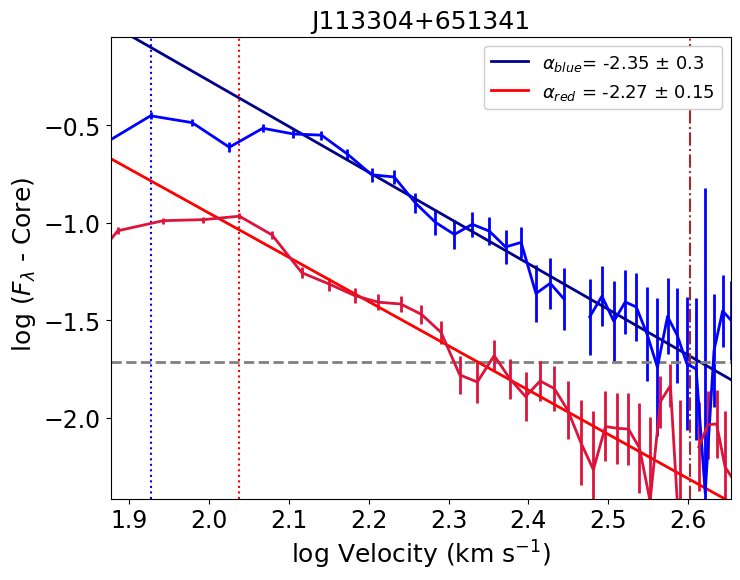}{0.48\textwidth}{}%
          \fig{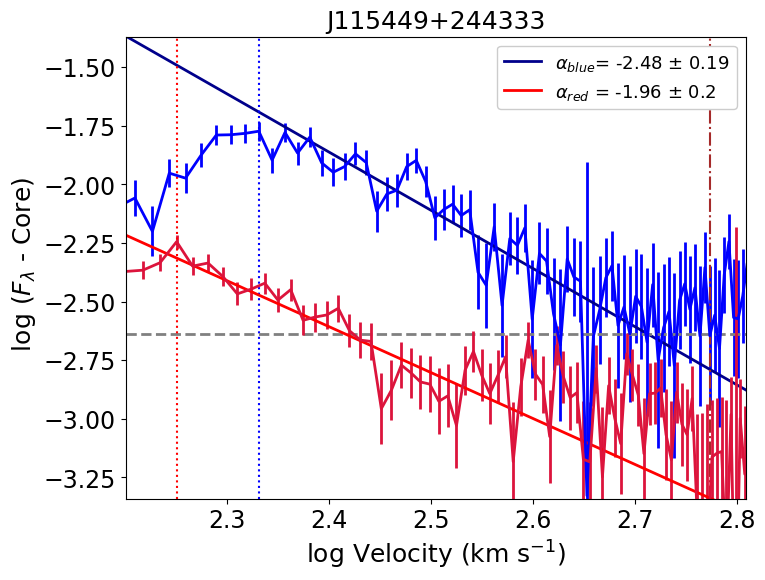}{0.48\textwidth}{}}
\gridline{\fig{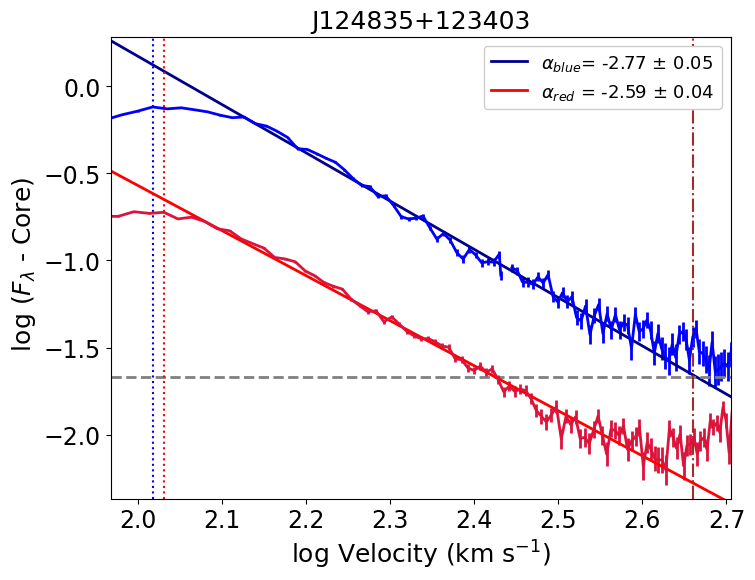}{0.48\textwidth}{}%
          \fig{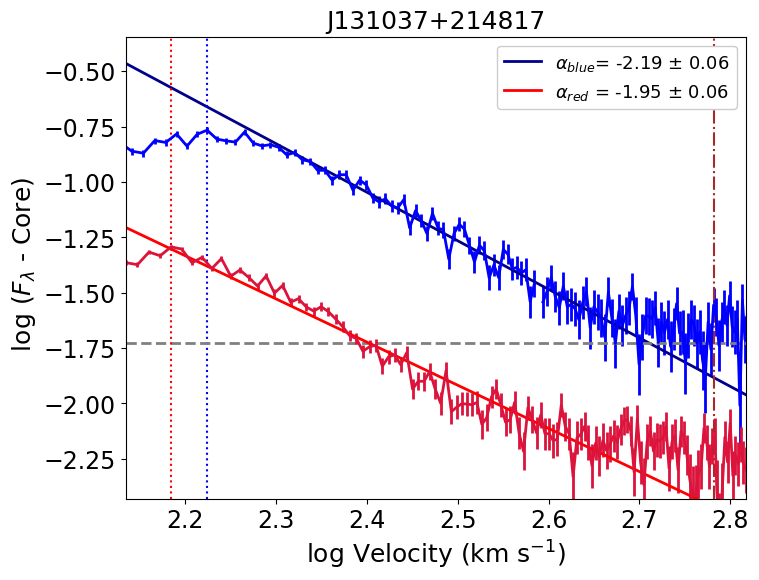}{0.48\textwidth}{}}
\caption{Continued.}
\end{figure*}
\clearpage

\setcounter{figure}{19}
\begin{figure*}
\centering
\gridline{\fig{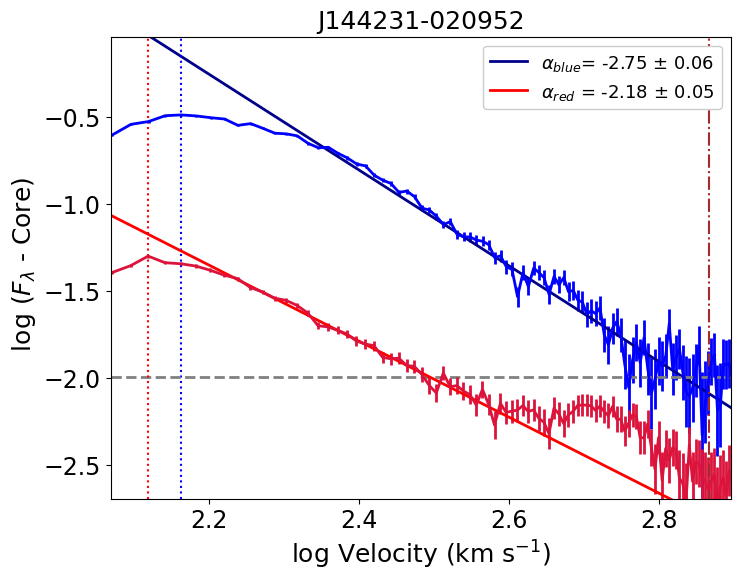}{0.48\textwidth}{}%
          \fig{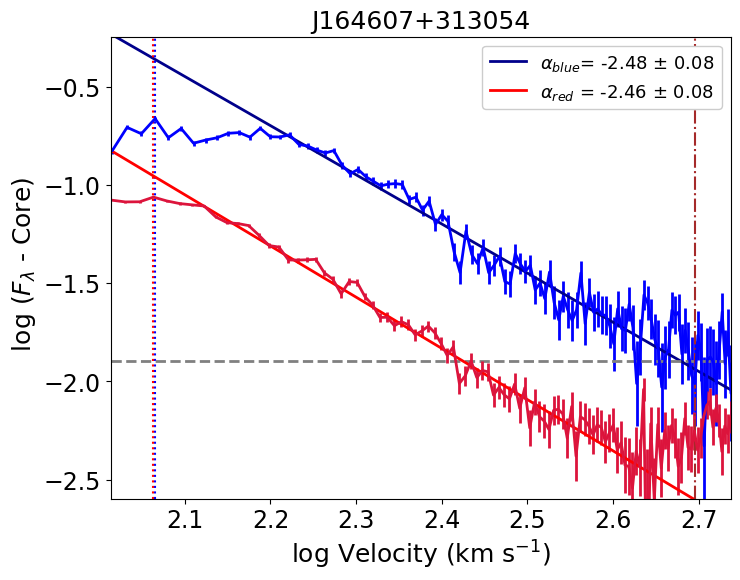}{0.48\textwidth}{}}
\caption{Continued.}
\end{figure*}

\clearpage

\section{Ambiguous Wing Morphologies}
\label{sec:App_ambig}
In Figure~\ref{fig:ambig_A}, we show the tested multi-Gaussian and power-law wing models for the eight galaxies classified with ambiguous wing morphologies: J081409+211459, J091703+315221, J115205+340050, J115855+312559, J123519+063556, J124423+021540, J134559+112848, and J144010+461937. 
Neither model is ultimately adopted for the shown objects, due to lack of statistically significant differences in the fit quality for the broad-wing region (Section~\ref{sec: analysis}). 

\begin{figure*}
\centering
\gridline{\fig{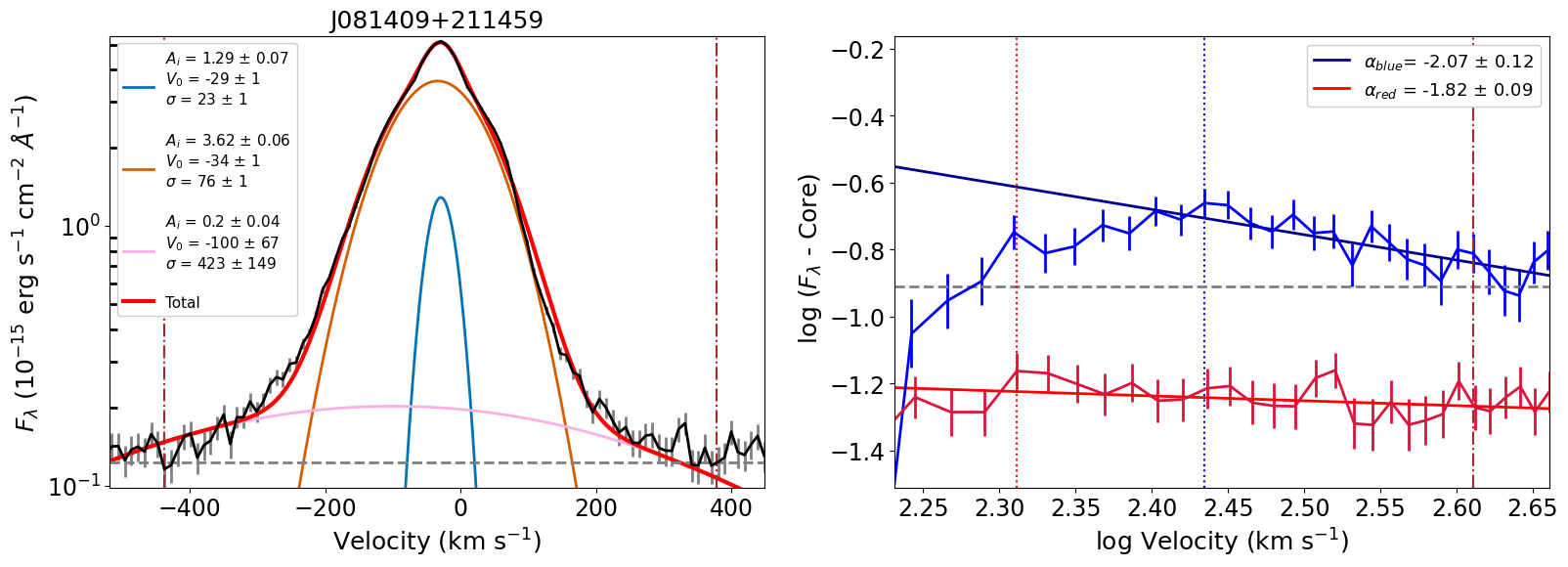}{0.99\textwidth}{}}
\gridline{\fig{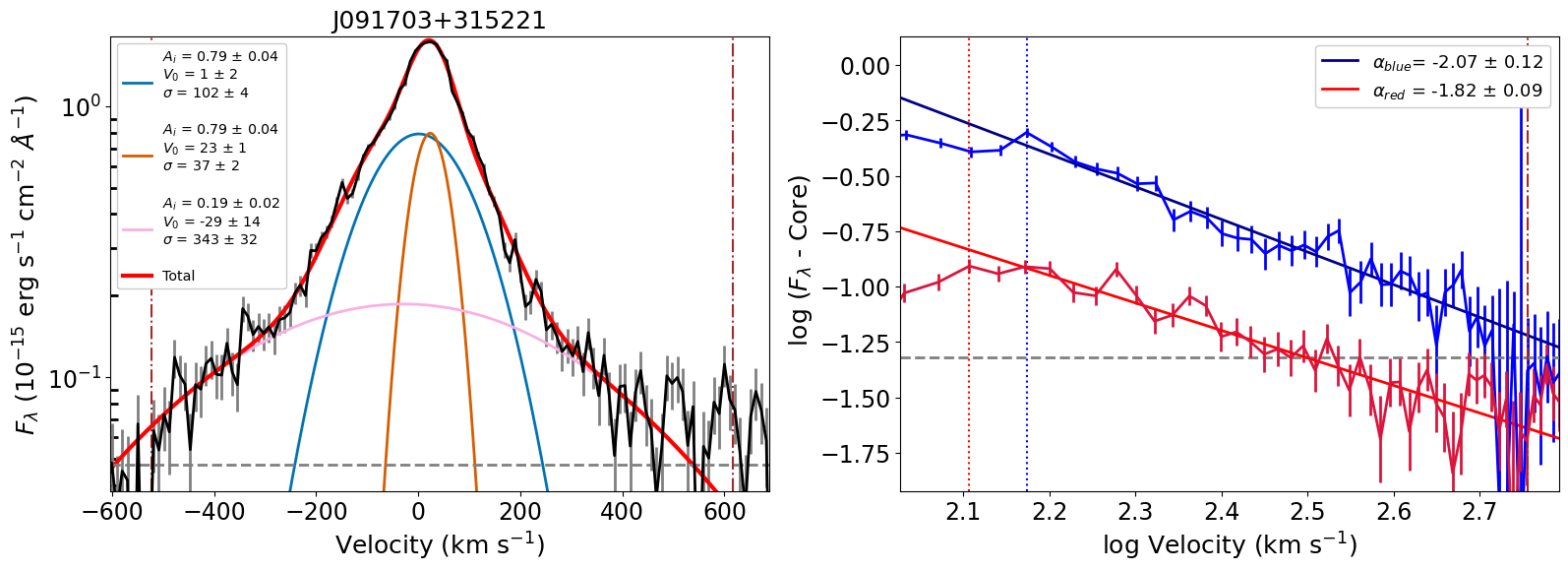}{0.99\textwidth}{}}
\gridline{\fig{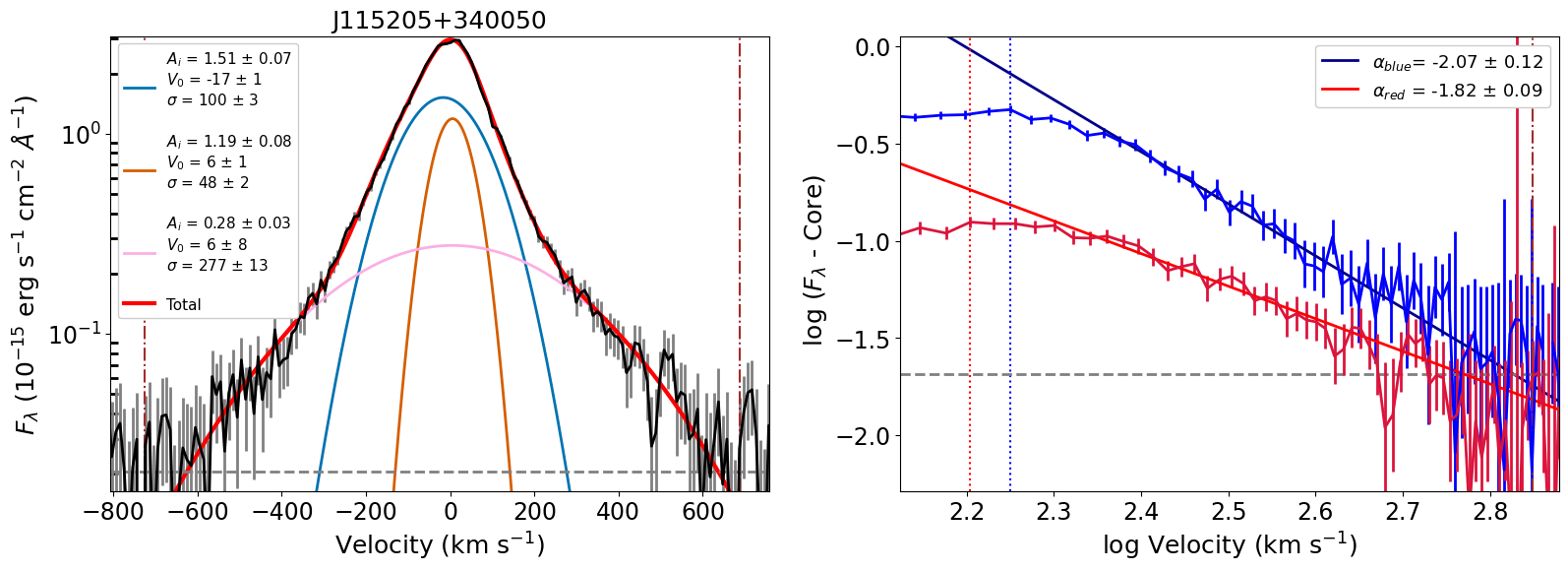}{0.99\textwidth}{}}
\caption{Power-law and multi-Gaussian [O~III]$\lambda$5007 fits for galaxies classified with ambiguous wing morphologies. The Gaussian parameters shown in the legend are in the following units: amplitudes $A$ in $10^{-15}$ erg s$^{-1}$ cm$^{-2}$ \AA$^{-1}$, centroids $V_0$ and widths $\sigma$ in km~$\rm s^{-1}$. The blue and red wing power-law fits are shown with their corresponding colors, and the respective best-fit slopes are shown in the legend. The red wing is offset to arbitrarily lower fluxes for display purposes. The best-fit local continuum is shown as a gray dashed line, and the maximum wing velocity $V_{\rm max}$ as dash-dotted brown lines. }
\label{fig:ambig_A}
\end{figure*}
\clearpage

\setcounter{figure}{20}
\begin{figure*}
\centering
\gridline{\fig{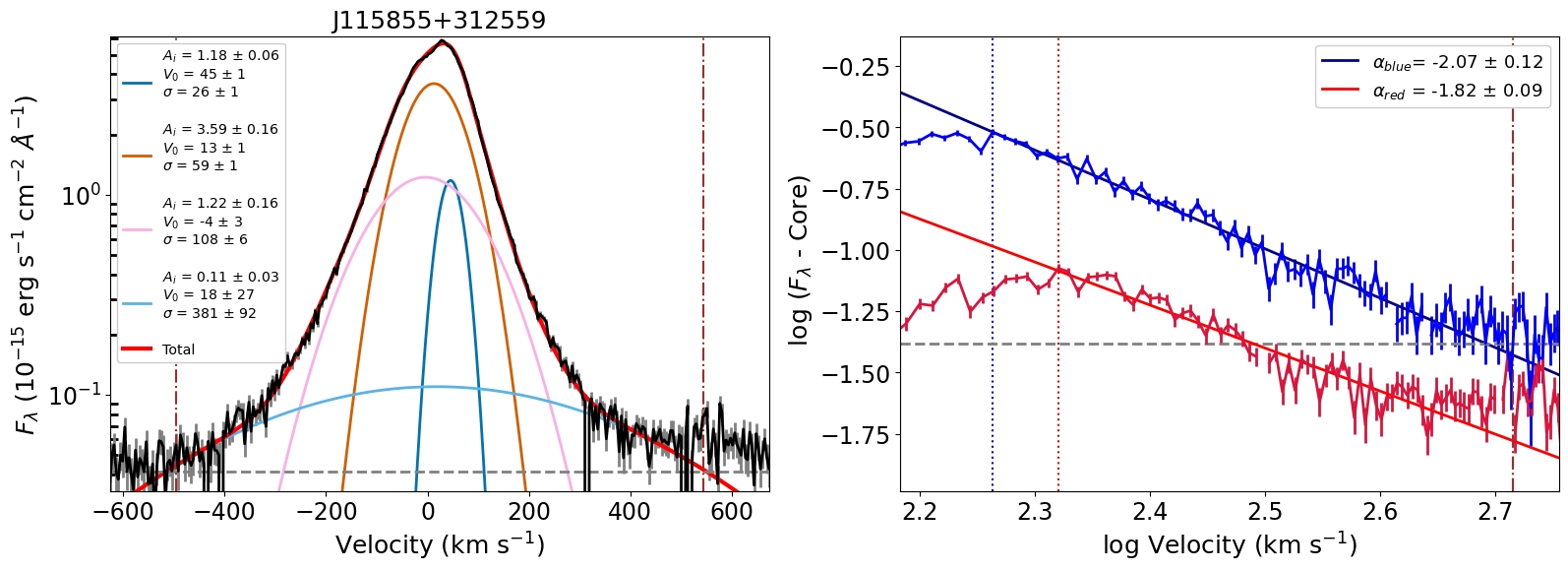}{0.99\textwidth}{}}
\gridline{\fig{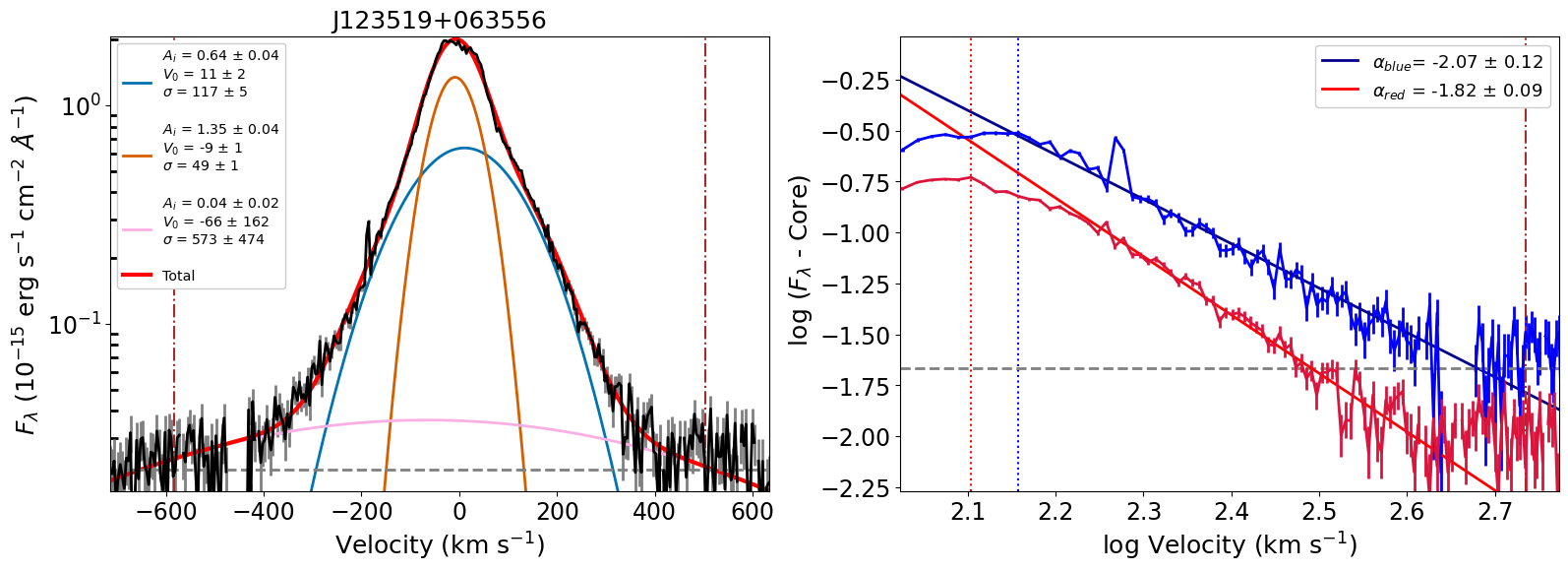}{0.99\textwidth}{}}
\gridline{\fig{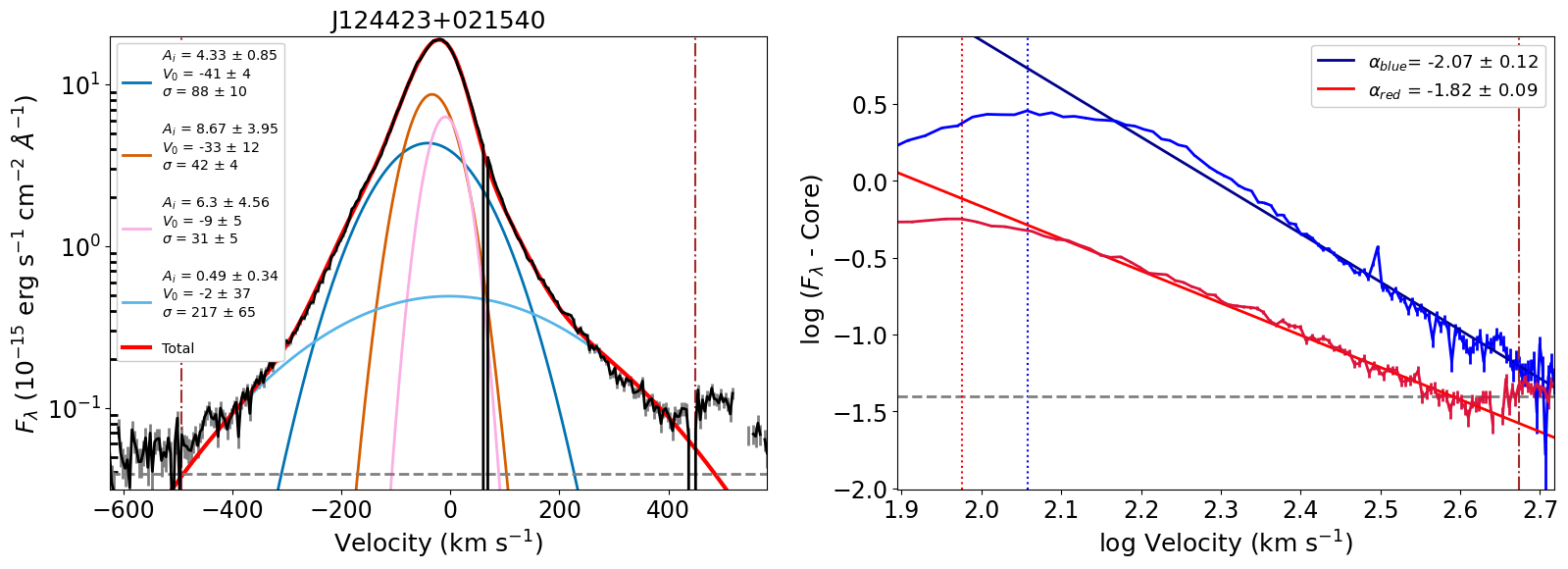}{0.99\textwidth}{}}
\caption{Continued.}
\end{figure*}
\clearpage

\setcounter{figure}{20}
\begin{figure*}
\centering
\gridline{\fig{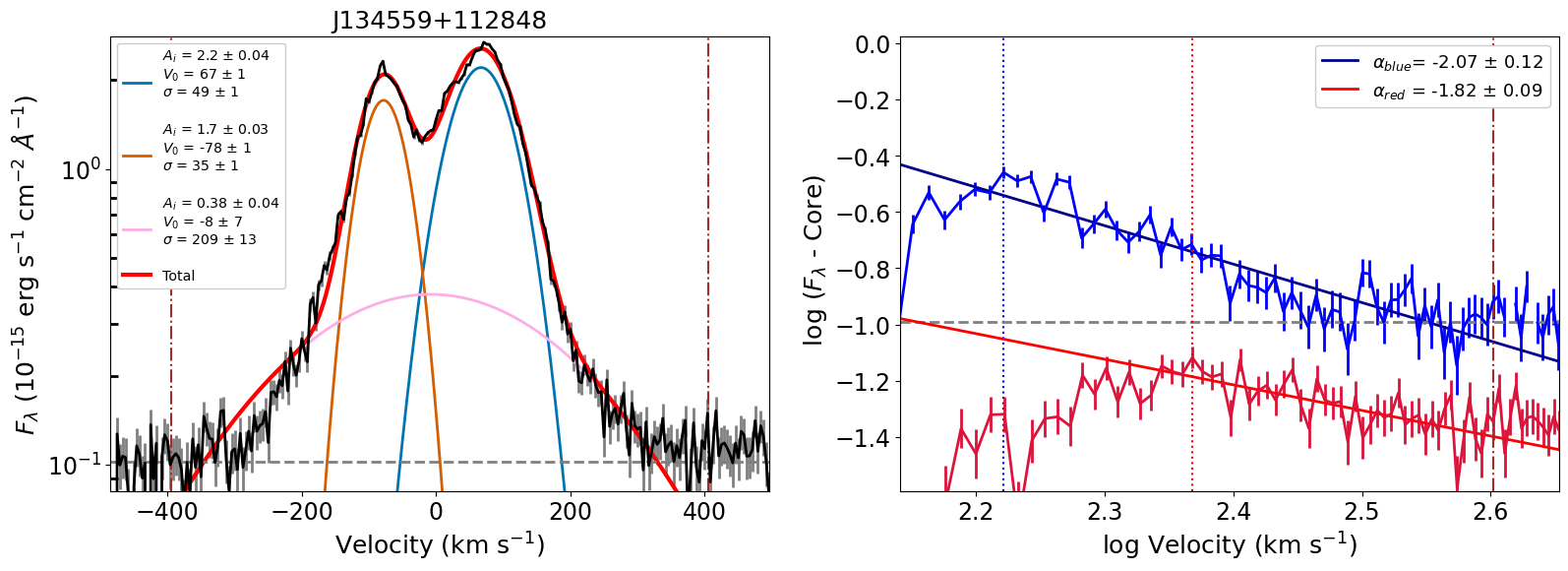}{0.99\textwidth}{}}
\gridline{\fig{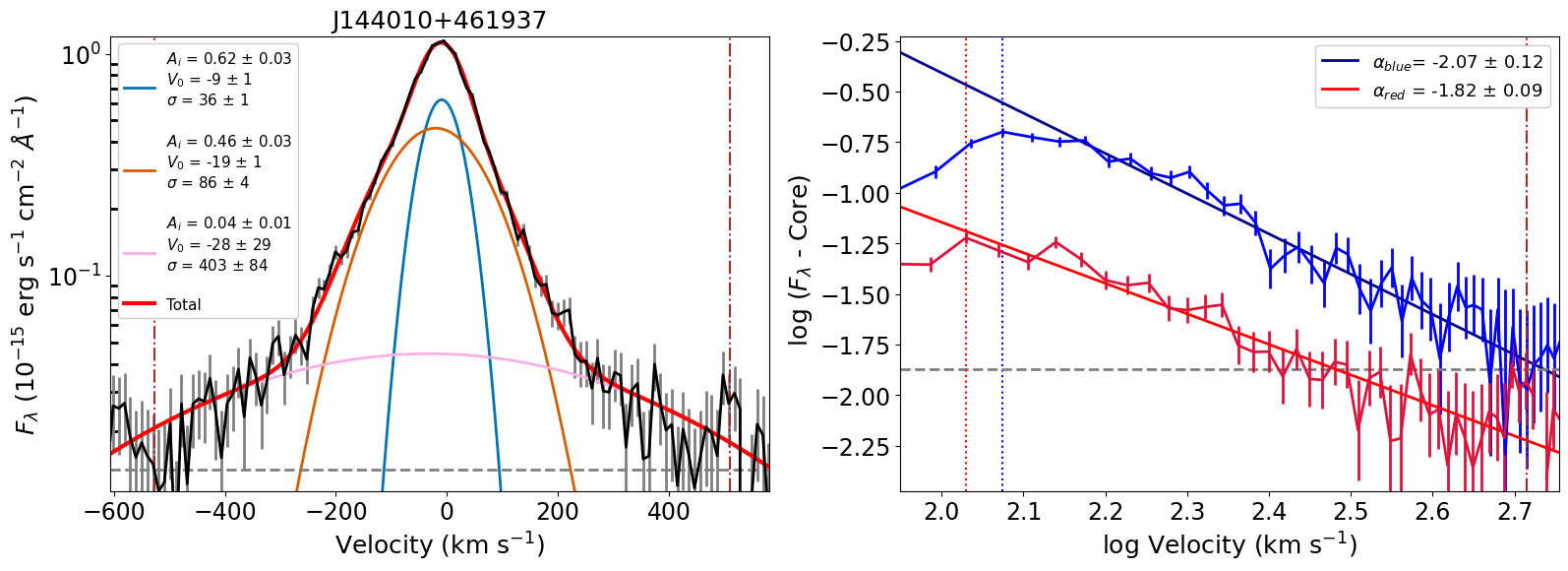}{0.99\textwidth}{}}
\caption{Continued.}
\end{figure*}

\bibliographystyle{aasjournal}

\end{document}